 \documentclass[final,5p,times,twocolumn]{elsarticle}

\usepackage{amssymb}
\usepackage{lipsum}

\usepackage{amsmath}
\usepackage[most]{tcolorbox}
\usepackage{xcolor}
\usepackage{subcaption}
\usepackage{placeins}
\usepackage{multirow}
\usepackage{cuted}
\usepackage{tikz}
\usetikzlibrary{arrows.meta, shapes.geometric, positioning}
\usepackage[breaklinks=true]{hyperref}

\usepackage{cleveref}

\newtcolorbox{important}[1][]{
  colback=red!5!white,
  colframe=red!75!black,
  coltitle=red!75!black,
  fonttitle=\bfseries,
  title=\Large\textbf{$\blacktriangle$ Important},
  sharp corners,
  boxrule=0.8pt,
  #1
}
\newcounter{bla}

\usepackage{lineno}

\journal{High Energy Astrophysics}

\begin{document}

\begin{frontmatter}

\title{SoPlasmaFoam: an OpenFOAM-based solver for streamer and dielectric barrier discharges with adaptive mesh refinement}

\author[label1]{Rention Pasolari\corref{cor1}}
\ead{r.pasolari@certh.gr} 
\author[label2,label1]{Konstantinos Kourtzanidis}

\address[label1]{Advanced Renewable Technologies for Energy \& Materials Integrated Systems (ARTEMIS) Laboratory, Chemical Process \& Energy Resources Institute (CPERI), Centre for Research \& Technology, Hellas (CERTH), 57001, Thessaloniki, Greece}
\address[label2]{Department of Mechanical Engineering, University of Western Macedonia, 50100, Kozani, Greece}

\cortext[cor1]{Corresponding author.}

\begin{abstract}
This work presents \texttt{SoPlasmaFoam}, an open-source, multi-region plasma–dielectric solver built on OpenFOAM and integrated with the PETSc linear-algebra suite (with CPU and GPU back-ends), the blastAMR adaptive-mesh-refinement library (which, unlike Cartesian-only AMR frameworks, operates on hexahedral and arbitrary polyhedral meshes and can therefore conform to curved surfaces and complex geometries), and the ROUND family of high-resolution convective schemes. It solves the drift-diffusion-reaction transport equations for charged species, coupled self-consistently to the Poisson equation either explicitly or through a semi-implicit formulation, with plasma and dielectric regions joined by a monolithic multi-domain coupling that supports an arbitrary number of curved dielectric interfaces. Beyond presenting the solver, this work makes three contributions of broader use to plasma modeling. First, a systematic assessment of convective flux schemes on a stiff scalar-advection problem and on the positive-streamer benchmark shows that the Scharfetter-Gummel scheme is highly stable but excessively diffusive, over-predicting the field and propagation speed on coarse meshes, while the ROUNDF scheme outperforms all tested TVD limiters and is recommended for streamer transport. Second, an analysis of the Poisson-transport coupling demonstrates that the number of fixed-point correction loops per time step critically controls accuracy, that a semi-implicit Poisson formulation does not by itself remove this requirement, and that the coupling must be tightened even when the convective Courant number and the dielectric-relaxation ratio are well below unity. Third, we introduce a drift-robust wall boundary condition that acts directly on the discretized matrix coefficients and remains accurate in the drift-dominated limit, where the conventional mixed-boundary mapping fails. The solver is validated against a low-pressure DC glow discharge and the positive-streamer benchmark, and its multi-region capability is demonstrated on a nanosecond surface dielectric barrier discharge actuator, capturing streamer propagation and surface charging along the dielectric. A performance analysis confirms the expected memory-bound behavior of a finite-volume code, with good single-node scaling, and shows that with adaptive mesh refinement the solver is competitive with the fastest reported plasma codes on the streamer benchmark. The framework provides a modular foundation for future multiphysics simulations in emerging applications, including plasma-assisted combustion of alternative fuels, plasma processing/conversion and subsonic/supersonic plasma-based flow control.  

\end{abstract}


\end{frontmatter}

%% main text
\clearpage
\section{Introduction} \label{introduction}

Non-thermal plasmas (NTPs) are partially ionized gases characterized by a strong thermodynamic non-equilibrium. While the electrons reach high temperatures (on the order of several $\text{eV}$), the bulk gas typically remains near or slightly above ambient temperature and in any case orders of magnitude below the electron temperature. This unique property enable reaction pathways in NTPs far from the thermodynamic equilibrium, generating for example highly reactive species (e.g. $O$, $OH$, and $NO$), radicals, ions and dissociation products in environments that cannot withstand high temperatures or under conditions that would otherwise require external thermal energy input.

During the last few decades, NTPs have evolved to essential industrial tools while showing promise in various emerging and urgent research fields. They have been used in fields such as medicine \cite{Thomas2024}, biology \cite{Sreedevi2023}, aerodynamic flow control \cite{Peng_Avallone_Kotsonis_2024}, sterilization \cite{Svarnas_Sanitaty-effect}, plasma-assisted combustion \cite{Bao2023}, flame stabilization \cite{Barleon2023}, and surface processing and modification \cite{Chiper2023}. Simulation tools are essential towards not only a deep understanding of the underlying physics but also the design and optimization of plasma-based technologies. 

However, the complexity of NTPs, makes their modeling a challenging task. NTPs span a wide range of discharge regimes, each characterized by distinct morphological and kinetic behaviors. The specific state of a discharge depends on a combination of parameters, mainly related to the geometry of the electrodes, the gas composition, and the electrical excitation parameters. Common regimes include diffusive discharges, such as Townsend and glow modes, strongly localized corona discharges with large ion drift zones as well as non-uniform filamentary discharges, like anode and cathode directed streamers and surface ionization waves.

The classification of these regimes is linked to the discharge configuration and the electrical source used. For example, Dielectric Barrier Discharges (DBDs) use dielectric layers to limit the current growth and prevent arc-transition, Surface DBDs (SDBDs) generate surface ionization waves that propagate along the dielectric interfaces, and needle-to-plane configurations are used to generate highly non-uniform fields. Moreover, the choice of excitation scheme, such as DC, AC, microwave or nanosecond pulsed power, affects the energy distribution and chemical reactivity of the plasma. These facts, supplement the inherent multiphysics nature of plasma-based processes and render NTPs extremely complex and diverse, which makes their modeling a challenging task that requires careful consideration of the specific discharge regime, multiphysics coupling and application.The development of robust and efficient numerical solvers for NTP simulations, facilitating a strong-coupling with other physics, is therefore crucial for advancing both fundamental research and practical applications in the field.

\subsection{Computational complexity}

Plasma simulation is uniquely difficult and computational plasma physics remains a field with much left to discover. Unlike pure Computational Fluid Dynamics (CFD), which has been studied for many decades, and has well-established commercial tools, plasma modeling is still highly case-sensitive. Every simulation setup (discharge regime and configuration) is so different, making it hard to create a universal tool. As a result, many research groups still rely on their own developed in-house solvers to meet their specific needs.

To understand these needs, one has to look at the hierarchy of the different modeling approaches exist, which vary significantly in fidelity and computational cost. Starting with the fastest methods but those with the most limited validity, we find analytical models \cite{Orlov2006}, which consist of simple formulas valid only for a specific range of conditions. Moving forward, 0D chemical kinetics models, or global models \cite{ChemPlasKin}, solve balance equations for all species based on their production and loss through chemical reactions. In these models, electron impact reaction rates are typically obtained from Boltzmann solvers. While these models are very fast and can handle detailed chemistry, they assume the plasma is spatially uniform and therefore do not account for transport.

Next are fluid models \cite{Dufour2015}, which solve the moments of the Boltzmann equation, most commonly the first two or three moments, which represent the mass, momentum, and energy equations. These are often solved under the drift-diffusion approximation which result to a system of non-linear continuity equations for each species. Fluid models are significantly faster than the detailed kinetic models described below and can incorporate detailed chemistry while remaining self-consistent with the electric field by solving the Poisson equation. However, their main disadvantage is that they assume particles are in a local equilibrium state due to frequent collisions. This makes them a good fit for collisional plasmas, such as those in atmospheric or near-atmospheric pressure regimes.

Beyond fluid models, statistical Monte Carlo (MC) models \cite{Hagelaar_2000} are highly accurate because they account for non-equilibrium behavior and can be applied to all species. However, they require long calculation times and are typically not self-consistent on their own because they usually track particles in a predefined electric field rather than updating the field based on the particles' motion. This is addressed by Particle-in-Cell / Monte Carlo Collision (PIC/MCC) models \cite{Tian_2026}. Similar to MC, these calculate the electric field through the Poisson equation at every time step. This makes them both accurate and self-consistent, but also very computationally demanding. Hybrid models \cite{Eylenceoglu_2015} offer a balance between accuracy and computational load, by combining these approaches. For instance, using Monte Carlo for fast non-equilibrium species and fluid models for slower species and the electric field. Lastly, full kinetic models involve the direct numerical solution of the Boltzmann transport equation in phase space, providing the most complete physical description at the highest computational cost.

However, even with the faster fluid models (such as the one developed in this work), several significant computational challenges remain. A prime example is the ionization front at the streamer head, which is extremely thin, yet must propagate across gaps that are several orders of magnitude larger than the front thickness itself. This requires a huge number of grid points to capture the steep gradients, which makes simulations computationally demanding. Adaptive Mesh Refinement (AMR) becomes a necessary tool to increase the resolution locally and automatically in the regions of interest. Moreover, in many applications such as DBDs, plasma interacts with dielectric materials, requiring specialized treatment of the charges that accumulate on the interface. Solving the Poisson equation across these multiple regions is a challenge for some computational tools. The Poisson equation can be solved in a partitioned way until the solution on the interface converges, but such a method can be slow. A monolithic solution of both domains is generally faster, but results in larger and more complex matrices.

Temporal constraints are equally demanding, as the physics changes so rapidly that very short time steps are essential for tracking those changes. The time step size is restricted by factors such as the dielectric relaxation time, the speed of chemical reactions, and the electron drift and diffusion velocities. To overcome these stability barriers, semi-implicit Poisson solvers~\cite{crispel2007asymptotic,hagelaar2011model, ventzek1993two} and implicit transport schemes are frequently employed. These solutions must also manage the fact that the transport equations and the electric field are highly non-linearly coupled, necessitating a tight coupling between the Poisson solver and the transport equations at each time step. Small errors in the particle density can distort the electric field, which in turn perturbs the local chemistry and species transport.

The selection of appropriate numerical flux schemes for the divergence of the drift flux is another critical factor in ensuring the accuracy and stability of streamer simulations. The choice is often a balance between numerical stability and the preservation of the steep gradients characteristic of the ionization front. Consequently, a careful consideration of the physical regime is required.

Finally, the transition from 2D to 3D simulations requires immense computational power and high-performance computing (HPC) scaling. Because plasma models involve numerous species and large, complex matrices, solvers must be highly parallelizable across CPU architectures, with a growing need to leverage GPU acceleration for future-proofing. At the hardware level, these fluid models are typically limited by memory bandwidth rather than raw compute power or total capacity. The overhead involved in the communication between the CPU and the memory (storing and accessing fields for many species simultaneously) often creates a bottleneck. To ensure the code is fast enough it must be designed with an awareness of cache efficiency and memory-access patterns to overcome these communication constraints while scaling effectively across modern hardware technologies.

\subsection{Existing numerical frameworks and related modeling studies}

To date, numerous codes have been developed for NTP simulations, ranging from open-source frameworks to proprietary software. Each tool has its own set of advantages and disadvantages, mainly regarding speed, customizability, supported geometries, and the range of cases it can resolve. Here, the most prominent codes currently used in the field are reviewed alongside selected modeling studies that have shaped their development and application. The information provided below reflects the best of the authors knowledge at the time of writing, and is not meant to be an exhaustive review of all available tools, capabilities and studies.
 
Among the open-source frameworks, Afivo-streamer \cite{Teunissen_2017, afivostreamer_repo} is a fast FVM tool for 1D-3D streamer research that utilizes quadtree/octree AMR, a geometric multigrid Poisson solver, and OpenMP-based parallelism. The solver has demonstrated great performance \cite{Bagheri2018} compared to other software, and the same group has produced extensive streamer works, including 3D PIC-MCC simulations in reactive mixtures \cite{Bouwman_2022} and the standard topical review on streamer phenomena \cite{Nijdam2020}. However, Afivo-streamer is restricted to shared-memory parallelism and does not scale across multiple nodes via MPI. In addition, as the developers acknowledge, extending the source code to accommodate new physics is rather difficult, and the solver does not support curved dielectric geometries. Another open-source effort is chombo-discharge \cite{Marskar2023}, an AMR-based gas-discharge code built on the Chombo library that supports embedded boundaries for complex geometries and has recently been extended to a fully kinetic 3D Monte Carlo description of streamer-tree discharges \cite{Marskar2024}. SOMAFOAM \cite{SOMAFoam} is another open-source software package, based on the foam-extend version of OpenFOAM. It employs the FVM, exploits the unstructured-mesh and MPI capabilities of OpenFOAM, has demonstrated great scalability across mTorr-to-atmospheric conditions, and shows good agreement with experimental results. It relies on the built-in numerical schemes and solvers of OpenFOAM, and no adaptive mesh refinement capability is mentioned. Recently, Vidyut3d \cite{Vidyut3d}, an AMReX-based solver that introduces adaptive mesh refinement and hybrid CPU/GPU support, has been published. The KAUST Clean Energy Research Platform has independently introduced reactPlasFOAM \cite{Shao_2025}, another OpenFOAM-based unified plasma-combustion solver that couples the ChemPlasKin gas-plasma kinetics library \cite{ChemPlasKin} to OpenFOAM, providing AMR, dynamic load balancing, and on-the-fly electron Boltzmann solves. The solver has been benchmarked against streamer cases \cite{Bagheri2018}, but its primary focus is plasma-assisted combustion.
 
Several established solvers are not publicly available but are distributed within their respective research groups or through collaboration agreements. AVIP \cite{AVIP} is an unstructured, vertex-centered FVM weakly ionized plasma simulation code developed in Fortran at CERFACS and tightly coupled to the AVBP combustion solver for plasma-assisted combustion and nanosecond repetitively pulsed discharge studies. The solver incorporates an improved Scharfetter-Gummel flux scheme shown to provide stable results and has been validated against \cite{Bagheri2018}. PASSKEy \cite{PASSKEy2021} is a 2D-axisymmetric drift-diffusion-Poisson solver with detailed plasma kinetics for nanosecond pulsed discharges, fast ionization waves, and streamer-to-filament/spark transitions, and is available upon request from the authors. The COPAIER solver \cite{COPAIER,kourtzanidis2021self}, originally developed at ONERA (the author of this work, K. Kourtzanidis is a co-developer), is a research code which combines an FEM Poisson solver with FV Scharfetter-Gummel transport, implicit time integration and MPI parallelism. The HPEM and nonPDPSIM modeling platforms, developed by Kushner and co-workers at the University of Michigan \cite{Kushner_2009, Norberg_2015}, target different pressure regimes. HPEM focuses on low-pressure plasma-processing reactors, while nonPDPSIM is a 2D unstructured-grid solver for medium- and high-pressure dielectric-barrier discharges, microplasmas, and plasma jets. Another fluid-model code is that of \cite{Simonovic_2024}, who have implemented an axisymmetric drift-diffusion streamer solver directly within the AMReX block-structured AMR library, and have recently extended it to fully three-dimensional configurations and curvilinear boundaries \cite{Simonovic_2025}. The code uses a first-order fluid model with the Local Field Approximation (LFA).
 
Beyond these named tools, much of the streamer physics now considered canonical was established with in-house codes that have not been publicly released. Notable examples include the foundational fluid models of Babaeva and Naidis \cite{Babaeva} for atmospheric positive- and negative-streamer dynamics, the 2D-axisymmetric drift-diffusion model of Boeuf, Yang and Pitchford \cite{Boeuf_2013} for plasma bullets in atmospheric helium jets, and the 2D fluid simulations of Komuro, Ono and co-workers \cite{Komuro_2013, Komuro_2018}. Dedicated in-house 2D/3D fluid frameworks have also been developed for plasma-actuator applications, in particular by Nishida and co-workers \cite{Nishida_2014}.

In addition to open-source and research codes, several proprietary simulation tools are widely used. PLASIMO \cite{PLASIMO} is a dedicated plasma simulation framework that has been applied extensively to low-temperature and atmospheric-pressure plasma discharges, with a strong focus on detailed plasma chemistry. COMSOL Multiphysics \cite{COMSOL} provides a FEM-based plasma module that allows coupling of plasma dynamics with gas flow, heat transfer and complex chemistry within a single interface, with recent versions adding stabilization schemes suitable for streamer and corona configurations. VizGlow \cite{VizGlow} is another commercial FVM-based plasma solver covering pressures from a few mTorr to atmospheric on 3D unstructured meshes.

\subsection{Proposed solver}
\label{proposed_solver}

This work presents the first release of \texttt{SoPLASMA} (Simulations of Plasma), a new open-source, collisional plasma simulation suite built entirely on the OpenFOAM distribution maintained by OpenCFD Ltd. The developed suite includes solvers ranging from simple single- and multi-region electrostatic solvers to more complex plasma solvers, along with dedicated boundary conditions, numerical schemes, and supporting utilities. The main focus of this paper is \texttt{SoPlasmaFoam} (Simulations of Plasma, OpenFOAM-based), a multi-region plasma-dielectric solver for the consistent solution of the drift-diffusion-reaction transport equations coupled with the Poisson equation.
 
The solver is built on OpenFOAM but integrates several state-of-the-art third-party libraries to make it suitable for large-scale high-performance computing. Matrix systems can be solved either with the native OpenFOAM linear solvers or through the petsc4Foam framework \cite{Petsc4Foam}, which incorporates the PETSc library into OpenFOAM, providing access to a broad range of linear and ODE solvers with both CPU and GPU support, and enabling efficient scaling on HPC clusters. Adaptive mesh refinement is provided through the blastAMR library \cite{blastAMR}, which supports both hexahedral and polyhedral meshes and operates natively in 1D, 2D, 2D-axisymmetric and 3D configurations, in contrast to the standard OpenFOAM AMR implementation which is limited to 3D. The solver also supports dielectric regions of arbitrary number and shape, and the Poisson equation can be solved monolithically across all regions simultaneously, eliminating the need for iterative, segregated inter-region coupling until convergence. On the discretization side, the ROUND family of Total Variation Diminishing (TVD) schemes \cite{ROUNDSchemes_theory, ROUNDSchemes_implementation} has been incorporated, providing a well-tested set of convective flux limiters. In addition, the Scharfetter-Gummel scheme \cite{Scharfetter1969} for the coupled drift-diffusion flux has been developed and integrated into the solver.
 
% These features make \texttt{SoPlasmaFoam} a distinct contribution rather than just another OpenFOAM-based plasma solver. The combination of the mature OpenFOAM CFD infrastructure with targeted third-party extensions addresses critical requirements for accurate and scalable plasma simulations that are not met by the standard framework alone.

It is worth stating precisely how \texttt{SoPlasmaFoam} differs from existing OpenFOAM-based plasma solvers and from the closely related open-source adaptive codes. Relative to SOMAFOAM \cite{SOMAFoam}, which spans a wide pressure range but mentions no AMR capability beyond what the underlying OpenFOAM framework natively provides (3D hexahedral cells only), and has not been applied to streamer dynamics, the present solver is oriented toward streamer and surface-discharge propagation and brings a full-dimensional AMR capability (1D through 3D and axisymmetric), PETSc back-end and ROUND limiters into a single OpenFOAM-native
framework. Relative to the AMReX-based Vidyut3d \cite{Vidyut3d}, it operates on both structured and unstructured meshes and resolves curved, conforming plasma-dielectric interfaces of arbitrary number through a monolithic multi-region Poisson assembly, which is difficult to achieve on purely block-structured grids. Relative to reactPlasFOAM \cite{Shao_2025}, whose emphasis is the coupling of gas-plasma kinetics for plasma-assisted combustion, the focus here is on the numerics of the
transport-Poisson system itself. Beyond these architectural differences, the distinguishing contributions of this work are methodological and transferable to other plasma-fluid codes: a systematic benchmark of convective flux schemes that identifies ROUNDF as superior to standard TVD limiters for streamer transport; a quantitative study of how tightly the Poisson and transport equations must be coupled, including the finding that a semi-implicit Poisson treatment does not
remove the need for outer correction loops; and a drift-robust wall boundary condition that overcomes the P\'eclet-dependent failure of the conventional mixed-boundary formulation. Taken together, and released as an open-source solver, these make \texttt{SoPlasmaFoam} a distinct contribution rather than just another OpenFOAM-based plasma solver.
 
Because OpenFOAM is inherently three-dimensional, \texttt{SoPlasmaFoam} naturally handles 1D, 2D, 2D-axisymmetric, and fully 3D configurations without code modifications, unlike many existing solvers that are restricted to 2D or axisymmetric geometries. OpenFOAM is also fully MPI-enabled, allowing the solver to scale efficiently across multiple nodes on high-performance computing clusters. Combined with the PETSc CPU/GPU solver back-end, this yields a solver that is well-positioned for large-scale parallel simulations.
 
OpenFOAM is an open-source and highly customizable finite volume software, allowing seamless integration with third-party libraries, especially those written in C++. Its solvers, numerical schemes, and boundary conditions have been extensively verified and validated across a wide range of applications and scientific fields, providing confidence in the robustness and reliability of the underlying numerics. A further advantage is the large and active global OpenFOAM community, which makes the proposed suite immediately familiar to a broad user base. For this reason, the suite follows the OpenFOAM coding style and design philosophy as closely as possible, ensuring that the code remains readable and maintainable. Extensive documentation and a large number of online resources are available for OpenFOAM, which lowers the barrier for adoption and further development of the toolkit.
 
The code is built in a modular fashion, making extensive use of the object-oriented features of C++ and the OpenFOAM run-time selection mechanism. Base classes are defined for the core physical components, like electromagnetics, plasma species, and plasma transport, from which specific models are derived through inheritance and polymorphism. For example, the electromagnetic base class gives rise to electrostatic models, while the transport base class yields drift-diffusion models, with the architecture designed so that new models (e.g.\ full-momentum or electromagnetic formulations) can be added later without modifying the existing code. Boundary conditions follow the same pattern. The run-time selection mechanism allows the user to choose between available models at run time through dictionary entries, without recompilation, making the solver both flexible and computationally efficient.
 
The plasma solver can be seamlessly coupled with a standard OpenFOAM fluid solver for coupled plasma-fluid-thermodynamic simulations, a capability that is particularly relevant for atmospheric-pressure plasma applications such as aerodynamic flow control, electrohydrodynamics, plasma processing and plasma-assisted combustion. An additional advantage is the availability of the preCICE adapter for OpenFOAM \cite{OpenFOAMpreCICE}, which enables coupling with a wide range of third-party solvers and libraries.
 
The aims of this paper are multifold. First, we validate \texttt{SoPlasmaFoam} against canonical test cases, including a low-pressure DC glow discharge and the positive streamer benchmark of \cite{Bagheri2018}, and demonstrate its application to a surface dielectric barrier discharge (SDBD) simulation. Second, a detailed comparison of different flux discretization schemes is carried out on a simple transport problem and on the positive streamer case to assess which schemes perform best for plasma transport. Third, we analyze the coupling between the Poisson and transport equations, examining the number of correction loops required in a positive streamer case and how this requirement varies with increasing time step size. Finally, we present a performance analysis of the solver, identifying the main computational bottlenecks of the finite volume approach and evaluating the parallel scaling of the code.

\subsection{Paper Structure}

The paper is organized as follows. Section~\ref{Equations} presents the governing equations and the physical models employed in this work. The numerical methodology and implementation is described in Section~\ref{Numerical Methodology}. Section~\ref{flux schemes} assesses the performance of different flux discretization schemes on a stiff scalar advection problem. Sections~\ref{DC glow} and~\ref{streamer} present the validation of the proposed solver against canonical benchmark cases. The former covers a low-pressure DC glow discharge, while the latter addresses the positive streamer configuration of \cite{Bagheri2018}, including a detailed assessment of flux discretization schemes and a performance comparison with other software. The application of the solver to a surface ionization wave driven by a nanosecond SDBD actuator is presented in Section~\ref{sdbd}.  The parallel scalability of the solver is analyzed in Section~\ref{scalability}. Finally, Section~\ref{Discussion} discusses the main findings and future directions, and Section~\ref{Conclusions} summarizes the conclusions of this work.

\section{Governing equations and physical model} \label{Equations}

The physical model considered in this work is based on the self-consistent solution of the Poisson equation for the electric potential (in the plasma region and in any dielectric materials, if present) coupled with drift-diffusion-reaction transport equations for the charged species. The source term $S_j$ is left unspecified at this stage and will be defined for each case according to the adopted computational strategy.

\subsection{Poisson equation}

The electric field is obtained from the electric potential $\phi$, which is computed by solving the Poisson equation,
\begin{equation} \label{eq:explicit_Poisson}
    \nabla^2 \phi = -\frac{\rho}{\epsilon}
\end{equation}
where $\rho$ is the space charge density and $\epsilon$ is the permittivity of the gas.

Zero-gradient (Neumann) boundary conditions are applied to the electric potential in the far-field regions, while Dirichlet boundary conditions are imposed on the electrode surfaces. In simulations involving multiple domains, such as plasma regions coupled with dielectric materials, continuity of the electric potential and normal component of the electric displacement field is enforced through coupled (jump) boundary conditions at material interfaces. These interface conditions are handled within a monolithic solution framework, as described later.

\subsubsection{Semi-Implicit treatment}

The explicit formulation of the Poisson equation in Eq.~(\ref{eq:explicit_Poisson}) imposes severe restrictions on the time step, which must be smaller than the dielectric relaxation time of the electrons. To alleviate this constraint, a semi-implicit formulation of the Poisson equation is adopted, following the approach proposed in \cite{Villa2013}, with the additional inclusion of the diffusive charge source term on the right-hand side, which was not considered in the original formulation. The resulting semi-implicit Poisson equation reads

\begin{equation}
    \nabla \cdot \left[ \left( \epsilon + \Delta t \, \sigma^k \right) \nabla \phi^{k+1} \right]
    =
    - \rho^k
    - \Delta t \sum_{j} q_j \nabla \cdot \left( D_j \nabla n_j^k \right)
\end{equation}
where $\Delta t$ is the time step, $k$ denotes the current time level, and $k+1$ the next time level. The plasma conductivity $\sigma^k$ is defined as
\begin{equation}
    \sigma^k = \sum_{j} |q_j| \, n_j^k \, \mu_j
\end{equation}
where the index $j$ runs over all charged species (electrons and ions). In these expressions, $q_j$, $n_j$, $\mu_j$, and $D_j$ denote the charge, number density, mobility, and diffusion coefficient of species $j$, respectively.

\subsection{Electric field evaluation}

The electric field can be obtained using different techniques. The cell-centred electric field at the centroid $P$ is straightforwardly computed by taking the gradient of the electric potential:
\begin{equation}
    \mathbf{E}_P = -\left(\nabla \phi\right)_P
\end{equation}
which in OpenFOAM is handled via the \texttt{fvc::grad} operator as:

\begin{equation}
    \mathbf{E} = -\texttt{fvc::grad}(\phi)
\end{equation}

The face-normal electric field, required for the evaluation of drift fluxes, is computed from the surface-normal gradient of the potential rather than from interpolation of the cell-centered field. This avoids the need to prescribe explicit boundary conditions for the electric field at electrode faces, where such conditions are not straightforward to define. The normal electric flux through face $f$ is therefore obtained as:

\begin{equation} \label{eq:electric_flux}
\phi_{E,f} = -\left(\frac{\partial \phi}{\partial n}\right)_f \left|\mathbf{S}_f\right|
\end{equation}
which in OpenFOAM is evaluated via the \texttt{fvc::snGrad} operator. For internal faces and coupled interfaces shared between regions, the gradient accounts for cell values on both sides. For boundary faces, it is computed consistently from the prescribed potential boundary condition.

From these face-normal fluxes, the cell-centred electric field vector can be reconstructed by summing the flux contributions over all faces of the cell and dividing by the cell volume:
\begin{equation} \label{electric_field_4}
    \mathbf{E}_P 
    = 
    \frac{1}{V_P}\sum_f \left(\phi_{E,f}\, \hat{\mathbf{n}}_f\right) 
    = 
    -\frac{1}{V_P}\sum_f \left(\frac{\partial \phi}{\partial n}\right)_f \mathbf{S}_f
\end{equation}
which is achieved through the \texttt{fvc::reconstruct} operator as:
\begin{equation} \label{electric_field_5}
    \mathbf{E} = \texttt{fvc::reconstruct}(\texttt{Ef})
\end{equation}

In the present solver, both approaches are available. However, the reconstruction-based method is generally recommended, as it is more consistent with the flux-conservative nature of the FVM.

\subsection{Transport equation}

The transport of plasma species is described using the drift-diffusion approximation. In addition, selected species can optionally be treated as immobile when appropriate. For a generic species $j$, the continuity equation reads
\begin{equation}
\frac{\partial n_j}{\partial t} + \nabla \cdot \boldsymbol{\Gamma}_j = S_j
\end{equation}
where $n_j$ is the number density, $\boldsymbol{\Gamma}_j$ is the particle flux, and $S_j$ represents the source term. Under the drift-diffusion approximation, the total particle flux is given by the sum of drift and diffusion contributions:
\begin{equation}
\boldsymbol{\Gamma}_j = \pm \mu_j n_j \mathbf{E} - D_j \nabla n_j
\end{equation}
where $\mu_j$ and $D_j$ are the mobility and diffusion coefficient of species $j$, respectively, and the $\pm$ sign accounts for the sign of the species charge.

Substituting the flux expression into the continuity equation yields the standard drift-diffusion form,
\begin{equation}
\frac{\partial n_j}{\partial t}
+
\nabla \cdot \left( \pm \mu_j n_j \mathbf{E} - D_j \nabla n_j \right)
=
S_j
\end{equation}
The source term $S_j$ accounts for physical and chemical processes such as ionization, recombination, attachment, detachment, and other plasma chemical reactions.

\subsection{Transport equation boundary conditions} \label{subsec:transport_equation_boundary_conditions}

At open boundaries, inlet-outlet boundary conditions are typically employed for the species transport equations. When the drift flux directs species out of the domain, they are allowed to leave freely, whereas when the flux is directed inwards, species enter with a specified value, typically set equal to the clamping density of the respective species defined in the simulation.

At wall boundaries, the treatment differs. In the vicinity of the wall, species leave with their thermal velocity, defined as:
\begin{equation}
    v_{\text{th},s} = \sqrt{\frac{8 k_B T_s}{\pi m_s}}
\end{equation}
so that the thermal flux of species $s$ towards the wall is:
\begin{equation}
    \Gamma_{\text{th},s} = \frac{1}{4} n_s v_{\text{th},s}
\end{equation}

Ions and electrons are treated differently from one another. Electrons always leave the domain with their thermal flux. Ions leave with their thermal flux, plus their drift flux, but only when the latter is directed outwards, i.e., towards the wall \cite{Hagelaar2008}. Neutral species leave solely with their thermal flux. These conditions are summarised as follows.

For positive and negative ions:
\begin{equation}
    \Gamma_{\pm} = \Gamma_{\text{th},\pm} + \max\left(0, \, \Gamma_{\text{drift},\pm}\right)
\end{equation}

For neutral species:
\begin{equation}
    \Gamma_{n} = \Gamma_{\text{th},n}
\end{equation}

For electrons:
\begin{equation}
    \Gamma_{e} = \Gamma_{\text{th},e} - \gamma_{\text{see}} \, \Gamma_{i}
\end{equation}
where $\gamma_{\text{see}}$ is the secondary electron emission (SEE) coefficient, accounting for the electrons released from the wall upon ion bombardment, and $\Gamma_i$ is the total ion flux reaching the wall.

Finally, the nature of the wall material determines the fate of the species that reach the surface. On conducting walls, the charged species that arrive at the wall are absorbed into the material. On dielectric surfaces, however, they accumulate as surface charge, which is subsequently taken into account in the solution of the Poisson equation.

\subsection{Transport coefficients and source terms}

The transport coefficients and source terms can be evaluated using different approaches within the developed solver. These quantities may be prescribed through analytical expressions, as adopted in the positive streamer benchmark cases presented in this work, or assumed to be constant. Alternatively, they can be obtained from precomputed lookup tables or coupled to third-party software. The latter approach, however, are beyond the scope of the present paper.

\section{Numerical methodology and implementation} \label{Numerical Methodology}

\subsection{Discretization schemes} \label{Discretization schemes}

All schemes available in OpenFOAM can be used within the developed solver. Here, we present the main schemes used for the simulations discussed in this paper.

For time integration, a second-order backward scheme is mainly employed throughout this work. Spatial gradients, such as those required for the calculation of the electric field, are evaluated using a second-order least-squares scheme. Regarding interpolation, mobility and diffusivity are interpolated using harmonic interpolation, while all other quantities are interpolated linearly.

In plasma simulations, the choice of flux discretization for particle transport is particularly important due to the strong coupling between drift and diffusion processes. Two approaches are available within the developed solver. For coupled advection-diffusion fluxes, the Scharfetter-Gummel (SG) scheme \cite{Scharfetter1969}, implemented in OpenFOAM is used. Alternatively, diffusion and advection can be discretized separately. In this uncoupled approach, the diffusion term is discretized using a second-order central scheme.

\begin{figure} [h!]
    \centering
    \includegraphics[scale=1]{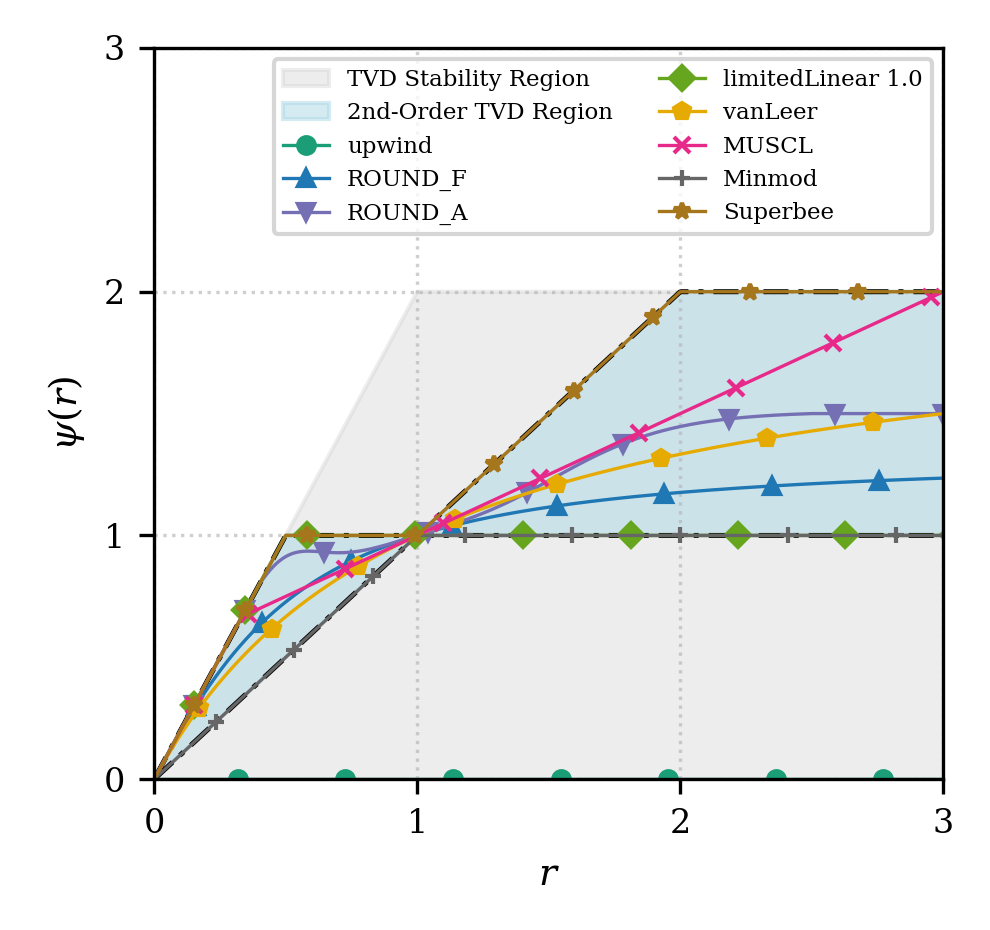}
    \caption{The Total Variation Diminishing (TVD) diagram of the schemes used in this work. The function $\psi(r)$ denotes the flux limiter of the scheme, which dynamically adjusts the numerical dissipation based on the local gradient ratio $r$. Schemes within the TVD stability region prevent the generation of non-physical oscillations at steep gradients, while those falling inside the blue region also maintain second-order accuracy.}
    \label{fig:tvd_diagram_scheme_limiters}
\end{figure}

Several flux schemes for the drift term are available, including built-in OpenFOAM schemes and the ROUND schemes developed in \cite{ROUNDSchemes_implementation}. The schemes tested in the next section are limitedLinear, MUSCL, ROUNDF, ROUNDA, SuperBee, Minmod, vanLeer, upwind and SG. Apart from the SG and the upwind schemes, the rest are TVD schemes, with the TVD diagram of their limiters displayed in \Cref{fig:tvd_diagram_scheme_limiters} (the figure also includes the upwind scheme for comparison).

\subsection{Time stepping}

An adjustable time-stepping scheme is employed to optimize computational efficiency. Although the implicit treatment of drift and diffusive fluxes bypasses the strict stability limits imposed by the drift and diffusive Courant numbers, time-step control  remains essential for maintaining temporal accuracy. Consistent with established CFD practices, the electron drift and diffusive Courant numbers can be constrained to ensure that the physical transport is well-resolved:
\begin{equation}
    Co_{e}^{c} = \frac{u_e \, \Delta t}{\Delta x} \leq Co_{max}^{c}, \qquad Co_{e}^{d} = \frac{D_e \, \Delta t}{\Delta x^2} \leq Co_{max}^{d}
\end{equation}
where $u_e$ is the electron drift velocity, $\Delta x$ the local cell size and $D_e$ the electron diffusion coefficient. In scenarios where a high Courant number is permitted, additional outer iterations, such as fixed-point iterations (outer PIMPLE loop iterations), are utilized to preserve the self-consistency and accuracy of the coupled system, as will be demonstrated in later sections.

Beyond transport constraints, the dielectric relaxation time $\tau_{dr}$ is a strict limit in plasma simulations, particularly when involving explicit Poisson solvers. To ensure numerical stability in highly conductive regions where charge relaxation occurs rapidly, the time step must satisfy:

\begin{equation}
    \Delta t \leq \tau_{dr} = \frac{\varepsilon}{\sigma}
\end{equation}
where $\varepsilon$ is the permittivity, and $\sigma$ is the electrical conductivity of the medium.  While the implementation of a semi-implicit Poisson scheme allows the solver to remain stable even when the ratio $\Delta t / \tau_{dr}$ exceeds $1.0$, this numerical flexibility comes at a cost. Similar to exceeding the Courant limit, bypassing the dielectric relaxation limit can lead to a significant loss of temporal accuracy.

In the present formulation, the reaction source terms are treated explicitly, which imposes a strict limitation on the simulation time step based on the fastest chemical time scales. Therefore, we monitor a dimensionless chemical Courant number $Co_{chem}$ to ensure numerical stability:

\begin{equation}
    Co_{chem} = k_{max} \Delta t \leq Co_{chem,max}
\end{equation}
where $k_{max}$ represents the fastest reaction rate in the system.

\subsection{Implementation of thermal-drift wall flux condition}

The physical flux conditions described in \Cref{subsec:transport_equation_boundary_conditions} are generally implemented within the OpenFOAM framework as in \cite{SOMAFoam}, by mapping them onto the standard mixed boundary condition, whose general form reads:
\begin{equation}
    n_f = f_v \, n_r + (1 - f_v) \left( n_C + \frac{g_r}{d} \right)
\end{equation}
where $n_r$ is the reference value, $g_r$ is the reference gradient, $f_v$ is the value fraction, $n_C$ is the internal cell value, and $d$ is the distance from the cell centre to the boundary face. A straightforward approach is to set $n_r = 0$ and $g_r = 0$, so that the face value reduces to:
\begin{equation}
    n_f = (1 - f_v) \, n_C
\end{equation}
The total normal flux through the boundary face is then:
\begin{equation}
    \Gamma = \underbrace{- \frac{D}{d}\left(n_f - n_C\right)}_{\text{diffusive}} 
    + \underbrace{Z \mu E_n \, n_f}_{\text{drift}}
    = \frac{D}{d} f_v \, n_C + Z \mu E_n (1 - f_v) \, n_C
\end{equation}
Requiring this to match the prescribed wall flux $\Gamma_{\text{wall}} = 
\frac{1}{4}v_{\text{th}}\,n_f + \max(0,\,Z\mu E_n)\,n_f$ and substituting 
$n_f = (1-f_v)\,n_C$:
\begin{equation}
    \frac{D}{d} f_v \, n_C + Z\mu E_n (1 - f_v) \, n_C 
    = \left[\frac{1}{4}v_{\text{th}} + \max(0,\,Z\mu E_n)\right](1 - f_v)\, n_C
\end{equation}
Dividing through by $n_C$, expanding and collecting $f_v$ terms:
\begin{align}
    f_v &\left(\frac{D}{d} - Z\mu E_n + \frac{1}{4}v_{\text{th}} 
    + \max(0,\,Z\mu E_n)\right) \notag \\
    &= \frac{1}{4}v_{\text{th}} + \max(0,\,Z\mu E_n) - Z\mu E_n
\end{align}
where the $\max(0,\,Z\mu E_n)$ term represents the ion drift contribution 
to the incoming wall flux. It applies to ions only and vanishes for electrons. Two cases arise naturally. When the drift is directed towards the wall 
($Z\mu E_n > 0$), $\max(0,\,Z\mu E_n) = Z\mu E_n$ and the numerator 
simplifies:
\begin{equation}
    f_v^{\text{ion,out}} = \frac{\dfrac{1}{4}v_{\text{th}}}{\dfrac{D}{d} 
    + \dfrac{1}{4}v_{\text{th}}}
\end{equation}
When the drift is directed away from the wall ($Z\mu E_n \leq 0$), 
$\max(0,\,Z\mu E_n) = 0$ and:
\begin{equation}
    f_v^{\text{ion,in}} = \frac{\dfrac{1}{4}v_{\text{th}} - Z\mu E_n}{\dfrac{D}{d} 
    - Z\mu E_n + \dfrac{1}{4}v_{\text{th}}}
\end{equation}
For electrons, the $\max(0,\,Z\mu E_n)$ term is absent from the wall flux, 
so the value fraction reduces to a single expression regardless of drift 
direction:
\begin{equation}
    f_v^{\text{e}} = \frac{\dfrac{1}{4}v_{\text{th}} - Z\mu E_n}{\dfrac{D}{d} 
    - Z\mu E_n + \dfrac{1}{4}v_{\text{th}}}
\end{equation}
The fundamental limitation of this approach becomes apparent when examining the limiting behaviour of the value fraction. For ions directed towards the 
wall, $f_v^{\text{ion,out}}$ depends only on the ratio of diffusive to 
thermal velocity coefficients:
\begin{equation}
    f_v^{\text{ion,out}} = \frac{\dfrac{1}{4}v_{\text{th}}}{\dfrac{D}{d} 
    + \dfrac{1}{4}v_{\text{th}}}
\end{equation}
When the drift is strong ($Z\mu E_n \gg D/d$), one might expect the flux to be drift-dominated. However, $f_v^{\text{ion,out}}$ is entirely independent of $Z\mu E_n$: the drift velocity has dropped out completely. As $D/d \to 0$ (large cell Péclet number $Pe_\Delta = Z\mu E_n d/D \gg 1$), $f_v^{\text{ion,out}} \to 1$, which means $n_f = (1-f_v)n_C \to 0$. The face density vanishes, and the drift flux $Z\mu E_n\,n_f \to 0$ regardless of how large $Z\mu E_n$ is. The boundary condition is therefore 
diffusion-limited: in the drift-dominated regime the mixed
formulation cannot correctly enforce the prescribed wall flux, as the diffusive stencil is the only mechanism available to set $n_f$.

For electrons ($Z = -1$), when repelled from the wall ($Z\mu E_n < 0$) the denominator always exceeds the numerator by $D/d > 0$, so $f_v^{\text{e}} \in [0,1)$ and the formulation is well-posed. When electrons drift towards the wall ($E_n < 0$, $Z\mu E_n > 0$), however, the numerator $\frac{1}{4}v_{\text{th}} - Z\mu E_n$ turns negative once $Z\mu E_n > \frac{1}{4}v_{\text{th}}$, giving $f_v^{\text{e}} < 0$ and 
therefore $n_f > n_C$, a spurious density enhancement at the wall face 
that renders the solution unphysical.

To overcome this limitation, we developed a custom boundary condition that acts directly on the matrix coefficients of the discretized transport equation, rather than prescribing a face value through the mixed framework. The discretized transport equation at a boundary cell reads:
\begin{equation}
    a_P n_P = \sum_{nb} a_{nb} n_{nb} + S
\end{equation}
The drift and diffusive contributions at the boundary face are treated 
separately and introduced directly as implicit coefficients. For the drift term, if the flux is directed outwards (towards the wall), its full contribution is retained implicitly; if it is directed inwards, it is zeroed out to prevent unphysical inflow:
\begin{equation}
    \Gamma_{\text{drift}}^{\text{wall}} = \max(0,\, Z\mu E_n)\, n_P
\end{equation}
For the diffusive term, the thermal flux condition is enforced by equating the diffusive flux to the prescribed thermal flux:
\begin{equation}
    \frac{D}{d}\left(n_P - n_f\right) = \frac{1}{4} v_{\text{th}} \, n_P
\end{equation}
which, upon rearrangement, yields the implicit diffusive contribution to the boundary cell coefficient:
\begin{equation}
    a_P^{\text{diff}} = \frac{D}{d} + \frac{1}{4} v_{\text{th}}
\end{equation}
with no explicit source contribution from the diffusive term. The total 
implicit coefficient at the boundary face is therefore:
\begin{equation}\label{eq:direct_flux_bc}
    a_P^{\text{wall}} = \frac{D}{d} + \frac{1}{4} v_{\text{th}} + \max(0,\, Z\mu E_n)
\end{equation}
This formulation directly populates the system matrix with the physically correct coefficients, without any dependence on the boundary Péclet number. In the drift-dominated limit, $\max(0,\,Z\mu E_n)$ simply dominates $a_P^{\text{wall}}$, and the correct outward flux is recovered regardless of the ratio $D/d$. For the electron equation, secondary electron emission is accounted for by adding the corresponding source term $\gamma_{\text{see}} \Gamma_i$ explicitly to the right-hand side.

\subsection{Monolithic solver}

In plasma applications, multidomain problems are frequently encountered where distinct physical domains share an interface (e.g. in plasma-dielectric discharges). These domains may possess different permittivities $\epsilon$ and may accumulate surface charges ($\sigma_S$) at the interface. Such configurations need a coupled (or jump) boundary condition, which enforces the continuity of the electric potential while allowing a discontinuity in the normal component of the electric field (a jump in the gradient) proportional to the local charge. Mathematically, this is expressed as:

\begin{subequations}\label{eq:coupled_boundary_condition}
    \begin{align}
        \phi_- &= \phi_+ \\
        \epsilon_- \left( \frac{\partial \phi}{\partial n} \right)_- &- \epsilon_+ \left( \frac{\partial \phi}{\partial n} \right)_+ = \sigma_S,
    \end{align}
\end{subequations}

While OpenFOAM includes similar logic for Conjugate Heat Transfer (CHT), the electrostatic case requires a specialized approach to account for the dynamic accumulation of surface charge. We have developed a custom boundary condition that maps these jump conditions onto the standard OpenFOAM mixed type. In this framework, the face value ($\phi_f$) is determined by a combination of a reference value ($\phi_r$), a reference gradient ($g_r$), and a value fraction ($f_v$)

\begin{equation}
    \phi_f = f_v \phi_r + (1 - f_v) (\phi_C +  \dfrac{g_r}{d})
\end{equation}
By rearranging the jump conditions in \cref{eq:coupled_boundary_condition}, we derive the following discretization coefficients for the interface:

\begin{equation}
    f_v = \frac{\displaystyle\sum \dfrac{\epsilon_{nb}}{d_{nb}}}{\dfrac{\epsilon_{p}}{d_{p}} + \displaystyle\sum \dfrac{\epsilon_{nb}}{d_{nb}}}, 
    \qquad 
    \phi_r = \frac{\displaystyle\sum \dfrac{\epsilon_{nb} \phi_{nb}}{d_{nb}} + \sigma_S}{\displaystyle\sum \dfrac{\epsilon_{nb}}{d_{nb}}}
\end{equation}
where the subscripts $p$ and $nb$ denote the local patch cell and the neighboring domain cells, respectively, and $d$ represents the distance between the cell center and the interface.

Standard OpenFOAM solvers typically employ explicit (segregated) coupling where regions are solved independently. Interface interactions enter only as explicit source terms based on the neighbor's previous iteration, requiring outer loops to ensure convergence. Consider two regions, $A^{1}$ and $A^{2}$, sharing a common interface. The system for region $A^{1}$, where the interface corresponds to cells $N_{1-k}$ up to $N_{1}$, is represented as \Cref{eq:segregated_system}:

\begin{figure*}
\begin{equation}\label{eq:segregated_system}
\begin{bmatrix}
a^{1}_{1,1} & a^{1}_{1,2} & \cdots & a^{1}_{1,\,N_{1}-1} & a^{1}_{1,\,N_{1}} \\[2pt]
a^{1}_{2,1} & a^{1}_{2,2} & \cdots & a^{1}_{2,\,N_{1}-1} & a^{1}_{2,\,N_{1}} \\[2pt]
\vdots      & \vdots      & \ddots & \vdots              & \vdots            \\[2pt]
a^{1}_{N_{1}-k-1,\,1} & a^{1}_{N_{1}-k-1,\,2} & \cdots & a^{1}_{N_{1}-k-1,\,N_{1}-1} & a^{1}_{N_{1}-k-1,\,N_{1}} \\[6pt]
a^{1}_{N_{1}-k,\,1}   & a^{1}_{N_{1}-k,\,2}   & \cdots & a^{1}_{N_{1}-k,\,N_{1}-1}   & a^{1}_{N_{1}-k,\,N_{1}}   \\[4pt]
a^{1}_{N_{1}-k+1,\,1} & a^{1}_{N_{1}-k+1,\,2} & \cdots & a^{1}_{N_{1}-k+1,\,N_{1}-1} & a^{1}_{N_{1}-k+1,\,N_{1}} \\[4pt]
\vdots                & \vdots                & \ddots & \vdots                      & \vdots                    \\[4pt]
a^{1}_{N_{1}-1,\,1}   & a^{1}_{N_{1}-1,\,2}   & \cdots & a^{1}_{N_{1}-1,\,N_{1}-1}   & a^{1}_{N_{1}-1,\,N_{1}}   \\[4pt]
a^{1}_{N_{1},\,1}     & a^{1}_{N_{1},\,2}     & \cdots & a^{1}_{N_{1},\,N_{1}-1}     & a^{1}_{N_{1},\,N_{1}}
\end{bmatrix}
\!
\begin{bmatrix}
\phi_{A^{1},1} \\
\phi_{A^{1},2} \\
\vdots \\
\phi_{A^{1},N_{1}-k-1} \\
\phi_{A^{1},N_{1}-k} \\
\phi_{A^{1},N_{1}-k+1} \\
\vdots \\
\phi_{A^{1},N_{1}-1} \\
\phi_{A^{1},N_{1}}
\end{bmatrix}
\! =
\begin{bmatrix}
b_{A^{1},1} \\
b_{A^{1},2} \\
\vdots \\
b_{A^{1},N_{1}-k-1} \\[4pt]
b_{A^{1},N_{1}-k}^{\text{int}}
\;\colorbox{blue!50!white}{$+\,S^{(A^{2})}_{A^{1},N_{1}-k}$} \\[4pt]
b_{A^{1},N_{1}-k+1}^{\text{int}}
\;\colorbox{blue!50!white}{$+\,S^{(A^{2})}_{A^{1},N_{1}-k+1}$} \\
\vdots \\
b_{A^{1},N_{1}-1}^{\text{int}}
\;\colorbox{blue!50!white}{$+\,S^{(A^{2})}_{A^{1},N_{1}-1}$} \\[4pt]
b_{A^{1},N_{1}}^{\text{int}}
\;\colorbox{blue!50!white}{$+\,S^{(A^{2})}_{A^{1},N_{1}}$}
\end{bmatrix}
\end{equation}
\[
S_{A^{1},i}
\;=\;
\alpha^{1}_{i}\,\phi^{\text{old}}_{A^{2},\,\text{nb}(i)},
\qquad
i = N_{1}-k,\dots,N_{1}.
\]
\end{figure*}

Similarly, the matrix equation is built for region $A^2$. To address the computational bottleneck of the Poisson equation (often the most demanding component of the entire simulation), it is critical to avoid redundant solves per iteration. Thus, we implemented an implicit/monolithic coupling based on recent ESI-Group developments \cite{Implicit_Coupling_OpenFOAM}. By assembilng the domains into a single linear system, interface terms are moved to the left-hand side as off diagonal blocks $C_{12}$ and $C{21}$, enabling a simultaneous solve:

\[
\begin{bmatrix}
A^{1} & \colorbox{blue!50!white}{$C_{12}$} \\[6pt]
\colorbox{orange!50!white}{$C_{21}$} & A^{2}
\end{bmatrix}
\begin{bmatrix}
\phi_{A^{1}} \\
\phi_{A^{2}}
\end{bmatrix}
=
\begin{bmatrix}
b^{1} \\
b^{2}
\end{bmatrix}.
\]

\subsection{Adaptive mesh refinement}

Adaptive mesh refinement (AMR) is a critical requirement for plasma simulations. Consider a positive streamer case, where the streamer head and channel must be resolved with high spatial precision. However, maintaining a uniform mesh of this resolution across the entire computational domain would be computationally prohibitive. Furthermore, in scenarios where the streamer propagation path is not predefined, static pre-refinement is not a viable strategy.

Standard OpenFOAM distributions provide a native AMR implementation. However, it is restricted to hexahedral cells and operates exclusively in 3D. This limitation complicates 2D or 2D-axisymmetric simulations, often requiring "pseudo-2D" setups that introduce unnecessary computational overhead.

To overcome these constraints, we integrated the blastAMR library \cite{blastAMR} into our solver. Developed for the ESI-OpenFOAM distribution, blastAMR supports AMR on both hexahedral and polyhedral meshes and functions natively in 2D and 2D-axisymmetric configurations. A primary advantage of this library is its dynamic load-balancing capability. As cells are refined or unrefined in localized regions, the workload is redistributed across processors to maintain parallel efficiency. This ensures that the simulation remains scalable even as the mesh topology changes significantly over time.

\subsection{PETSc solver and petsc4Foam integration}

OpenFOAM provides a robust suite of native linear solvers and preconditioners, including conjugate gradient variants (PCG, PBiCGStab), geometric-algebraic multigrid (GAMG), and various smoothers. While these offer competitive speed and scalability, more advanced options are often required for highly complex systems. PETSc \cite{petsc-web-page} is a comprehensive open-source suite designed for the parallel solution of large-scale scientific applications. It provides an extensive range of linear and nonlinear solvers, ordinary differential equation (ODE) integrators, and optimization tools, supporting C, C++, Fortran, and Python. PETSc is engineered for high-efficiency execution on both CPUs and GPUs using MPI and is widely recognized as a standard in computational mechanics.

Since version v1912, the OpenFOAM ESI-CFD distribution has integrated the petsc4Foam library \cite{Petsc4Foam}, which embeds PETSc and its external dependencies, such as Hypre. A core component of petsc4Foam is the ldu2csr matrix conversion tool. This utility transforms OpenFOAM’s native Lower-Diagonal-Upper (LDU) matrix format into the Compressed Sparse Row (CSR) format utilized by PETSc. 

The developers of petsc4Foam report excellent scalability in High-Performance Computing (HPC) environments, noting that the overhead associated with LDU-to-CSR conversion is negligible compared to the solver gains in large-scale simulations. Specifically, the PETSc-AMG-CG solver has demonstrated excellent linear scalability up to 64 nodes (2,304 cores) \cite{Petsc4Foam}. Such performance is critical for the present work, where the computational cost is driven by the large number of chemical species, the inherent stiffness of the Poisson equation, and the high cell counts required for 3D discretization.

\subsection{Numerical solution strategy}

The governing equations are solved using a segregated time-marching strategy within an outer PIMPLE loop, in which the transport equations for the plasma species are coupled with the solution of the Poisson equation. At each time step, the electric field and species densities are updated self-consistently. Two coupling strategies are supported: a fully explicit treatment and a semi-implicit treatment of the Poisson equation.

\paragraph{Explicit coupling} Starting from known values at time level $k$, the transport coefficients $\mu_j, D_j$ and the source terms $S_j$ are first updated for each species. The drift-diffusion-reaction equation is then advanced in time using explicit source terms and the electric field from the previous iteration. The updated species densities are used to evaluate the space charge density $\rho = \sum_j q_j n_j$, which closes the Poisson equation. Solving the Poisson equation yields the electric potential at the new time level, from which the electric field is reconstructed using the \Cref{electric_field_4,electric_field_5}. The procedure is then repeated within the outer PIMPLE loop, and the simulation advances to the next time step.

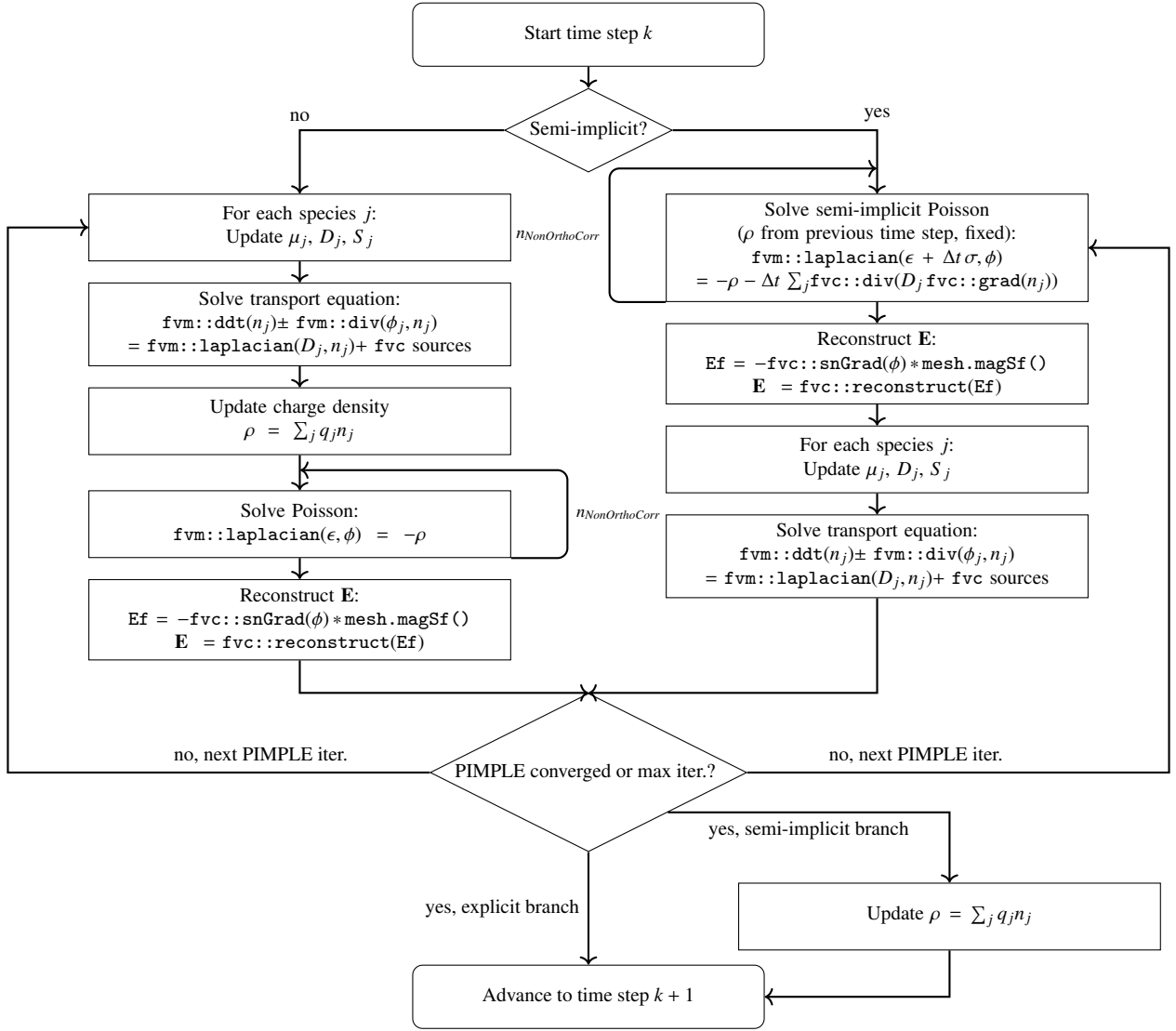
\begin{figure*}[h!]
    \centering
    \begin{tikzpicture}[
        node distance=1.6cm and 3.cm,
        every node/.style={font=\footnotesize},
        startstop/.style={
            rectangle, rounded corners,
            draw, align=center,
            minimum width=5cm,
            minimum height=0.9cm
        },
        process/.style={
            rectangle, draw, align=center,
            minimum width=6cm,
            minimum height=0.95cm,
            text width=5.5cm
        },
        decision/.style={
            diamond, draw, align=center,
            aspect=2,
            inner sep=1pt
        },
        arrow/.style={->, thick}
        ]

        \node (start) [startstop] {Start time step $k$};

        \node (coupling) [decision, below=0.3cm of start] {Semi-implicit?};
        
        % Explicit branch
        \node (updateTransportExp) [process, below left=0.6cm and 0.5cm of coupling] {
            For each species $j$:\\ Update $\mu_j$, $D_j$, $S_j$
        };
        
        \node (speciesSolveExp) [process, below=0.3cm of updateTransportExp] {
            Solve transport equation:\\
            \texttt{fvm::ddt}$(n_j) \pm$ \texttt{fvm::div}$(\phi_j, n_j)$\\
            $=$ \texttt{fvm::laplacian}$(D_j, n_j) +$ \texttt{fvc} sources
        };
        
        \node (rhoExp) [process, text width=3cm, below=0.3cm of speciesSolveExp] {
            Update charge density\\
            $\rho = \sum_j q_j n_j$
        };
        
        \node (poissonExp) [process, below=0.5cm of rhoExp] {
            Solve Poisson:\\
            \texttt{fvm::laplacian}$(\epsilon, \phi) = -\rho$
        };
        
        \node (EfieldExp) [process, text width=5.5cm, below=0.3cm of poissonExp] {
            Reconstruct $\mathbf{E}$:\\
            \texttt{Ef} $= -$\texttt{fvc::snGrad}$(\phi)\,*\,$\texttt{mesh.magSf()}\\
            $\mathbf{E} =$ \texttt{fvc::reconstruct}(\texttt{Ef})
        };
        
        % Semi-implicit branch
        \node (poissonSI) [process, below right=0.6cm and 0.5cm of coupling] {
            Solve semi-implicit Poisson\\
            ($\rho$ from previous time step, fixed):\\
            \texttt{fvm::laplacian}$(\epsilon + \Delta t\,\sigma, \phi)$\\
            $= -\rho - \Delta t\,\sum_j$\texttt{fvc::div}$(D_j\,$\texttt{fvc::grad}$(n_j))$
        };
        
        \node (EfieldSI) [process, text width=5.5cm, below=0.3cm of poissonSI] {
            Reconstruct $\mathbf{E}$:\\
            \texttt{Ef} $= -$\texttt{fvc::snGrad}$(\phi)\,*\,$\texttt{mesh.magSf()}\\
            $\mathbf{E} =$ \texttt{fvc::reconstruct}(\texttt{Ef})
        };
        
        \node (updateTransportSI) [process, below=0.3cm of EfieldSI] {
            For each species $j$:\\ Update $\mu_j$, $D_j$, $S_j$
        };
        
        \node (speciesSolveSI) [process, below=0.3cm of updateTransportSI] {
            Solve transport equation:\\
            \texttt{fvm::ddt}$(n_j) \pm$ \texttt{fvm::div}$(\phi_j, n_j)$\\
            $=$ \texttt{fvm::laplacian}$(D_j, n_j) +$ \texttt{fvc} sources
        };
        
        % Convergence check (shared)
        \node (pimple) [decision, below=7.4cm of coupling] 
        {
            PIMPLE converged or max iter.?
        };
        
        \node (rhoFinalSI) [process, text width=3cm, below right=1cm and 1cm of pimple] {
            Update $\rho = \sum_j q_j n_j$
        };
        
        \node (advance) [startstop, below=1.6cm of pimple] {Advance to time step $k+1$};
        
        % Arrows
        \draw[arrow] (start) -- (coupling);
        
        % Explicit branch
        \draw[arrow] (coupling) -| node[midway, above]{no} (updateTransportExp);
        \draw[arrow] (updateTransportExp) -- (speciesSolveExp);
        \draw[arrow] (speciesSolveExp) -- (rhoExp);
        \draw[arrow] (rhoExp) -- (poissonExp);
        \draw[arrow] (poissonExp) -- (EfieldExp);
        \draw[arrow, rounded corners=4pt] (poissonExp.south east) 
            -- ++(0.8cm, 0) 
            -- ++(0, 1.25cm)
            node[midway, right, font=\scriptsize\itshape] {$n_{\text{NonOrthoCorr}}$}
            -- ++(-3.8cm, 0);
        \draw[arrow] (EfieldExp.south) |- (pimple.north);
        
        % Semi-implicit branch
        \draw[arrow] (coupling) -| node[midway, above]{yes} (poissonSI);
        \draw[arrow, rounded corners=4pt] (poissonSI.south west) 
            -- ++(-0.8cm, 0) 
            -- ++(0, 1.9cm)
            node[midway, left, font=\scriptsize\itshape] {$n_{\text{NonOrthoCorr}}$}
            -- ++(3.8cm, 0);
        \draw[arrow] (poissonSI) -- (EfieldSI);
        \draw[arrow] (EfieldSI) -- (updateTransportSI);
        \draw[arrow] (updateTransportSI) -- (speciesSolveSI);
        \draw[arrow] (speciesSolveSI.south) |- (pimple.north);
        
        % No convergence or max iter.
        \draw[arrow] (pimple.west) 
        -- node[above, pos=0.4] {no, next PIMPLE iter.} ++(-6cm, 0) 
        |- (updateTransportExp.west);

        \draw[arrow] (pimple.east) 
        -- node[above, pos=0.4] {no, next PIMPLE iter.} ++(+6cm, 0) 
        |- (poissonSI.east);
        
        \draw[arrow] (pimple.south) -- node[midway, left]{yes, explicit branch} (advance.north);
        
        \draw[arrow] (pimple.south east) -| node[pos=0.25, below]{yes, semi-implicit branch} (rhoFinalSI.north);
        \draw[arrow] (rhoFinalSI.south) |- (advance.east);
        
    \end{tikzpicture}
    \caption{Flowchart of the numerical solver. Two coupling strategies are 
    available: an explicit Poisson treatment (left branch) and a 
    semi-implicit treatment (right branch). Both are wrapped in an outer 
    PIMPLE loop for sub-iteration within each time step. In the 
    semi-implicit branch, $\rho$ is held fixed at the previous time-step 
    value throughout the PIMPLE iterations, since the semi-implicit 
    formulation already accounts for the new-time-step advancement.}
    \label{fig:solver_flowchart}
\end{figure*}

\paragraph{Semi-implicit coupling} In the semi-implicit scheme, the Poisson equation is solved first using the charge density $\rho$ from the previous time step. The charge density $\rho$ already includes the implicit advancement of the species densities through the semi-implicit Poisson formulation and must not be updated within the PIMPLE loop, as doing so would double-count the time advancement. The remaining terms in the Poisson equation (transport coefficients, conductivity, drift fluxes) are updated at each PIMPLE iteration. Once the potential is obtained, the electric field is reconstructed as in the explicit case. The transport coefficients and source terms are then updated, and the drift-diffusion-reaction equation is solved with explicit sources. The procedure iterates within the PIMPLE loop. At the end of the loop, the charge density is finally updated and the simulation advances to the next time step.

\section{Assessement of flux schemes} \label{flux schemes}

\begin{figure}[!htbp]
    \centering
    \includegraphics[scale=0.98]{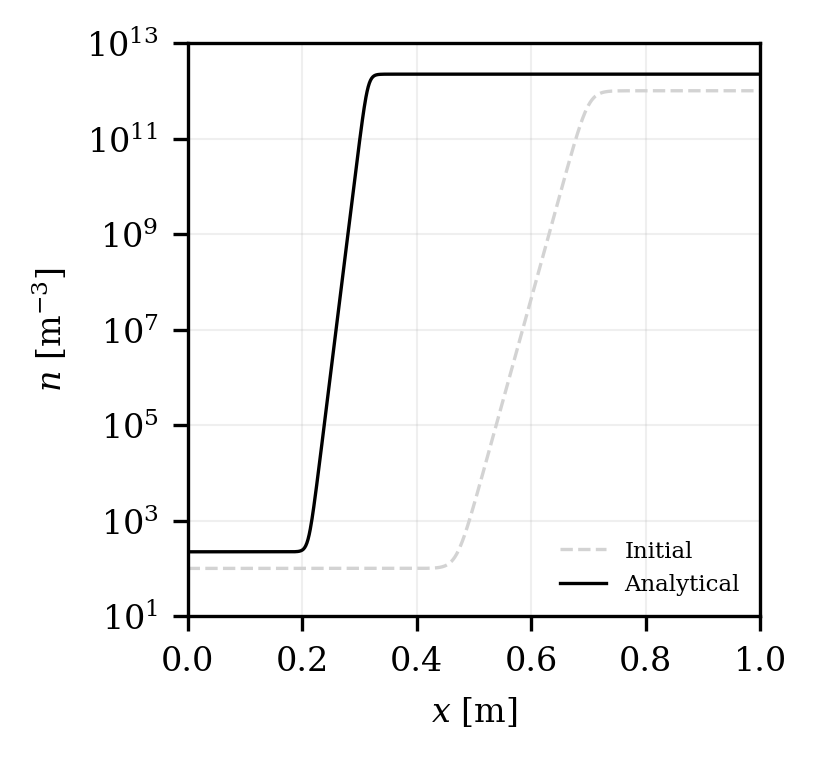}
    \caption{Initial distribution and analytical solution for the 1D stiff scalar advection problem under a linear electric field.}
    \label{fig:initial_and_analytical_solution_stiff_scalar_advection}
\end{figure}

In plasma simulations, the flux schemes used for the drift of the particles is highly important. As mentioned in \Cref{Discretization schemes}, different flux schemes are available in the developed solver. In order to assess their robustness and accuracy, the schemes are tested on a stiff scalar advection problem, similar to the one presented in \cite{AVIP}. It is a $1D$ domain $[0, 1]$, where a linear electric field $E(x) = A x$ is applied, with $A = 10^4$ in order to get an advection dominated problem. The mobility and diffusivity are set to $-1$ and $1$, while reactions are not considered. The equation that governs the problem can be written as:

\begin{equation}
    \frac{\partial n}{\partial t} - \frac{\partial A x n}{\partial x} - \frac{\partial^2 n}{\partial x^2} =0
\end{equation}
An initial distribution of particles is considered:

\begin{equation}
    n_0(x) = n_1 + \frac{1}{2}\left[ 1 + tanh\left( \frac{x - x_0}{\sigma} \right) \right] n_2
\end{equation}
where $n_1 = 10^2$ is the lowest density, $n_2 = 10^{12}$ the highest density, $x_0 = 0.7$ and $\sigma = 0.02$. The initial profile can be seen in \Cref{fig:initial_and_analytical_solution_stiff_scalar_advection} along with the analytical solution, which is governed by the following equation:

\begin{equation}
    n(x,t) = n_0 (x e^{At}) e^{At}
\end{equation}

Simulations were performed for three different mesh resolutions, $100, 200$ and $400$ cells, with corresponding time steps of $\Delta_t = 4.0 \cdot 10^{-7}, 2.0 \cdot 10^{-7}$ and $1.0 \cdot 10^{-7}$ respectively. Each simulation was evolved until a final time of $t_{end} = 8.0 \cdot 10^{-5}$. 

\begin{figure*}[!htbp]
    \centering
    \begin{subfigure}{0.45\textwidth}
        \centering
        \includegraphics[scale=1]{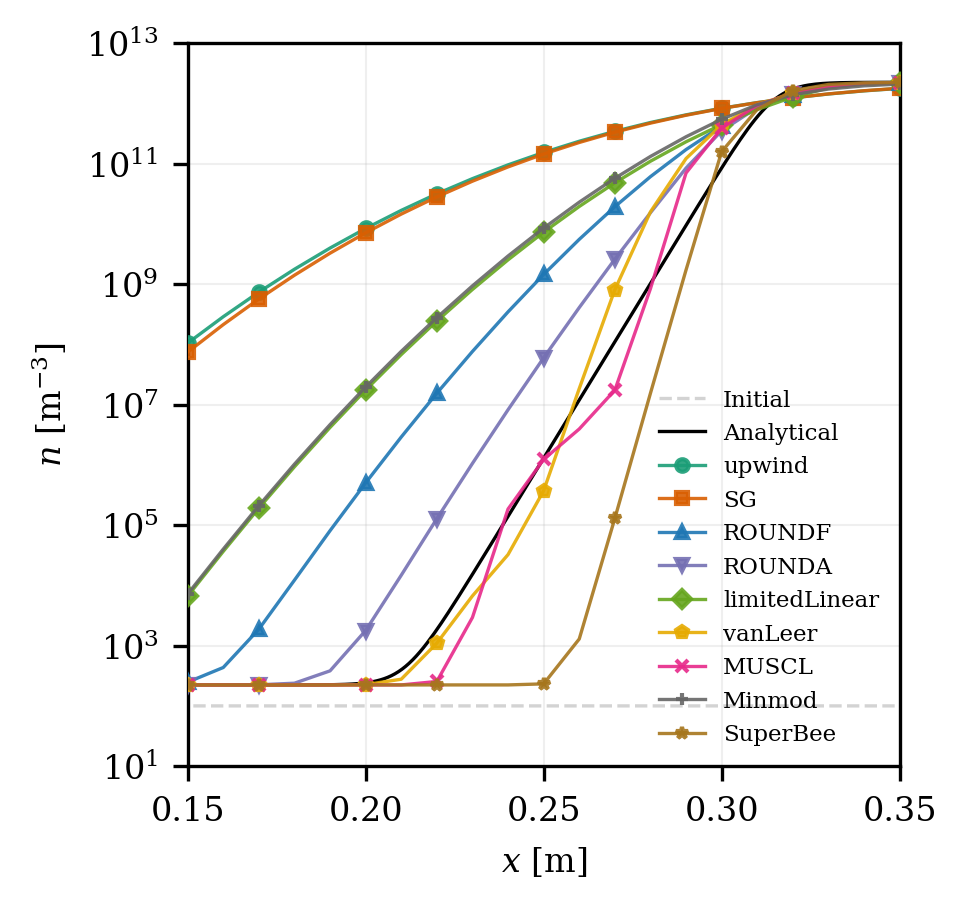}
        \caption{$100$ cells}
        \label{fig:stiff_scalar_advection_100_cells_comparison}
    \end{subfigure}
    \hspace{0.05\textwidth}
    \begin{subfigure}{0.45\textwidth}
        \centering
        \includegraphics[scale=1]{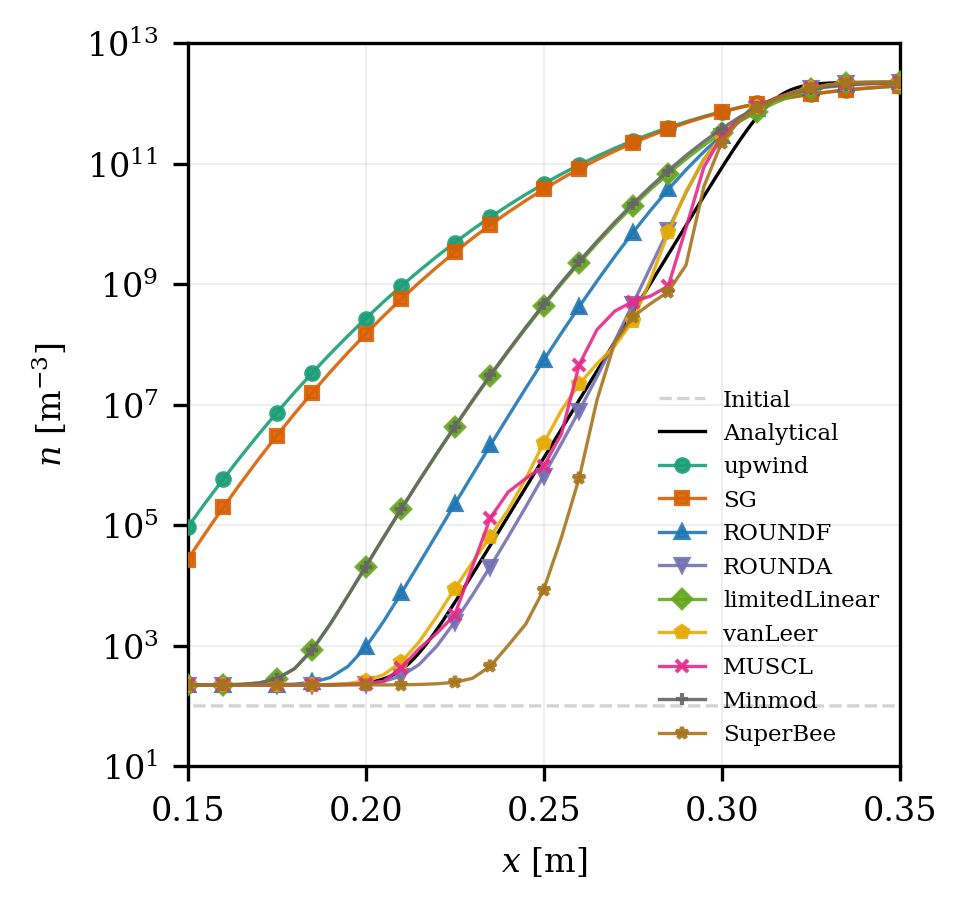}
        \caption{$200$ cells}
        \label{fig:stiff_scalar_advection_200_cells_comparison}
    \end{subfigure}

    \vspace{0.2cm} 

    \begin{subfigure}{0.45\textwidth}
        \centering
        \includegraphics[scale=1]{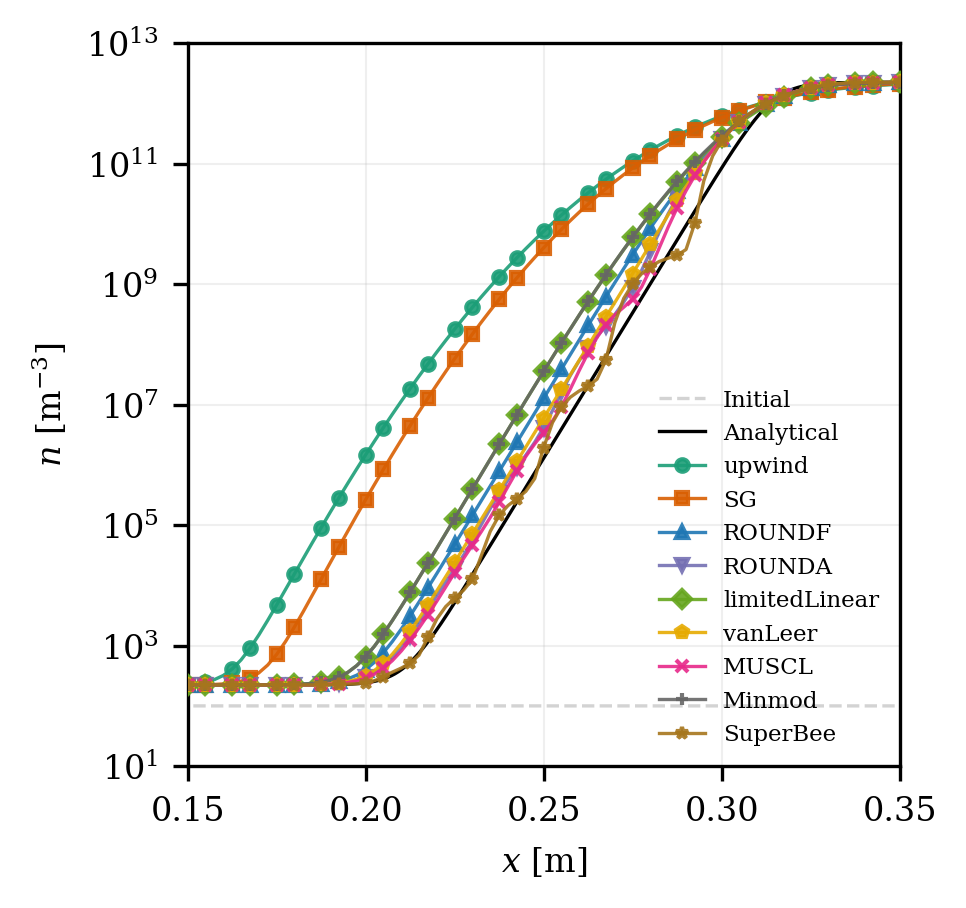}
        \caption{$400$ cells}
        \label{fig:stiff_scalar_advection_400_cells_comparison}
    \end{subfigure}

    \caption{Comparison of numerical flux schemes for the 1D stiff scalar advection problem under a linear electric field at $t_{end} = 8.0 \times 10^{-5}$~s. Results are presented for three mesh resolutions: (a) 100 cells, (b) 200 cells, and (c) 400 cells. The solid black line represents the analytical solution, illustrating the convergence and oscillatory behavior of the various TVD and ROUND schemes as the grid is refined.}
    \label{fig:stiff_scalar_advection_comparison}
\end{figure*}

\Cref{fig:stiff_scalar_advection_comparison} illustrates the numerical results for the three mesh resolutions across all evaluated flux schemes. As expected, all solutions converge toward the analytical solution as the grid is refined. The Scharfetter-Gummel (SG) scheme slightly outperforms the standard upwind method, though it maintains an accuracy order near unity. This is consistent with its behavior in high-field regions where the drift component dominates.

Among the remaining TVD schemes, the van Leer, MUSCL, and Superbee schemes exhibit oscillatory profiles. While these oscillations diminish with grid refinement for van Leer and MUSCL, they persist for the Superbee limiter. This behavior is expected, as the Superbee limiter operates at the upper boundary of the TVD stability region (\Cref{fig:tvd_diagram_scheme_limiters}), making it highly compressive and prone to "over-sharpening" gradients. In contrast, the limitedLinear (limiter=1.0) and Minmod schemes produce identical results. This overlap is anticipated, as their limiter functions are mathematically equivalent over a wide range of the gradient ratio $r$, as shown in the TVD diagram.

The highest performance is observed with the ROUNDF and ROUNDA schemes, which produce no artificial oscillations. On the coarsest mesh, ROUNDA demonstrates superior accuracy compared to ROUNDF. In the medium-resolution case, the ROUNDA profile exhibits a slight lag behind the analytical front, while both schemes show nearly identical performance on the finest mesh.

From these results, we can conclude that the ROUND schemes currently offer the highest consistency across varying resolutions. Furthermore, while the SG scheme performs according to theoretical expectations, standard TVD schemes only achieve comparable accuracy to the ROUND schemes when the mesh is sufficiently refined.

\section{Validation - DC glow discharge} \label{DC glow}

The first case for the validation of the proposed solver is the DC glow discharge regime, as presented by \cite{Derzsi_2009}. In that work, fluid models are employed for DC glow discharges, and we compare our results against their corresponding fluid model results. The configuration consists of a glow discharge occurring between a DC-powered electrode and a grounded electrode across a gap $L$. We consider two species: electrons and positive argon ions. We adopt a simple fluid approach under the Local Field Approximation (LFA), where we do not solve for the electron energy equation. The extension of the solver to the Local mean Energy Approximation (LMEA or LEA) is rather straightforward, but it is left out of the scope of this work.

We solve the drift-diffusion equations for both species, coupled with the semi-implicit Poisson equation. The ions are considered to be at a constant temperature $T_{gas}$. Their mobility is a function of the reduced electric field $E/N$ ($N$ being the gas density), while their diffusivity is obtained by the Einstein relation as $D_i / \mu_i = k T_{gas}/q$, where $k$ is the Boltzmann constant, $q$ is the electric charge of the ions and $kT_{gas}$ is their characteristic energy (here $T_{gas}$ in $K$). For electrons, the mobility is given as a constant from an experimental value, and the diffusivity is again obtained by the Einstein relation $D_e / \mu_e = k T_{e}$. Two different values of $kT_{e}$ are used, specifically $0.1$~eV and $1.0$~eV.

The source terms are calculated using a flux-based approach, $S(x) = \alpha \left( E/N \right) |\Gamma_e(x)|$, where $\alpha$ is Townsend's first ionization coefficient as a function of the reduced electric field, and $\Gamma_e$ is the total flux of the electrons. The Townsend's first ionization coefficient $\alpha$ is obtained from the work of \cite{Phelps_1999} as:

\begin{equation}
    \begin{split}
        \alpha/N = &1.1 \times 10^{-22} \exp \left[ \frac{-72}{E/N} \right] \\
        + &5.5 \times 10^{-21} \exp \left[ \frac{-187}{E/N} \right] \\
        + &3.2 \times 10^{-20} \exp \left[ \frac{-700}{E/N} \right] \\
        - &1.5 \times 10^{-20} \exp \left[ \frac{-10000}{E/N} \right]
    \end{split}
\end{equation}

For the boundary conditions, we apply Dirichlet $\phi = 0$ at the cathode and Dirichlet $\phi = V$ at the anode. To remain consistent with the original paper, the electron and ion densities are considered to be zero at the anode. At the cathode, the gradient of the ion density is taken to be zero (Neumann), while for the electrons only the secondary electron emission is taken into account on this boundary, based on the secondary electron emission coefficient $\gamma_i$ of the ions. The conditions involve low pressure, and two distinct cases are considered. The operational parameters of these cases are summarized in Table \ref{tab:glow_discharge_params}

\begin{table}[!htbp]
    \centering
    \caption{Operational parameters for the DC glow discharge validation cases.}
    \label{tab:glow_discharge_params}
    \begin{tabular}{l c c c}
    \hline
    \textbf{Parameter} & \textbf{Symbol [Unit]} & \textbf{Case 1} & \textbf{Case 2}\\ 
    \hline
    Pressure & $P$ $[Pa]$ & 133 & 40\\ 
    Gas temperature & $T_{gas}$ $[K]$ & 273 & 300\\ 
    Applied voltage & $V$ $[V]$ & 250 & 441\\ 
    Electrode gap & $L$ $[m]$ & 0.01 & 0.03\\ 
    SEE coefficient & $\gamma_i$ $[-]$ & 0.06 & 0.033\\ 
    \hline
    \end{tabular}
\end{table}

\subsection{Case 1 - Results}
\begin{figure}[!htbp]
    \centering
    \begin{subfigure}{0.45\textwidth}
        \centering
        \includegraphics[scale=0.9]{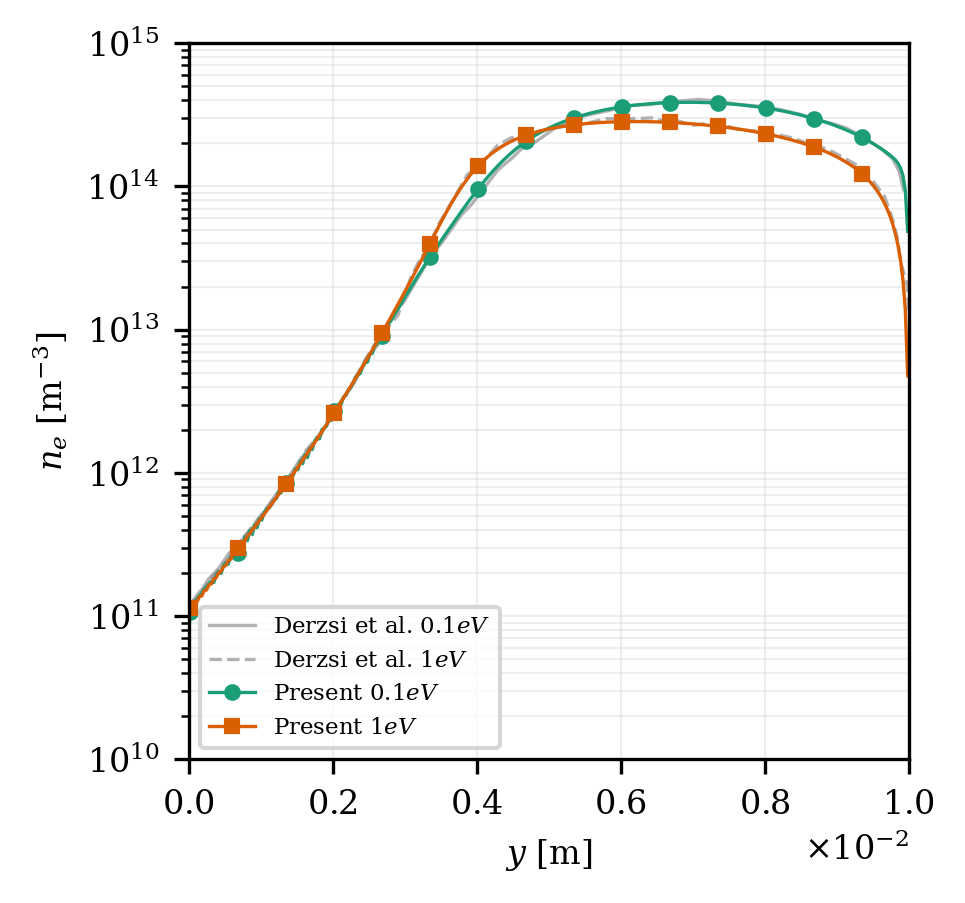}
        \caption{Electron number density}
        \label{fig:ne_gec_cell_p_133_Tg_273_V_250V_L_0.01_gamma_0.06}
    \end{subfigure}
    \begin{subfigure}{0.45\textwidth}
        \centering
        \includegraphics[scale=0.9]{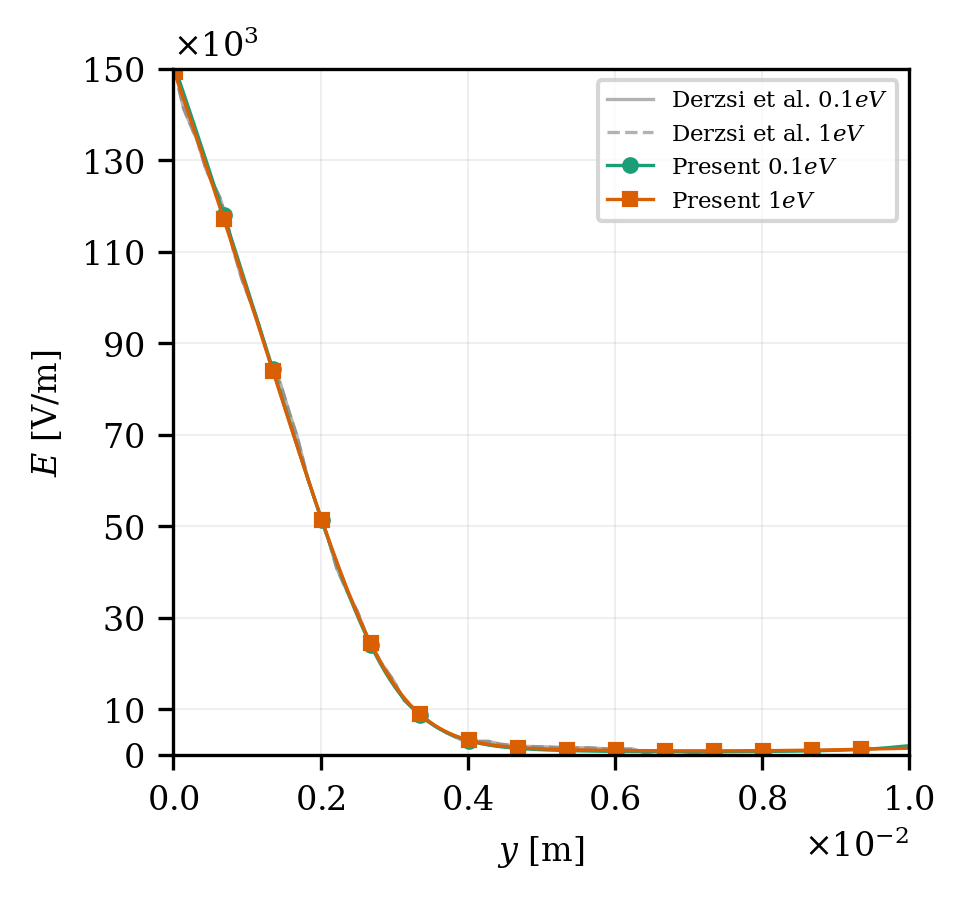}
        \caption{Electric field}
        \label{fig:E_gec_cell_p_133_Tg_273_V_250_L_0.01_gamma_0.06.png}
    \end{subfigure}
    \begin{subfigure}{0.45\textwidth}
        \centering
        \includegraphics[scale=0.9]{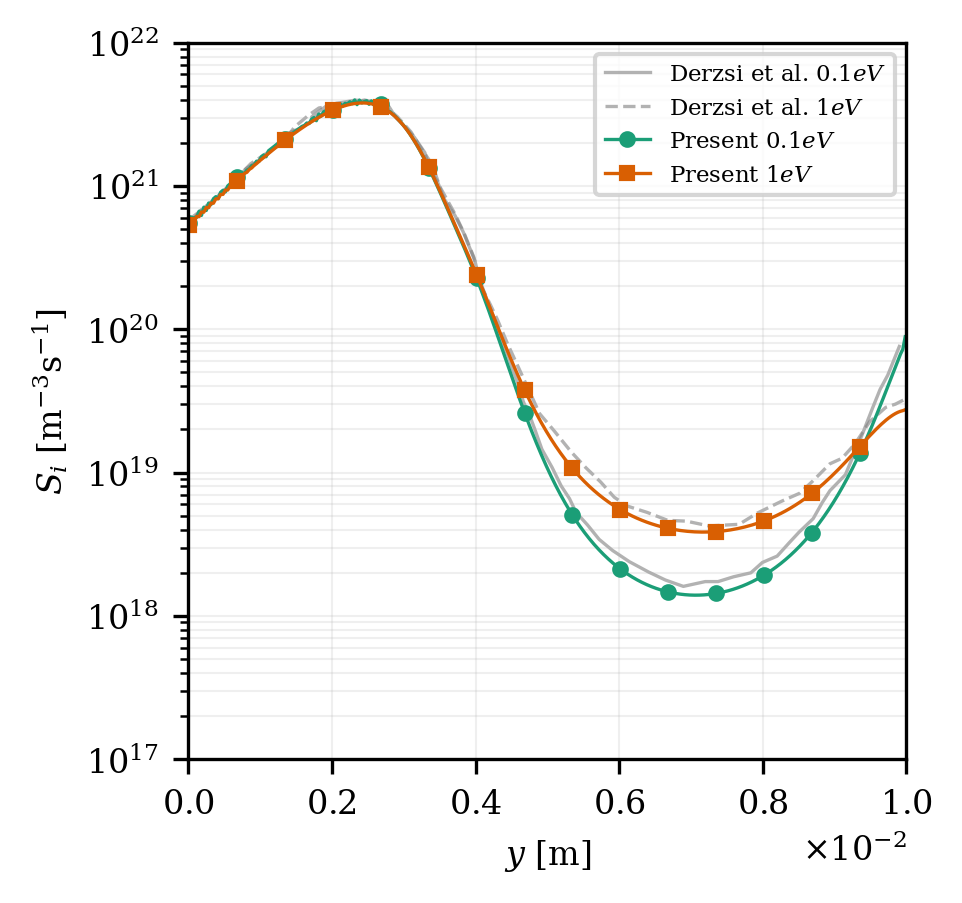}
        \caption{Ionization Source term}
        \label{fig:source_gec_cell_p_133_Tg_273_V_250_L_0.01_gamma_0.06}
    \end{subfigure}
    \caption{Comparison of (a) electron number density, (b) electric field, and (c) ionization source term for $kT_e = 0.1$~eV and $kT_e = 1.0$~eV, between the present work and the reference data from \cite{Derzsi_2009}. Results correspond to Case 1: $P = 133$~Pa, $T_{gas} = 273$~K, $V = 250$~V, $L = 0.01$~m, and $\gamma_i = 0.06$.}
    \label{fig:gec_cell_p_133_Tg_273_V_250_L_0.01_gamma_0.06}
\end{figure}

\Cref{fig:gec_cell_p_133_Tg_273_V_250_L_0.01_gamma_0.06} illustrates the comparison between the results from the present solver and the reference data for Case 1. The results are shown for the two different electron temperatures discussed previously. It can be seen that the obtained electron number densities and the electric field match the reference data perfectly. Regarding the source term values in the bottom plot, there are very small variations between our results and the reference. However, these can be attributed to the manual digitization process used to extract data from the original plots, as the raw numerical data were not available.

\subsection{Case 2 - Results}
\begin{figure}[!htbp]
    \centering
    \begin{subfigure}{0.45\textwidth}
        \centering
        \includegraphics[scale=0.9]{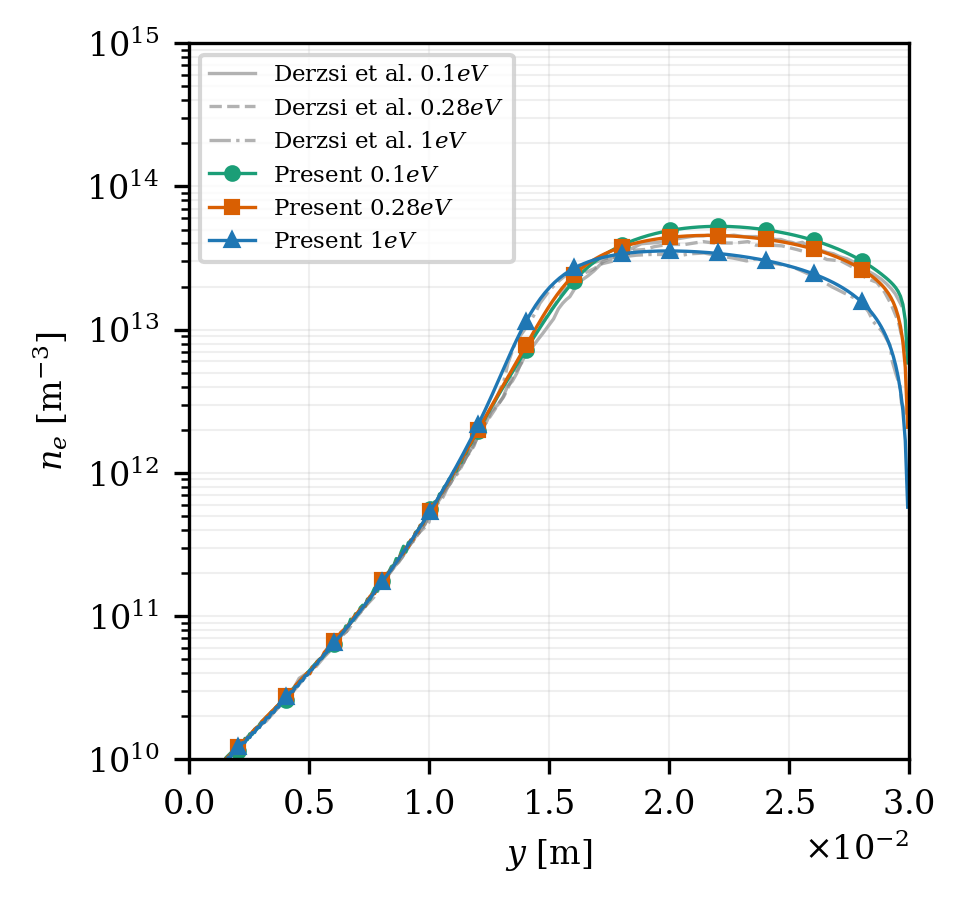}
        \caption{Electron number density}
        \label{fig:ne_gec_cell_p_40_Tg_300_V_441_L_0.03_gamma_0.033}
    \end{subfigure}
    \begin{subfigure}{0.45\textwidth}
        \centering
        \includegraphics[scale=0.9]{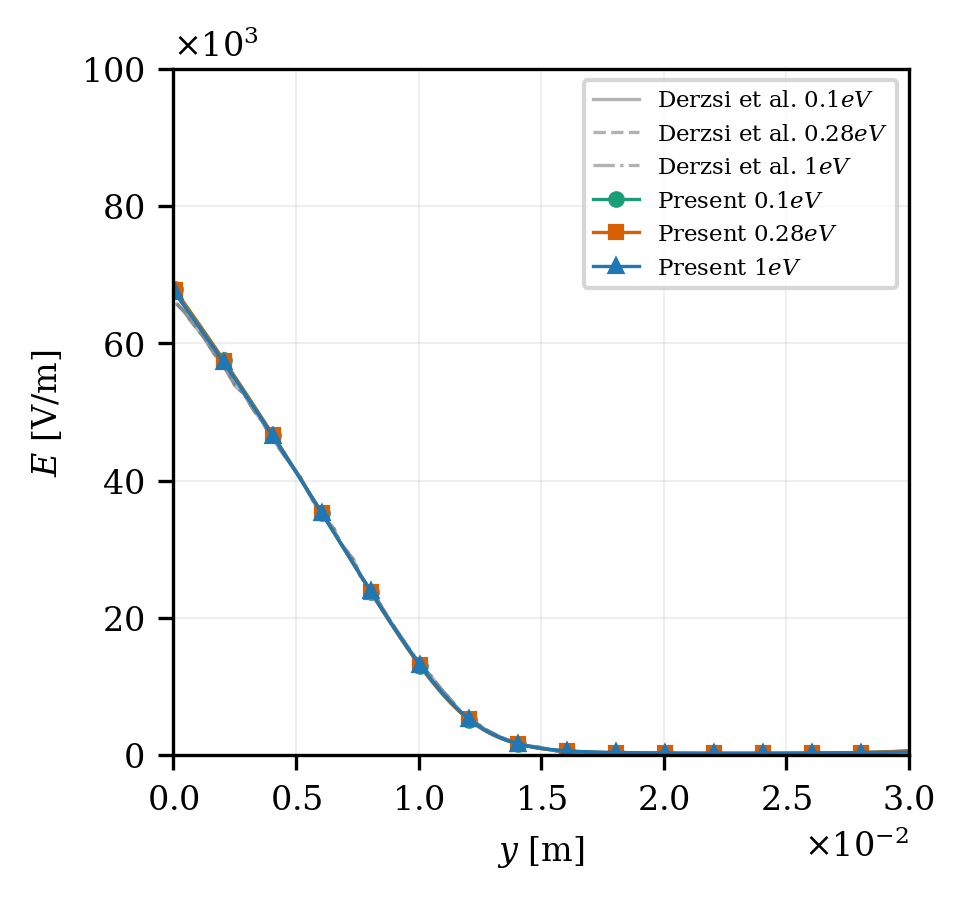}
        \caption{Electric field}
        \label{fig:E_gec_cell_p_40_Tg_300_V_441_L_0.03_gamma_0.033}
    \end{subfigure}
    \begin{subfigure}{0.45\textwidth}
        \centering
        \includegraphics[scale=0.9]{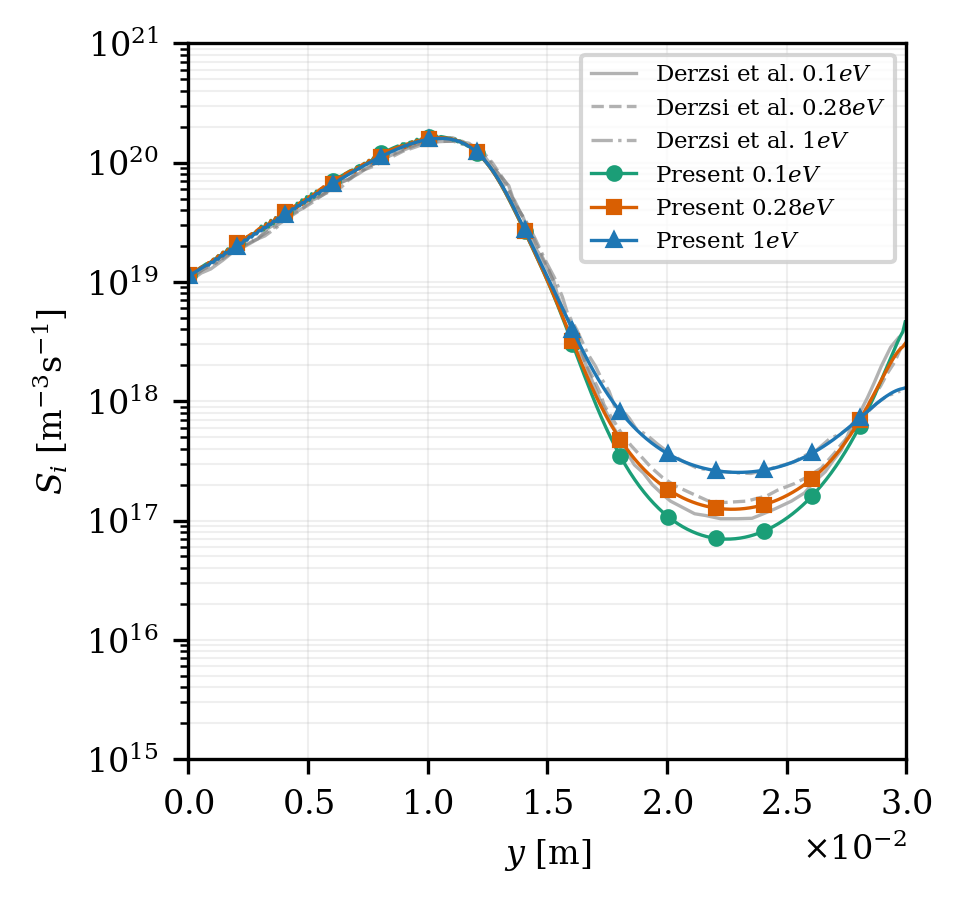}
        \caption{Ionization Source term}
        \label{fig:source_gec_cell_p_40_Tg_300_V_441_L_0.03_gamma_0.033}
    \end{subfigure}
    \caption{Comparison of (a) electron number density, (b) electric field, and (c) ionization source term for $kT_e = 0.1$~eV, $kT_e = 0.28$~eV and $kT_e = 1.0$~eV, between the present work and the reference data from \cite{Derzsi_2009}. Results correspond to Case 2: $P = 40$~Pa, $T_{gas} = 300$~K, $V = 441$~V, $L = 0.03$~m, and $\gamma_i = 0.033$.}
    \label{fig:gec_cell_p_40_Tg_300_V_441_L_0.03_gamma_0.033}
\end{figure}

\Cref{fig:gec_cell_p_40_Tg_300_V_441_L_0.03_gamma_0.033} illustrates the results for Case 2. Here, an additional case for $kT_e = 0.28$~eV is included following the reference study. As with the previous case, the obtained electron number densities and the electric field match the reference data perfectly. Regarding the source terms, there is a very small difference observed for the $0.28$~eV and $1.0$~eV cases, which can again be attributed to the manual digitization of the data from the original plots.

\section{Validation - Freely propagating positive streamer in air} \label{streamer}

Next, the developed solver is validated against a more demanding case, the positive streamer, which is widely documented in the literature. Specifically, we adopt the study by \cite{Bagheri2018} as primary reference. This publication investigates three distinct cases that vary in background plasma densities and the inclusion of photoionization. In the present work, we focus on the two cases involving varying background electron and ion densities. The photoionization case is considered out of scope for this paper (integration of photoionization models is under development). The only difference between Case A and Case B is the background ion and electron density. In Case A, the densities are set to $n_e = n_i = 10^{13} \text{ m}^{-3}$. This represents a relatively high-density background, providing a significant population of seed electrons to facilitate the ionization process. In contrast, Case B utilizes a lower background density of $n_e = n_i = 10^9 \text{ m}^{-3}$. Comparing these two cases allows us to evaluate the solver’s performance under varying levels of pre-ionization and its ability to handle the steeper gradients associated with lower background populations.

The primary objective of this case is to benchmark the solver against the established results in the literature. However, in this case, we also conduct a study on the performance of the different flux schemes discussed earlier. Furthermore, we investigate the effect of the number of outer correctors, utilizing PIMPLE-based fixed-point iterations, on the coupling between species transport and the Poisson equation. We also evaluate the computational performance of the solver relative to other software used in the benchmark, and test the AMR capability on the streamer problem.

\subsection{Case configuration}

\begin{figure}[!h]
    \centering
    \includegraphics[scale=1]{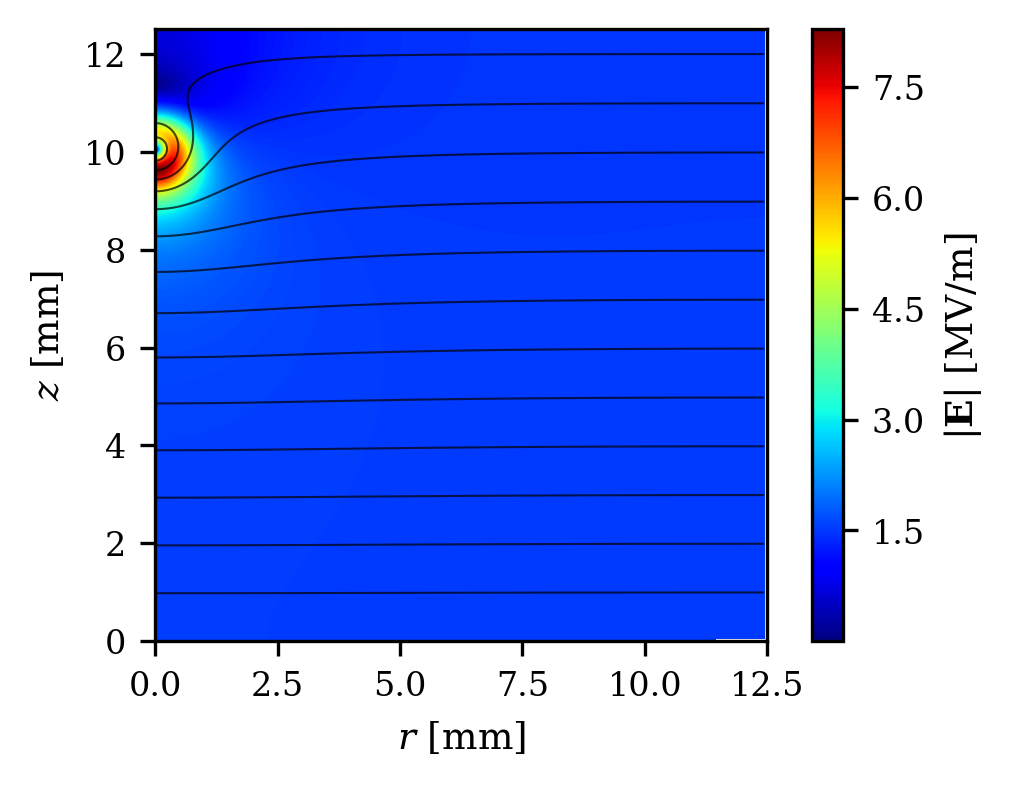}
\caption{Initial configuration for the positive streamer benchmark cases \cite{Bagheri2018}. Electric field magnitude $|\mathbf{E}|$ [kV/cm] with equipotential lines over the $1.25 \times 1.25$~cm$^2$ domain. The $-15$~kV/cm background field is set by an $18.75$~kV difference between the powered top and grounded bottom electrodes, and the axial field enhancement at $r=0$ comes from a Gaussian ion seed at $z_0 = 1$~cm.}
    \label{fig:initial_electric_field_positive_streamer}
\end{figure}

The problem is modeled using a 2D axisymmetric configuration. The computational domain is a square in the $(r, z)$ plane with dimensions $L_r = L_z = 1.25$ cm. The top boundary ($z = L_z$) acts as the powered electrode, while the bottom boundary ($z = 0$) is grounded. A constant potential of $18.75$ kV is applied to the top electrode, establishing a homogeneous background electric field of $-15$ kV/cm, which is kept below the breakdown threshold of the medium. To locally enhance the electric field and facilitate the initialization and directional control of the streamer, a localized electric field enhancement is induced by an initial seed of positive ions. Following the methodology in \cite{Bagheri2018}, the initial ion density profile is defined by a Gaussian distribution:

\begin{equation}
    n_i(r,z) = N_0 \exp \left[ -\frac{r^2 + (z-z_0)^2}{\sigma^2}\right]
\end{equation}
where the parameters are set to $N_0 = 5 \times 10^{18}$ m$^{-3}$, $\sigma = 0.4$ mm, and the seed center is located at $z_0 = 1$ cm. This initial patch of ions creates the necessary space-charge field to initiate the streamer propagation toward the grounded electrode. The initial electric field due to the applied voltage and the ion seed is illustrated in \Cref{fig:initial_electric_field_positive_streamer}.

\subsection{Transport coefficients and source terms}

In this simplified physical model, two species are considered: electrons and positive ions. Due to the short timescales associated with streamer propagation, the ions are assumed to be immobile. The governing transport equations are formulated as follows:

\begin{subequations}
    \begin{align}
        \frac{\partial n_e}{\partial t} + \nabla \cdot \left( -\mu_e n_e \mathbf{E} - D_e \nabla n_e \right) &= \bar{\alpha} \mu_e |\mathbf{E}| n_e \\
        \frac{\partial n_i}{\partial t} = \bar{\alpha} \mu_e |\mathbf{E}| n_e
    \end{align}
\end{subequations}
where $n_e$ and $n_i$ are the electron and ion number densities, respectively. The effective ionization coefficient, $\bar{\alpha} = \alpha - \eta$, accounts for both the Townsend ionization coefficient ($\alpha$) and the electron attachment coefficient ($\eta$).

The transport coefficients and source terms are derived from the analytical expressions provided by \cite{Bagheri2018}. These expressions describe a streamer discharge in dry air ($80\%~\text{N}_2$ and $20\%~\text{O}_2$) at a pressure of $1$~bar and a temperature of $300$~K. Given a gas number density of $N = 2.414 \times 10^{25}~\text{m}^{-3}$, the transport coefficients and source terms are expressed as functions of the electric field magnitude $ |\mathbf{E}|$. The field-dependent coefficients, with the electric field magnitude $|\mathbf{E}|$ expressed in $\text{V/m}$, are given by:

\begin{subequations}
\label{eq:transport_coefficients}
\begin{align}
    \mu_e  &= 2.398 |\mathbf{E}|^{-0.26} \hfill [\text{m}^2 \text{V}^{-1} \text{s}^{-1}] \\
    D_e    &= 4.3628 \times 10^{-3} |\mathbf{E}|^{0.22} \hfill [\text{m}^2 \text{s}^{-1}] \\
    \begin{split}
    \alpha &= \left( 1.1944 \times 10^6 + \frac{4.3666 \times 10^{26}}{|\mathbf{E}|^3} \right) \\
           & \quad \times \exp \left( \frac{-2.73 \times 10^7}{|\mathbf{E}|} \right) \hfill [\text{m}^{-1}]
    \end{split} \\
    \eta   &= 340.75 \hfill [\text{m}^{-1}]
\end{align}
\end{subequations}

\subsection{Case A - Results}

\begin{figure}[!h]
    \centering
    \includegraphics[scale=0.995]{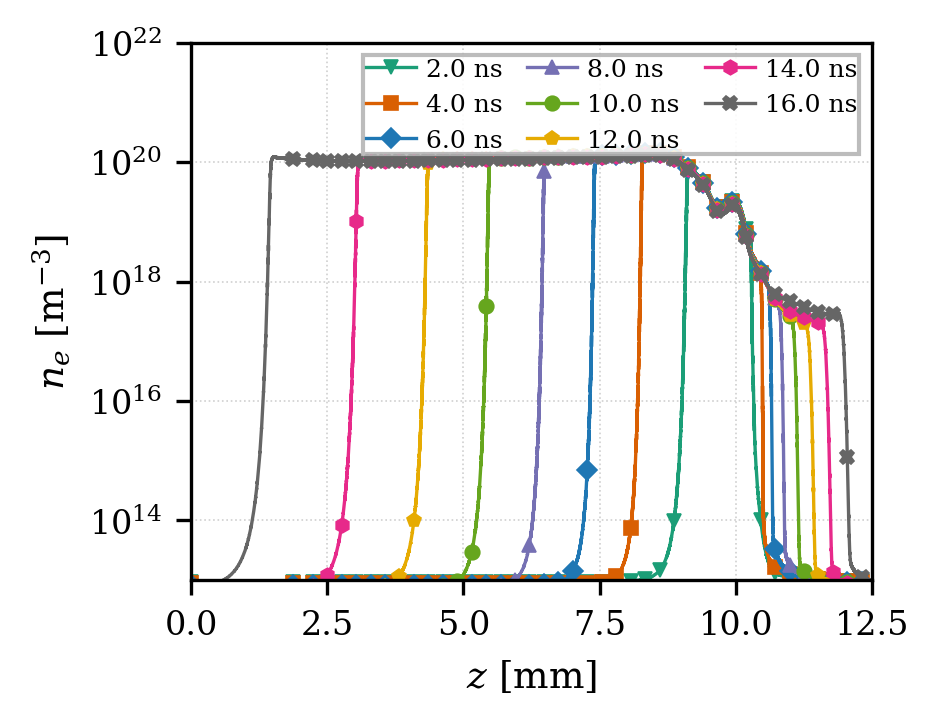}
    \caption{Electron number density profiles along the symmetry axis ($r = 0$)
    at eight time instances ($t = 2, 4, 6, 8, 10, 12, 14$ and $16$\,ns)
    for the positive streamer Case~A, with a background ionization of
    $10^{13}\,\text{m}^{-3}$ for both electrons and ions.}
    \label{fig:ne_vs_z_positive_streamer_high_ionization}
\end{figure}

\begin{figure}[!h]
    \centering
    \includegraphics[scale=0.98]{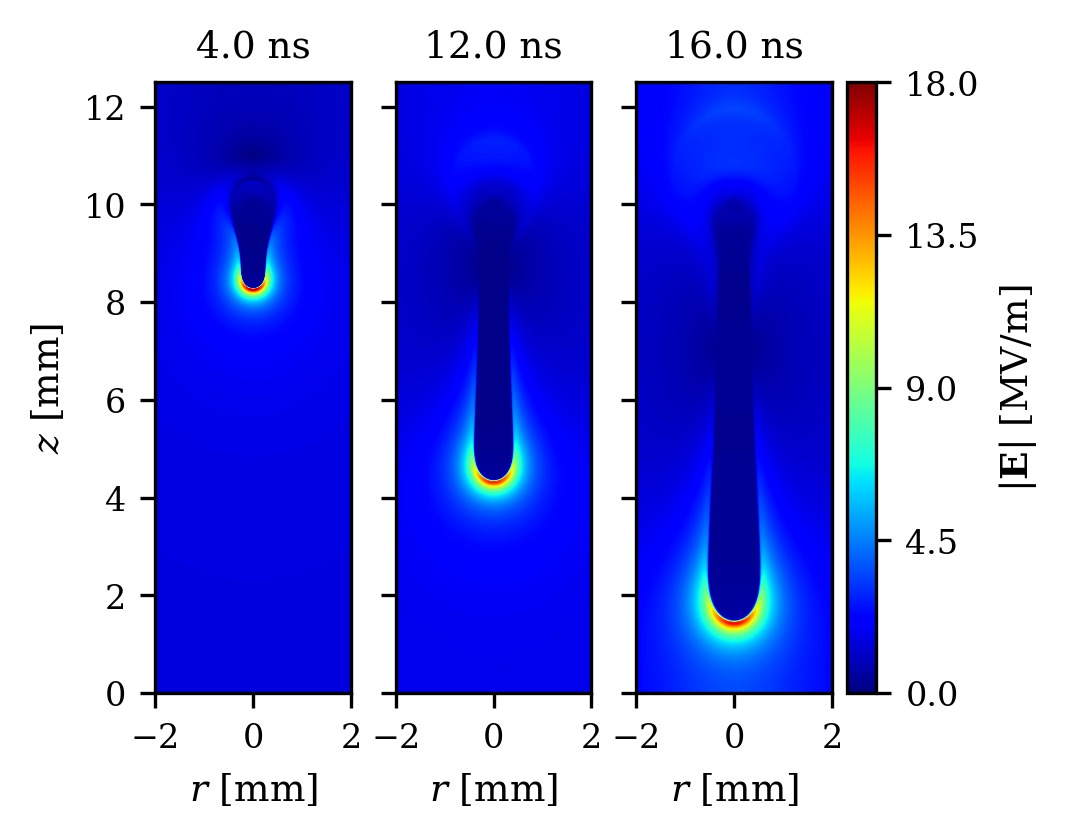}
    \caption{Electric field magnitude (MV/m) for the positive streamer Case A at $t = 4$, $12$ and $16$\,ns, with a background ionization of $10^{13}\,\text{m}^{-3}$ for electrons and ions. The axisymmetric slice is mirrored about the symmetry axis.}
    \label{fig:E_streamer_high_ionization_contours}
\end{figure}

Case A, characterized by a higher background ionization level, is compared against the streamer codes provided in \cite{Bagheri2018} as well as the results from \cite{AVIP}. \Cref{fig:ne_vs_z_positive_streamer_high_ionization} illustrates the electron number density profiles along the symmetric axis ($r =0$) at eight time instances. \Cref{fig:E_streamer_high_ionization_contours} shows the evolution of the electric field magnitude. At $t = 4$\,ns, the streamer has just started to develop, with the field enhancement concentrated in a compact region close to the anode. By $t = 12$\,ns, the streamer has propagated significantly downward along the symmetry axis, exhibiting the characteristic streamer head with a strong localized electric field at its tip and a much weaker field in the conductive channel left behind. At $t = 16$\,ns, the streamer has crossed most of the gap and the head is approaching the lower boundary, while the channel behind it remains quasi-neutral with a reduced internal field.

\begin{figure*}[t]
    \centering
    \begin{subfigure}{1\textwidth}
        \centering
        \includegraphics[scale=1]{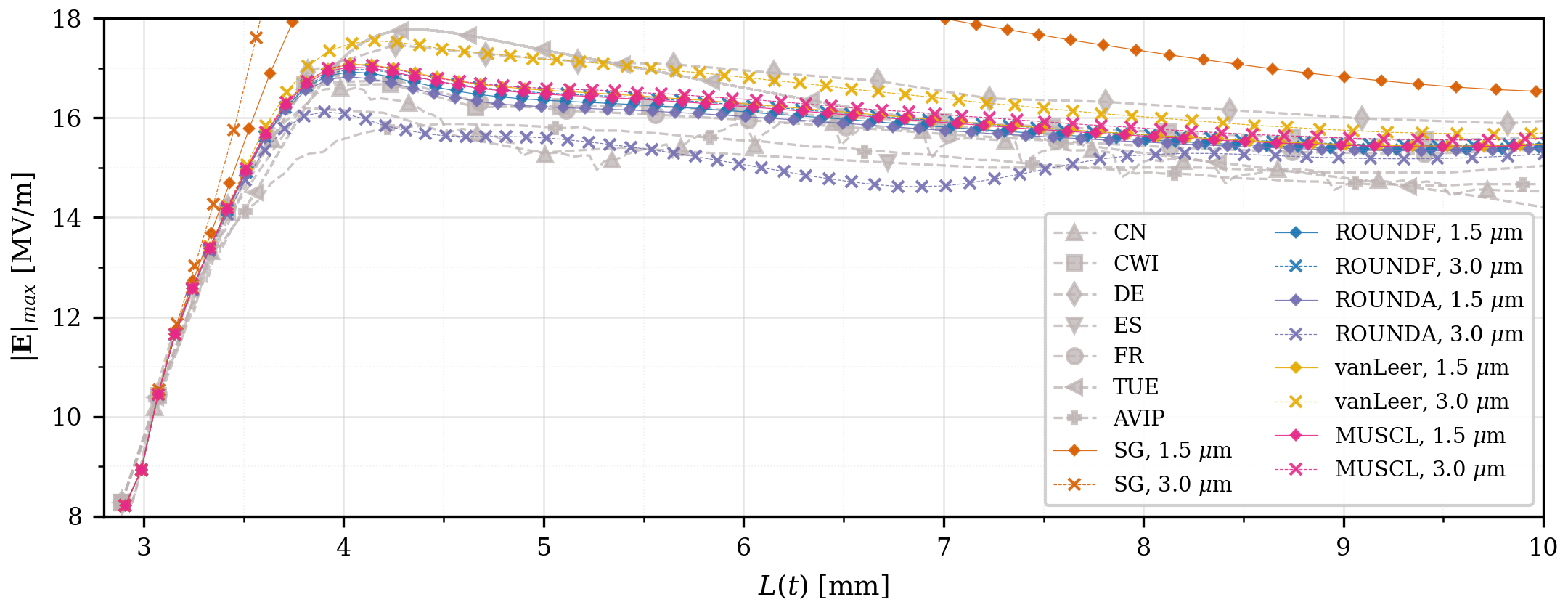}
        \caption{}
        \label{fig:Emax_vs_time_explicit_comparison}
    \end{subfigure}
    \begin{subfigure}{1\textwidth}
        \centering
        \includegraphics[scale=1]{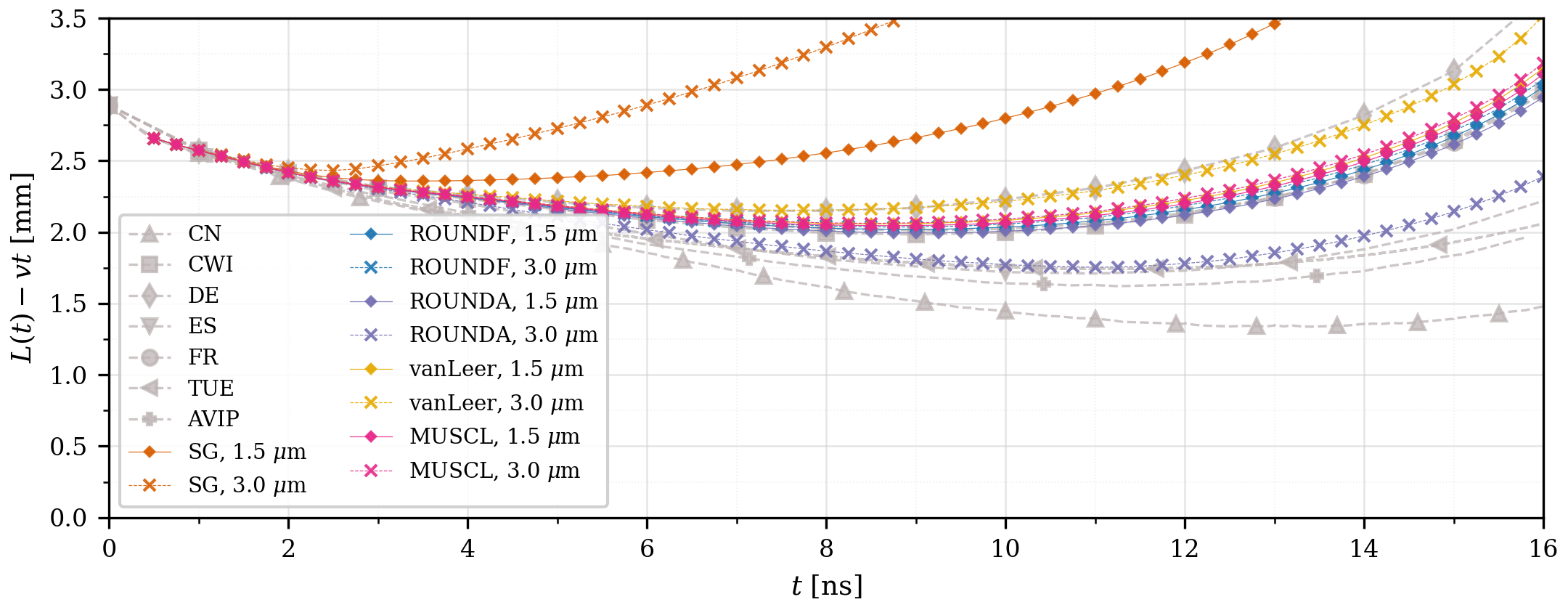}
        \caption{}
        \label{fig:streamer_length_minus_vt_vs_time_explicit_comparison}
    \end{subfigure}
    \caption{Validation of the high-background ionization streamer (Case A) against reference data from \cite{Bagheri2018} and \cite{AVIP}: (a) maximum absolute electric field, (b) reduced streamer length $z - v_0 t$ (where $v_0 = 0.5\,\text{mm\,ns}^{-1}$). The results compare the proposed solver using five flux schemes at mesh resolutions of $3~\mu\text{m}$ and $1.5~\mu\text{m}$ with benchmark data from several research groups.}
    \label{fig:positive_streamer_high_ionization_explicit_comparisons}
\end{figure*}

\Cref{fig:positive_streamer_high_ionization_explicit_comparisons} illustrates the maximum absolute electric field along the streamer length and the reduced streamer length ($z - v_0 t$, where $v_0 = 0.5\,\text{mm\,ns}^{-1}$ according to the reference paper). The comparison includes seven reference datasets and a collection of cases run with the present solver. These internal cases utilize five different flux schemes at two mesh resolutions: a coarse grid ($3~\mu\text{m}$) and a refined grid ($1.5~\mu\text{m}$). The results demonstrate that all TVD schemes tend to converge toward a single curve as the mesh is refined. Notably, the SG scheme moves closer to the other solutions as the mesh becomes denser; however, even at the highest resolution, it remains significantly far from the reference results, overestimating the electric field, and substantially increasing the streamer propagation speed. It is also worth mentioning that upon mesh refinement, some schemes increase their predicted electric field values while others decrease them, a behavior previously observed in \cite{Bagheri2018}. The TVD schemes, MUSCL, ROUNDA, ROUNDF, and vanLeer appear to converge on the same curve. However, on the coarser mesh, ROUNDA exhibits some oscillations, while ROUNDF and MUSCL provide stable and consistent results. Based on the comparison with the reference data, our solver aligns most closely with the results obtained by the CWI, DE, and ES research groups.

\subsubsection{Coupling loop sensitivity analysis}

\begin{figure*}[!htbp]
    \centering
    \includegraphics[scale=0.96]{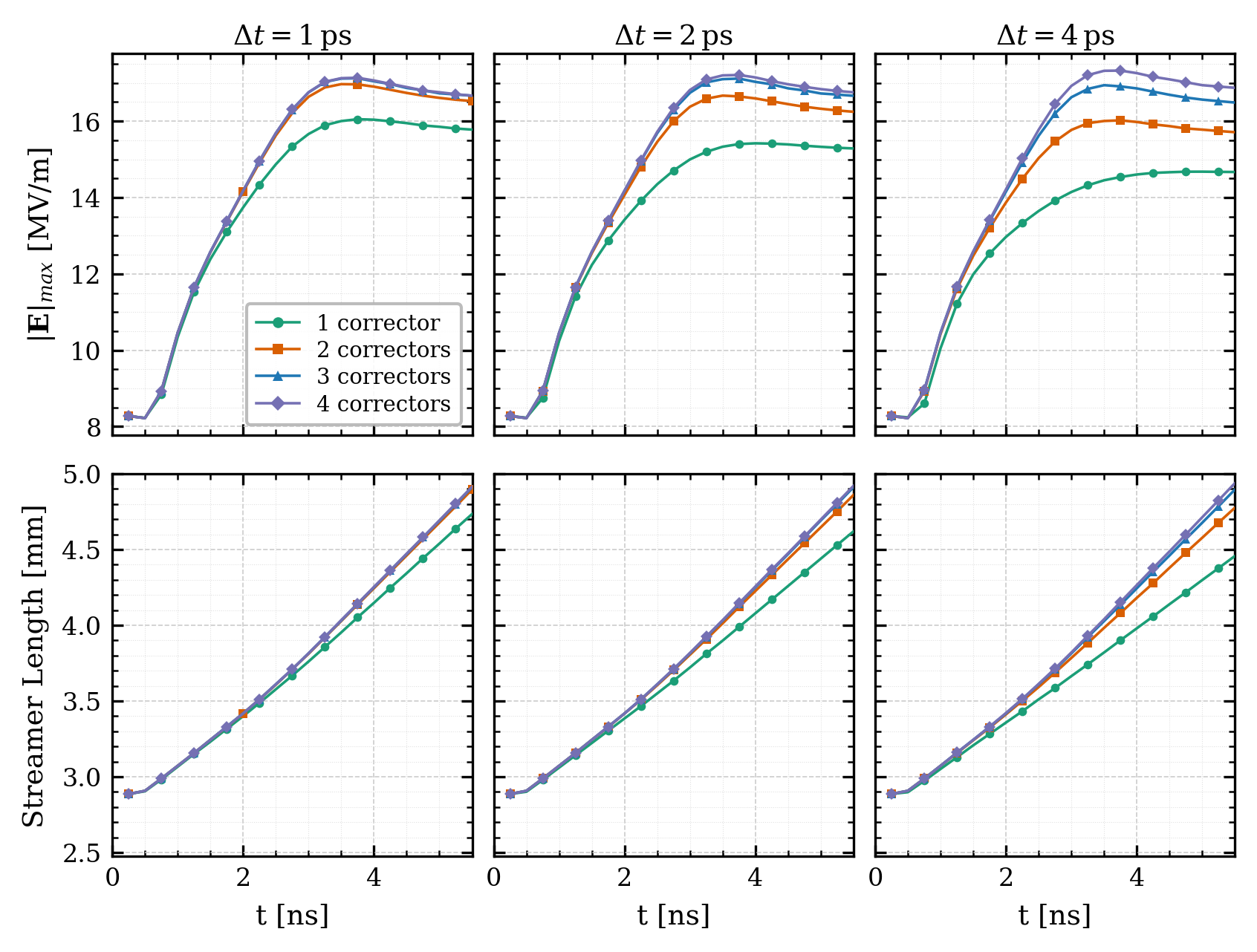}
    \caption{Convergence study of the outer PIMPLE iterations for the positive streamer Case~A: (a) maximum absolute electric field $|\mathbf{E}|_{\max}$ and (b) streamer length $L(t)$, both as a function of time up to 5.5\,ns. Results are shown for time steps $\Delta t = 1$\,ps, 2\,ps and 4\,ps, each with 1, 2, 3 and 4 outer PIMPLE correctors, on a 3\,$\mu$m mesh using the ROUNDF flux scheme.}
    \label{fig:maxE_and_length_1_2_4ps_1_2_3_4corr_explicit_3um_positive_streamer}
\end{figure*}

\begin{figure*}[!htbp]
    \centering
    \includegraphics[scale=0.96]{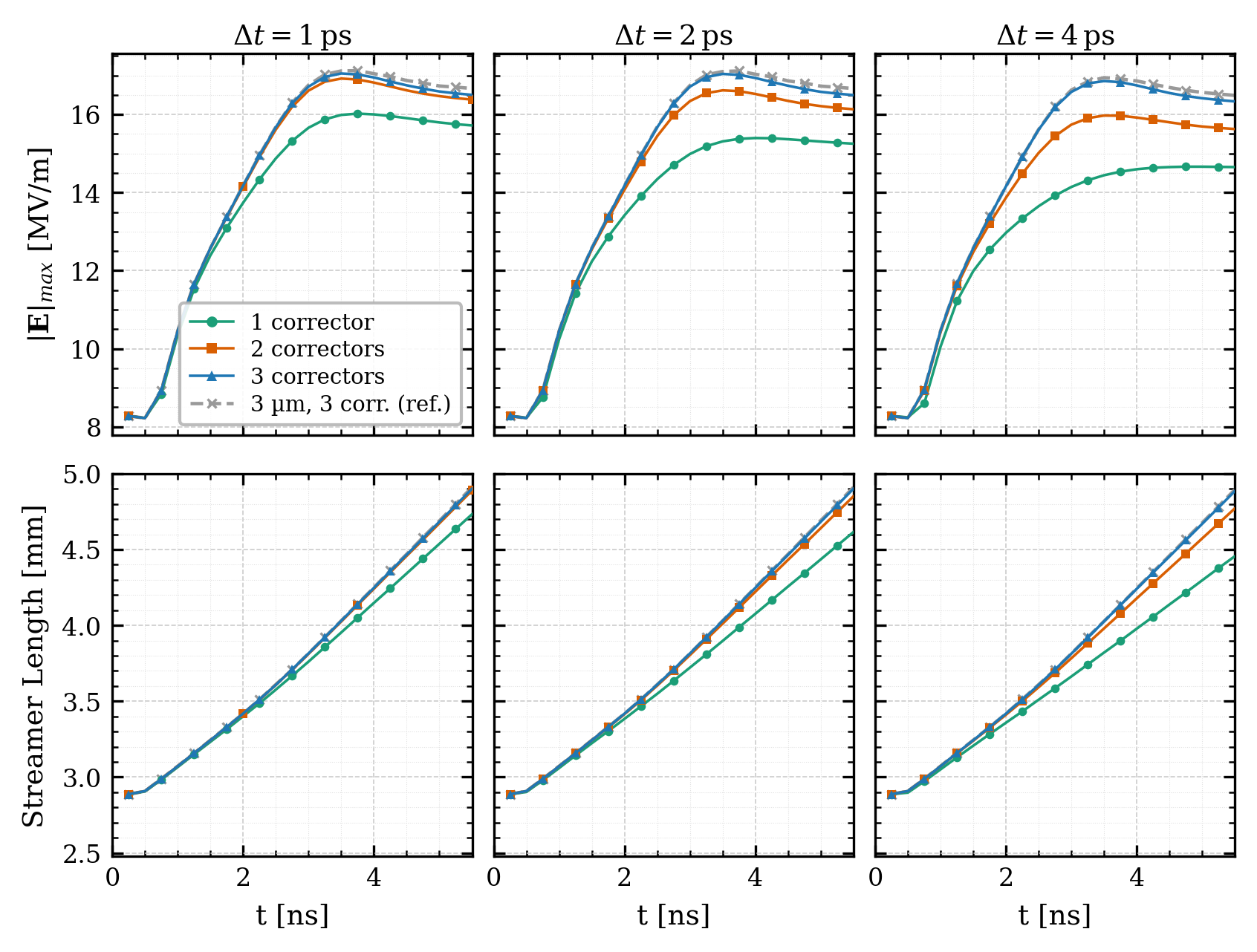}
    \caption{Convergence study of the outer PIMPLE iterations for the positive streamer Case~A: (a) maximum absolute electric field $|\mathbf{E}|_{\max}$ and (b) streamer length $L(t)$, both as a function of time up to 5.5\,ns. Results are shown for time steps $\Delta t = 1$\,ps, 2\,ps and 4\,ps, each with 1, 2 and 3 outer PIMPLE correctors, on a 1.5\,$\mu$m mesh using the ROUNDF flux scheme. The 3\,$\mu$m, 3-corrector result is included as a reference.}
    \label{fig:maxE_and_length_1_2_4ps_1_2_3_4corr_explicit_1.5um_positive_streamer}
\end{figure*}

\begin{figure*}[!htbp]
    \centering
    \includegraphics[scale=0.96]{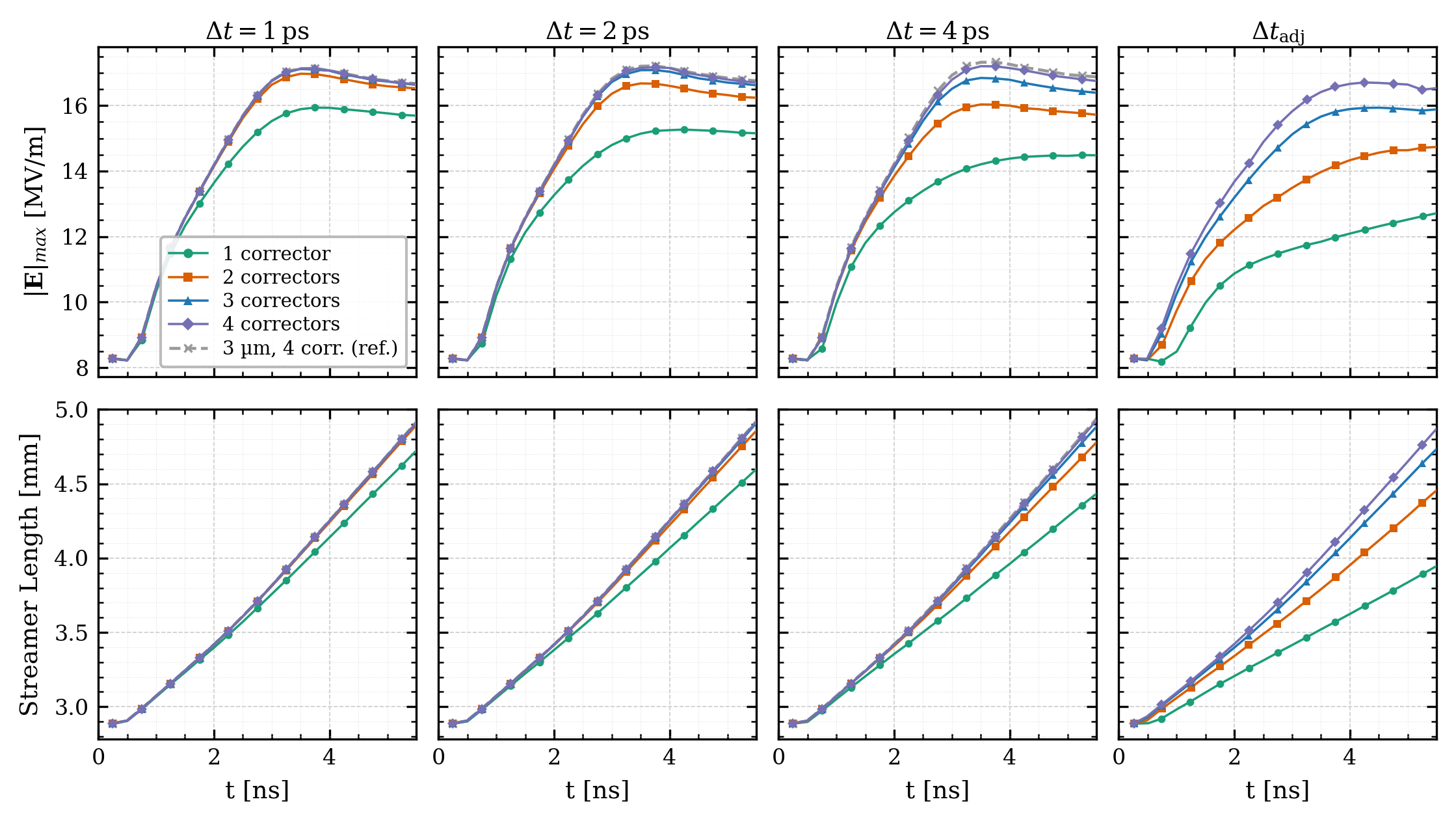}
    \caption{Convergence study of the outer PIMPLE iterations for the positive streamer Case~A using a semi-implicit Poisson scheme: (a) maximum absolute electric field $|\mathbf{E}|_{\max}$ and (b) streamer length $L(t)$, both as a function of time up to 5.5\,ns. Results are shown for time steps $\Delta t = 1$\,ps, 2\,ps and 4\,ps, and an adjustable time step with maximum dielectric relaxation ratio $C_{\varepsilon} = \text{2}$, each with 1, 2, 3 and 4 outer PIMPLE correctors, on a 3\,$\mu$m mesh using the ROUNDF flux scheme. The explicit Poisson, 4-corrector result is included as a reference.}
    \label{fig:maxE_and_length_1_2_4ps_adj_1_2_3_4corr_semiImplicit_3um_positive_streamer}
\end{figure*}

An important point of the present study is assessing the differences obtained in a streamer case using different numbers of outer PIMPLE corrections, which enhance the coupling between the Poisson equation and the species continuity equations. For this reason we performed the following tests. For Case~A, we ran the simulation from the beginning until the electric field reaches its maximum and starts dropping, on a 3\,$\mu$m mesh, with $\Delta t$ of 1\,ps, 2\,ps and 4\,ps, solving the two equations once per time step, and then with 2, 3 and 4 correctors (all using the ROUNDF scheme). \Cref{fig:maxE_and_length_1_2_4ps_1_2_3_4corr_explicit_3um_positive_streamer} shows the results for the maximum electric field magnitude and the streamer length against time (up to 5.5\,ns). The results are very telling.

For $\Delta t = 1$\,ps, solving the equations once underpredicts both the electric field and the streamer length, giving a maximum of 16\,MV/m, while 2, 3 and 4 correctors all converge to 17\,MV/m. Increasing the time step to $\Delta t = 2$\,ps, the 1-corrector case drops even lower to around 15.5\,MV/m, the 2-corrector case to around 16.5\,MV/m, while 3 and 4 
correctors converge to the same 17\,MV/m. For $\Delta t = 4$\,ps the coupling is much looser.  The 1-corrector case underpredicts substantially, also altering the profile of the maximum $|\mathbf{E}|$, and underpredicting the streamer length, with 4 corrections needed to reach the converged value. These results highlight how sensitive the positive streamer case is to the strength of the Poisson-continuity coupling. It is also worth noting that for the 
1\,ps case, the maximum dielectric relaxation ratio is around 0.3 and the maximum convective Courant number around 0.23, showing that even for values well below 1.0 the coupling still needs to be stricter.

The same cases are run for the 1.5\,$\mu$m mesh (up to 3 correctors, for computational economy). The corresponding results are shown in 
\Cref{fig:maxE_and_length_1_2_4ps_1_2_3_4corr_explicit_1.5um_positive_streamer}, where the 3\,$\mu$m, 3-corrector case is also included as a reference. For $\Delta t = 1$\,ps we obtain the same behavior as before: the 1-corrector case underpredicts both $|\mathbf{E}|_{\max}$ and the streamer length, while 2 and 3 correctors converge to the same value. For $\Delta t = 2$\,ps, 3 correctors are again needed for convergence, and for $\Delta t = 4$\,ps the results mirror those of the 3\,$\mu$m mesh, confirming that 4 correctors would be required. Crucially, the results are nearly identical to the 3\,$\mu$m case, demonstrating that refining the mesh does not compensate for the loss of temporal accuracy that arises when the Poisson and continuity equations are not tightly coupled.

Finally, the same cases are employed for the 3\,$\mu$m mesh using a semi-implicit Poisson scheme, in order to test whether the semi-implicit coupling can strengthen the temporal accuracy of the Poisson-continuity coupling. This time we use up to 4 correctors, and we additionally include a case with an adjustable time step, constrained by a maximum dielectric relaxation ratio of $C_{\varepsilon} = \text{2}$. The explicit Poisson results for the 3\,$\mu$m 
mesh with 4 correctors are included as a reference for the 1\,ps, 2\,ps and 4\,ps cases. From the results shown in \Cref{fig:maxE_and_length_1_2_4ps_adj_1_2_3_4corr_semiImplicit_3um_positive_streamer}, it is clear that the semi-implicit Poisson coupling alone cannot overcome the need for outer PIMPLE iterations: the same sensitivity to the number of correctors is observed as in the explicit case. For the adjustable time step this is even more apparent. The 1-corrector case severely distorts the electric field profile, failing to capture the dynamics correctly, and only the 4-corrector case approaches the correct behavior.

\subsubsection{AMR simulation}
\begin{figure}[!h]
    \centering
    \includegraphics[scale=1]{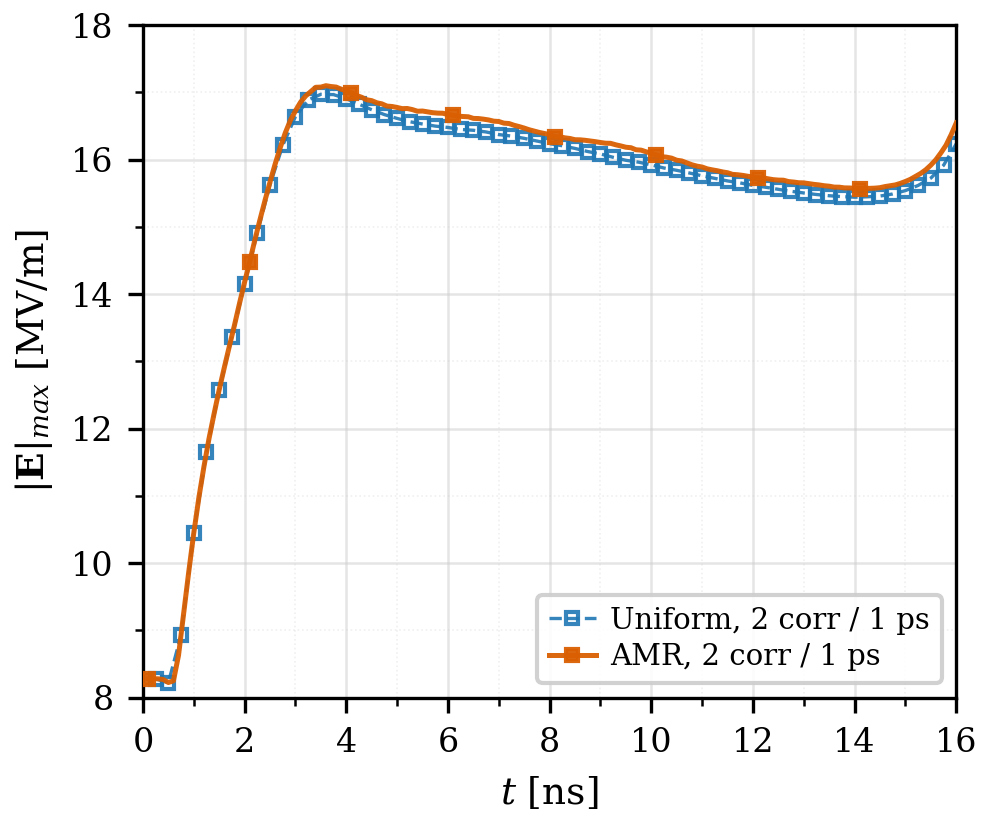}
    \caption{Maximum absolute electric field as a function of streamer length 
    for the AMR case compared against the uniform $3~\mu$m 
    mesh result. The AMR configuration produces results in very close 
    agreement with the uniform fine mesh case.}
    \label{fig:Emax_vs_time_AMR_comparison}
\end{figure}

The solver is also tested with the AMR capability, achieved through the 
blastAMR library \cite{blastAMR}. After testing numerous combinations of 
AMR parameters and initial mesh resolutions, we conclude that a very good 
configuration is to start with a cell size of $24~\mu$m in the streamer 
region and allow 3 levels of refinement with a refinement ratio of 2, 
yielding a minimum cell size of $3~\mu$m. The refinement criterion is based 
on the Townsend ionization coefficient $\alpha$, following the approach of 
\cite{Bagheri2018, Shao_2025}. Specifically, we refine when 
$\alpha \Delta x > 0.5$ and unrefine when $\alpha \Delta x < 0.1$. The 
refinement frequency is set to every 10 time steps, unrefinement to every 
100 time steps, and processor load rebalancing to every 80 time steps, with 
a maximum allowed imbalance of $20\%$. This configuration gives a maximum total cell count of approximately $1.5 \times 10^5$, slightly higher than the $1.2 \times 10^5$ cells reported by the CWI group using the Afivo code in \cite{Bagheri2018}. The results are in very good agreement with those of 
the uniform $3~\mu$m mesh cases, as shown in 
\Cref{fig:Emax_vs_time_AMR_comparison}.

\begin{figure}[!h]
    \centering
    \includegraphics[scale=1]{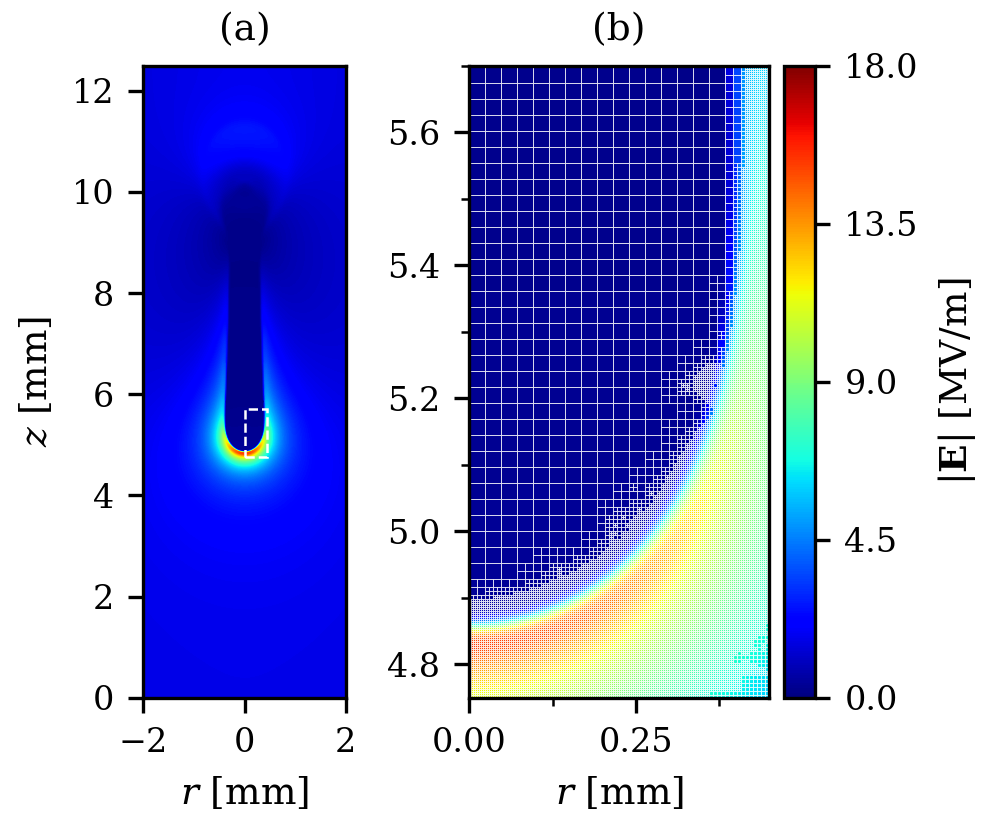}
    \caption{Electric field magnitude (MV/m) for the AMR case at 
    $t = 12$\,ns. (a) Full streamer structure, mirrored about the symmetry 
    axis. (b) Zoomed view of the highlighted region in (a), showing the 
    electric field together with the adaptive mesh. The mesh is refined to 
    the minimum cell size of $3~\mu$m at the streamer head and edges, while 
    the interior of the channel is unrefined to the base resolution of 
    $24~\mu$m.}
    \label{fig:AMR_E_field_12ns}
\end{figure}

\Cref{fig:AMR_E_field_12ns} shows the AMR result at $t = 12$\,ns. 
Subfigure (a) shows the electric field magnitude over the full streamer 
structure, mirrored about the symmetry axis for visualization, and 
subfigure (b) shows a zoomed view of the region highlighted by the white 
frame in (a), displaying both the electric field and the mesh at that time. 
It can be seen that inside the streamer body the mesh has been unrefined 
back to the original cell size, while at the streamer head and along the 
streamer edges the full 3 levels of refinement are active. For this case, 
10 buffer layers are used for refinement and 4 for unrefinement.

\subsubsection{Solver performance}

Here, the performance of \texttt{SoPlasmaFoam} is compared with the computational times reported by other solvers running the same case. It has to be mentioned that these simulations ran on one node of an AMD EPYC 7543P with 32 cores in total.

\begin{table*}[!htbp]
\centering
\caption{Computational performance comparison for the positive streamer benchmark case (high background ionization). Groups marked with $^*$ correspond to the participants of \cite{Bagheri2018}. ``Corr.'' denotes the number of outer Poisson-transport correction loops per time step.}
\label{tab:performance_comparison}
\setlength{\tabcolsep}{4pt}
\small
\begin{tabular}{l c c c c c c l c}
\hline
\textbf{Group / Code} & \textbf{AMR} & $\boldsymbol{\Delta x_{min}}$ & $\boldsymbol{N_{cells,max}}$ & $\boldsymbol{\Delta t}$ & \textbf{Corr.} & \textbf{Cores} & \textbf{Processor} & \textbf{Wall time} \\
 & & \textbf{(\textmu m)} & & & & & & \\
\hline
% Bagheri et al. groups
CWI$^*$ (Afivo)       & Yes & 3.0  & $1.2 \times 10^5$ & dyn.            & --- & 4   & Xeon E3-1271 v3, 3.6 GHz & 5 min     \\
CWI$^*$ (Afivo)       & No  & 3.05/1.53/0.76 & ---     & dyn.            & --- & 4   & Xeon E3-1271 v3, 3.6 GHz & several h to 3 d \\
ES$^*$                & Yes & 3.9  & $2.0 \times 10^6$ & 1 ps            & --- & 1   & Core i7-6700K, 4.0 GHz   & 20 h      \\
FR$^*$                & No  & 3.0  & $1.1 \times 10^6$ & dyn.            & --- & 1   & Xeon X5272, 3.4 GHz      & 6 h       \\
FR$^*$                & No  & 1.5  & ---               & dyn.            & --- & 1   & Xeon X5272, 3.4 GHz      & 32 h      \\
CN$^*$                & Yes & 2.0  & $6.5 \times 10^5$ & dyn., max $10^{-11}$ s & --- & 4 & Xeon E31225, 3.1 GHz   & 18 h      \\
TUE$^*$               & No  & ---  & $4.2 \times 10^6$ & dyn., max 1 ps  & --- & 1   & Core i7-5820K, 3.3 GHz   & 25 h      \\
DE$^*$                & No  & 4.2  & $5.1 \times 10^5$ & dyn., max 5 ps  & --- & 6   & Xeon E5-2690             & 15 h      \\
\hline
% Newer codes
reactPlasFOAM \cite{Shao_2025} & Yes & 3.125 & $1.1 \times 10^5$ & --- & --- & --- & ---                  & 20 min    \\
Vidyut3D \cite{Vidyut3d} & Yes & 1.5  & $\sim 2 \times 10^6$ & dyn., $\sim$ 2.5 ps & 2 & 64 & EPYC-Genoa + 4$\times$H100 (GPU) & 30 min \\
\hline
% Present solver
\texttt{SoPlasmaFoam} & No  & 1.5  & $3.5 \times 10^6$ & 1 ps & 1 & 20 & EPYC 7543P, 2.8 GHz & 6 h       \\
\texttt{SoPlasmaFoam} & No  & 3.0  & $1.2 \times 10^6$ & 1 ps & 2 & 4  & EPYC 7543P, 2.8 GHz & 5.9 h     \\
\texttt{SoPlasmaFoam} & No  & 3.0  & $1.2 \times 10^6$ & 1 ps & 2 & 8  & EPYC 7543P, 2.8 GHz & 3.6 h     \\
\texttt{SoPlasmaFoam} & No  & 3.0  & $1.2 \times 10^6$ & 1 ps & 2 & 12 & EPYC 7543P, 2.8 GHz & 2.8 h     \\
\texttt{SoPlasmaFoam} & No  & 3.0  & $1.2 \times 10^6$ & 1 ps & 2 & 20 & EPYC 7543P, 2.8 GHz & 2.3 h     \\
\texttt{SoPlasmaFoam} & No  & 3.0  & $1.2 \times 10^6$ & 1 ps & 1 & 20 & EPYC 7543P, 2.8 GHz & 1.2 h     \\
\texttt{SoPlasmaFoam} & No  & 3.0  & $1.2 \times 10^6$ & 2 ps & 2 & 20 & EPYC 7543P, 2.8 GHz & 1.2 h     \\
\texttt{SoPlasmaFoam} & Yes & 3.0  & $1.5 \times 10^5$ & 1 ps & 1 & 2  & EPYC 7543P, 2.8 GHz & 40 min    \\
\texttt{SoPlasmaFoam} & Yes & 3.0  & $1.5 \times 10^5$ & 1 ps & 1 & 4  & EPYC 7543P, 2.8 GHz & 18 min    \\
\texttt{SoPlasmaFoam} & Yes & 3.0  & $1.5 \times 10^5$ & 1 ps & 1 & 8  & EPYC 7543P, 2.8 GHz & 11 min    \\
\texttt{SoPlasmaFoam} & Yes & 3.0  & $1.5 \times 10^5$ & 1 ps & 1 & 12 & EPYC 7543P, 2.8 GHz & 8 min     \\
\texttt{SoPlasmaFoam} & Yes & 3.0  & $1.5 \times 10^5$ & 1 ps & 2 & 12 & EPYC 7543P, 2.8 GHz & 14 min    \\
\texttt{SoPlasmaFoam} & Yes & 3.0  & $1.5 \times 10^5$ & 2 ps & 2 & 12 & EPYC 7543P, 2.8 GHz & 8 min     \\
\hline
\end{tabular}
\end{table*}

\Cref{tab:performance_comparison} collects the results for the codes of \cite{Bagheri2018}, together with the results from \cite{Shao_2025} and \cite{Vidyut3d}, and a series of cases run with \texttt{SoPlasmaFoam}. For a dense mesh of around 3.5 million cells, with a 1 ps time step and 1 outer corrector on 20 cores, we obtain a simulation time of around 6 hours. On a comparable uniform mesh without AMR but with dynamic time stepping, the Afivo code reports a wide range, from several hours for the coarse 3.05 \textmu m grid up to around 3 days for the finest 0.76 \textmu m grid. For a similar resolution the FR code, also with dynamic time stepping, takes 32 hours on 1 core. The TUE team, with slightly more cells at 4.2 million and dynamic time stepping capped at 1 ps on 1 core, takes 25 hours. Vidyut3D gives very good results, running around 2 million cells with 2 correctors on 64 cores in 30 minutes. For our 3 \textmu m static mesh with a 1 ps time step and 2 correctors, we obtain 5.9, 3.6, 2.8, and 2.3 hours on 4, 8, 12, and 20 cores, respectively. Dropping to 1 corrector, which also gave very good results, brings this down to 1.2 hours on 20 cores. Other groups, at lower core counts, report between 6 and 20 hours. With these static meshes we cannot draw a strong conclusion about our solver relative to the rest, since there are many differences in time stepping and number of correctors, which roughly doubles the work per time step.

The big advantage of our code is the speedup we gain with AMR enabled. For a minimum cell size of 3 \textmu m and a maximum of $1.5 \times 10^5$ cells, with a 1 ps time step and 1 corrector, we run on 2 cores in 40 minutes, on 4 cores in 18 minutes, on 8 cores in 11 minutes, and on 12 cores in 8 minutes. For the 2 corrector case on 12 cores we run in 14 minutes and obtain the results presented in \Cref{fig:Emax_vs_time_AMR_comparison}. Finally, with a 2 ps time step and 2 correctors we run in 8 minutes. These results can be compared with CWI and KAUST, which report a similar number of cells. CWI, for a maximum of $1.2 \times 10^5$ cells but with dynamic time stepping on 4 cores, runs in 5 minutes, although we do not know whether they used a single corrector or more, or what their time step criterion was. The KAUST code, which is also built on OpenFOAM, reports a maximum of $1.1 \times 10^5$ cells, around 40\,000 fewer than ours, and runs in 20 minutes. We do not know their time stepping or other parameters such as the refinement and unrefinement frequencies.

These results show that the proposed solver is very fast at resolving the positive streamer case, placing it among the fastest available.

\subsubsection{Case B - Results}

\begin{figure}[h]
    \centering
    \includegraphics[scale=1]{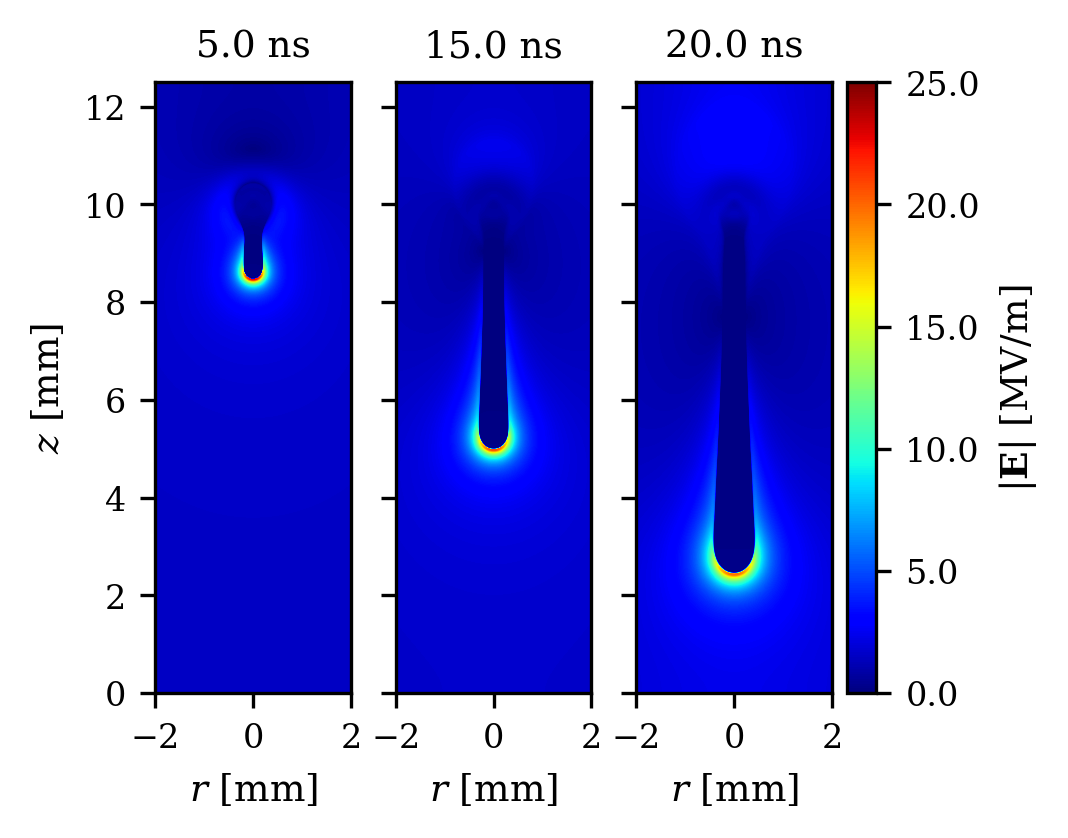}
    \caption{Contour plots of the electric field magnitude (V/m) at three 
    time instances ($t = 5$, $15$ and $20$\,ns) for the positive streamer 
    Case B, with a background ionization of $10^{9}\,\text{m}^{-3}$ for 
    both electrons and ions. The axisymmetric slice is mirrored about the symmetry axis.}
    \label{fig:E_streamer_lower_ionization_contours}
\end{figure}

For the second case, a background density of $n_e = n_i = 10^9\,\text{m}^{-3}$ 
for electrons and positive ions is set. Here, only one flux scheme (the ROUNDF) 
is presented, since it outperformed the others in the first test case. \Cref{fig:E_streamer_lower_ionization_contours} shows the evolution of the 
electric field magnitude during the streamer propagation for Case B. The 
main differences with respect to Case A are first the higher electric field magnitudes, which here reach values close to 25\,MV/m, and second the noticeably thinner streamer channel. Both observations are 
consistent with the expected physics. A lower background ionization makes 
the propagation more difficult to sustain, requiring stronger local field 
enhancement at the streamer head and producing sharper density gradients, 
which in turn results in a narrower streamer structure.

\begin{figure*}[t]
    \centering
    \begin{subfigure}{1\textwidth}
        \centering
        \includegraphics[scale=1]{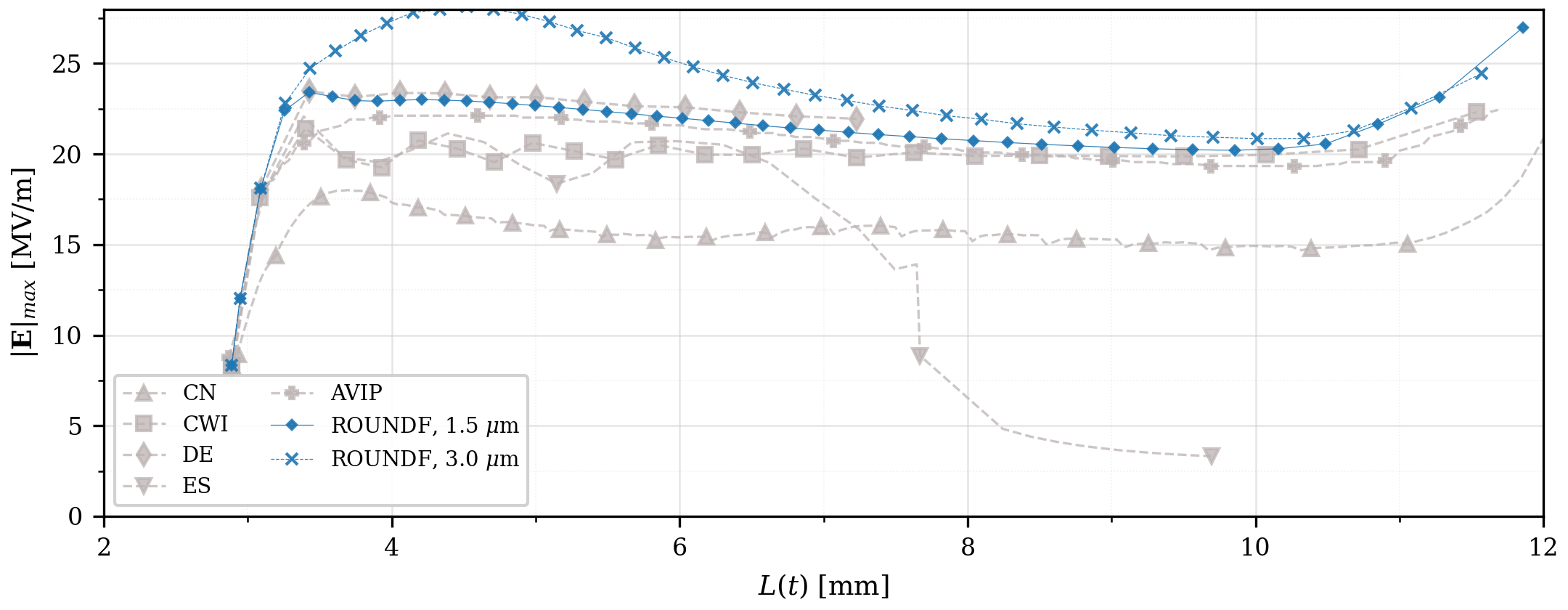}
        \caption{}
        \label{fig:Emax_vs_length_explicit_comparison_lower_ionization}
    \end{subfigure}
    \begin{subfigure}{1\textwidth}
        \centering
        \includegraphics[scale=1]{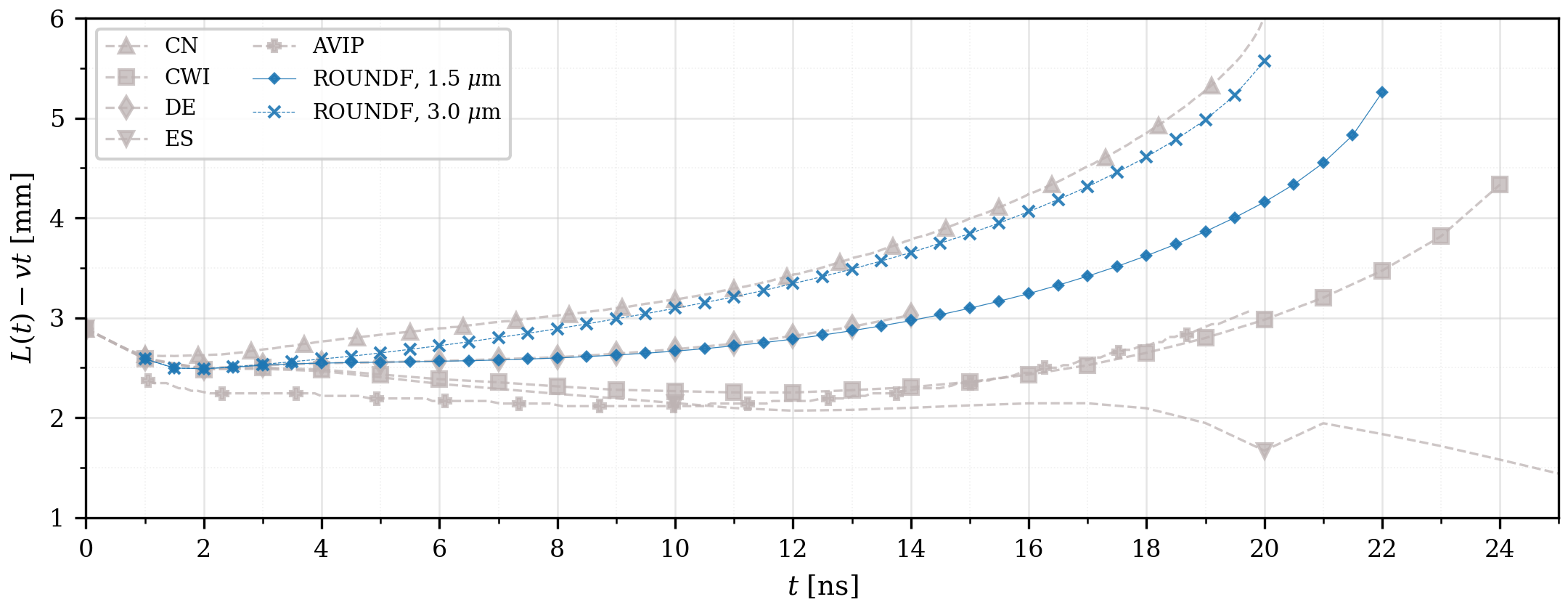}
        \caption{}
    \label{fig:streamer_length_minus_vt_vs_time_explicit_comparison_lower_ionization}
    \end{subfigure}
    \caption{Validation of the low-background ionization streamer (Case B) against reference data from \cite{Bagheri2018} and \cite{AVIP}: (a) maximum absolute electric field, (b) reduced streamer length $z - v_0 t$ (where $v_0 = 0.3\,\text{mm\,ns}^{-1}$). The results compare the proposed solver using the ROUNDF flux scheme at mesh resolutions of $3~\mu\text{m}$ and $1.5~\mu\text{m}$ with benchmark data from several research groups.}
    \label{fig:positive_streamer_lower_ionization_explicit_comparisons}
\end{figure*}

\Cref{fig:positive_streamer_lower_ionization_explicit_comparisons} shows the maximum absolute electric field across the streamer length (a) and the the reduced streamer length $z - v_0 t$ vs time, where here $v_0$ is set to $0.3 mm/s$. It seems that the differences between the two meshes is higher in this case, with the coarser mesh producing quite higher fields, and so the streamer travels faster. However, the resolved mesh, with resolution of cell size of $1.5 \mu m$ being close to reference data. Specifically, the electric field results are laying between the DE and the AVIP groups results, while the reduced streamer length results are very close to the DE results.

\section{Demonstration - nanosecond SDBD actuator} \label{sdbd}

\subsection{Case configuration}

\begin{figure*}[h!]
    \centering
    \includegraphics[scale=1.65]{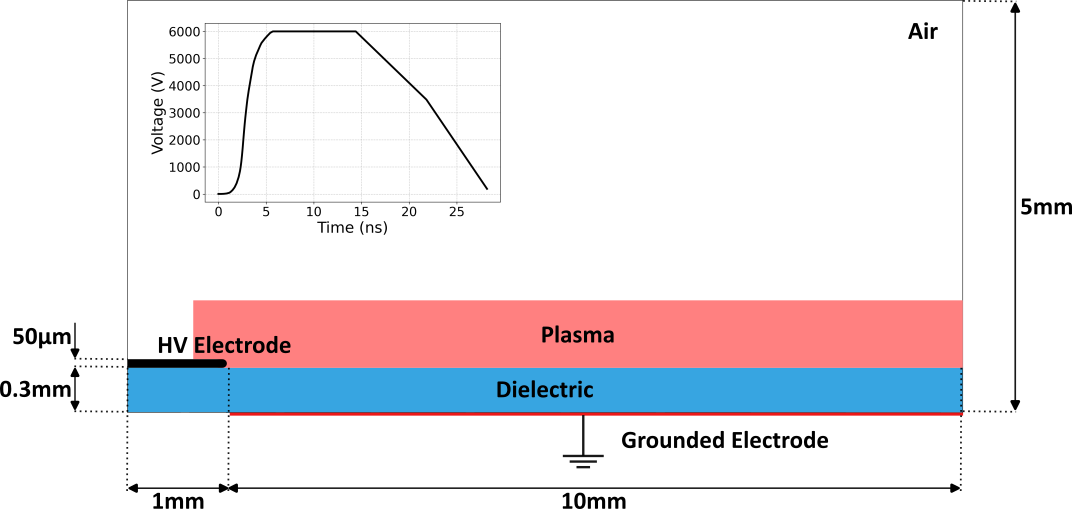}
    \caption{Computational domain of the nanosecond SDBD actuator. The powered electrode ($50\,\mu$m thick, $1\,$mm long) sits on top of a dielectric layer ($0.3\,$mm, $\varepsilon_r = 3.5$), above a $10\,$mm grounded electrode. The applied nanosecond voltage waveform is shown on the top left.}
    \label{fig:SDBD_domain}
\end{figure*}

Having validated the solver against other numerical solvers, we demonstrate its ability to handle multi-domain problems through the simulation of a nanosecond surface dielectric barrier discharge (SDBD) actuator, a surface ionization wave propagation case. The computational domain, illustrated in \Cref{fig:SDBD_domain}, consists of a powered electrode of $50\,\mu$m thickness and $1\,$mm length with a rounded tip, exposed to air. Beneath it lies a $0.3\,$mm dielectric layer with relative permittivity of $\varepsilon_r = 3.5$, above a $10\,$mm long grounded electrode. The applied nanosecond voltage waveform (rise time 2 ns, 20 ns duration at FWHM) is also shown in \Cref{fig:SDBD_domain}. The case mimics the experimental configuration of Ref.\cite{zhu2017nanosecond} but the applied voltage amplitude is set to 6 kV. 

In this case, we consider three species: electrons, positive ions, and negative ions, 
along with a background neutral number density of $N = 2.447 \times 10^{25}\,$m$^{-3}$. Four reaction mechanisms are taken into account, summarized in \Cref{tab:SDBD_reactions}.

\begin{table}[h!]
\centering
\caption{Reaction mechanisms considered in the SDBD simulation.}
\label{tab:SDBD_reactions}
\begin{tabular}{llll}
\hline
\textbf{\#} & \textbf{Reaction} & \textbf{Type}  \\
\hline
1 & $e + n \rightarrow i^+ + 2e$       & Ionization  \\
2 & $e + n \rightarrow i^-$            & Attachment   \\
3 & $i^+ + e \rightarrow n$            & Recombination \\
4 & $i^+ + i^- \rightarrow 2n$         & Ion-ion recombination \\
\hline
\end{tabular}
\end{table}

The transport properties (mobilities and diffusivities) and rate coefficients are taken from lookup tables exported from BOLSIG+ \cite{BOLSIG+} as a function of the reduced electric field $E/N$.

This case is treated under the Local Field Approximation (LFA). To stabilize the surface ionization wave, the full flux scheme is employed, retaining the diffusive flux only when it opposes the drift, together with the Townsend ionization coefficient applied exclusively to the ionization rate. This approach introduces a diffusive character in the sheath region, partially mimicking the effect of the Local Energy Approximation (LEA) \cite{Teunissen_2020}.

Initial and minimum densities are set to $10^{11}\,$m$^{-3}$ for electrons and positive ions, and to $10^{5}\,$m$^{-3}$ for negative ions. The simulation is carried out on a mesh of approximately $700{,}000$ triangular elements, over a physical time of $10\,$ns. The gas and dielectric regions are coupled implicitly for the Poisson equation, meaning both domains are assembled and solved simultaneously in a single matrix. A variable time step is employed, governed by the following criteria:a maximum chemistry Courant number of $5$, a maximum diffusive Courant number of $20$ (electrons), a maximum convective Courant number of $20$ (electrons), and a maximum dielectric relaxation time ratio of $10$. The numerical schemes are summarized in \Cref{tab:schemes}. Additionally, for the flux terms the Scharfetter-Gummel (SG) scheme was also employed.

\begin{table}[!htbp]
\centering
\caption{Numerical schemes used in the SDBD simulation.}
\label{tab:schemes}
\begin{tabular}{@{}p{3.2cm}p{4.4cm}@{}}
\hline
\textbf{Term} & \textbf{Scheme} \\
\hline
Time derivative          & backward \\
Divergence               & Gauss ROUNDF \\
Laplacian (transport)    & Gauss harmonic \newline limited corrected 1.0 \\
Laplacian (Poisson)      & Gauss linear \newline limited corrected 0.5 \\
Interp. mob. \& diff.    & Harmonic \\
Interp. densities        & Linear \\
snGrad potential         & limited corrected 0.5 \\
snGrad densities         & limited corrected 1.0 \\
\hline
\end{tabular}
\end{table}

\subsection{Results}

\begin{figure}[!htbp]
    \centering
    \includegraphics[scale=1]{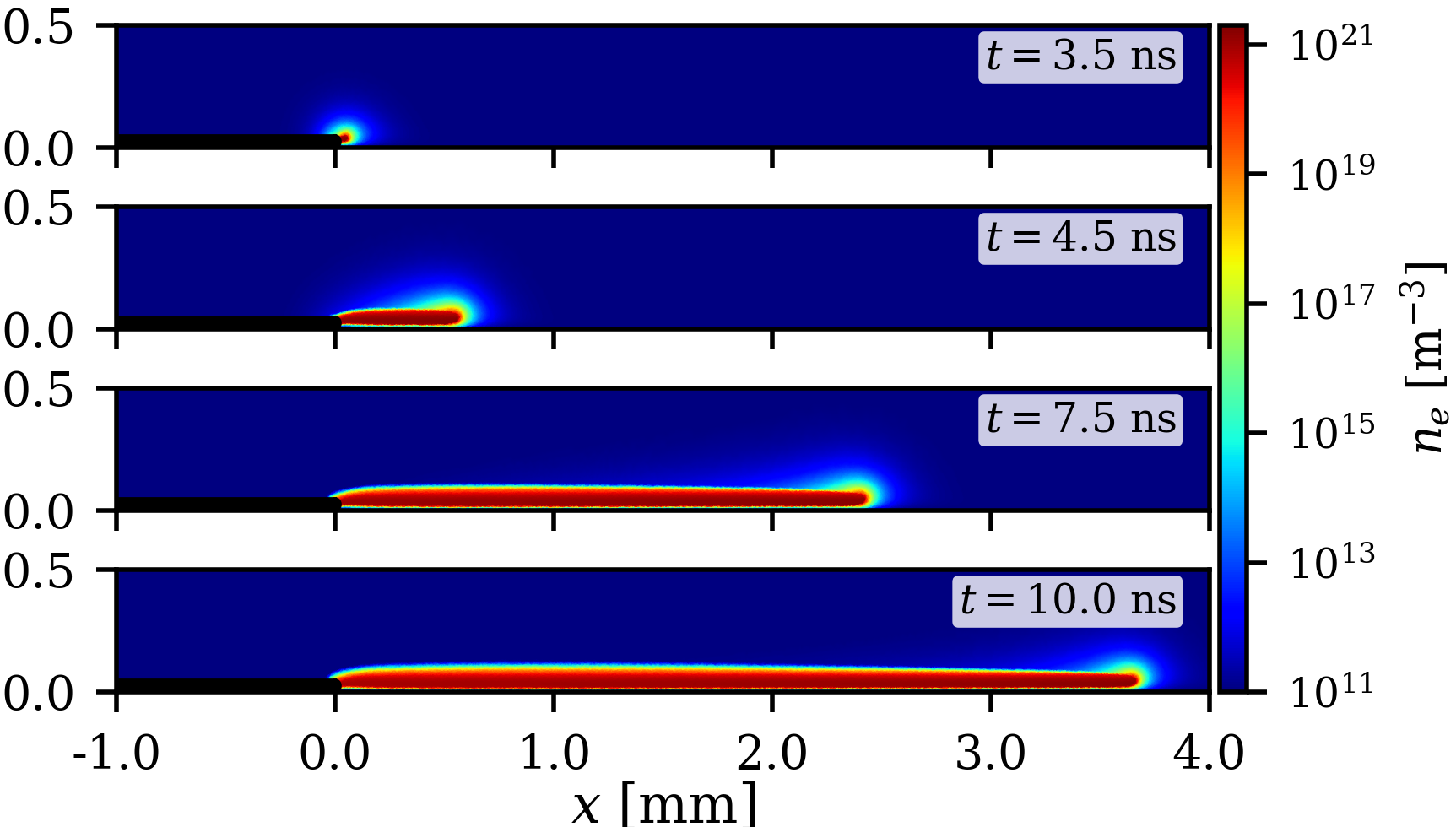}
    \caption{Electron density contours at $t = 3.5$, $4.5$, $7.5$, and $10\,$ns 
    for the nanosecond SDBD actuator simulation using the Scharfetter-Gummel (SG) scheme. The domain shown spans $x \in [-1, 4]\,$mm and $y \in [0, 0.5]\,$mm, 
    restricted to the gas region.}
    \label{fig:n_e_SDBD_contour}
\end{figure}

\begin{figure}[!htbp]
    \centering
    \includegraphics[scale=1]{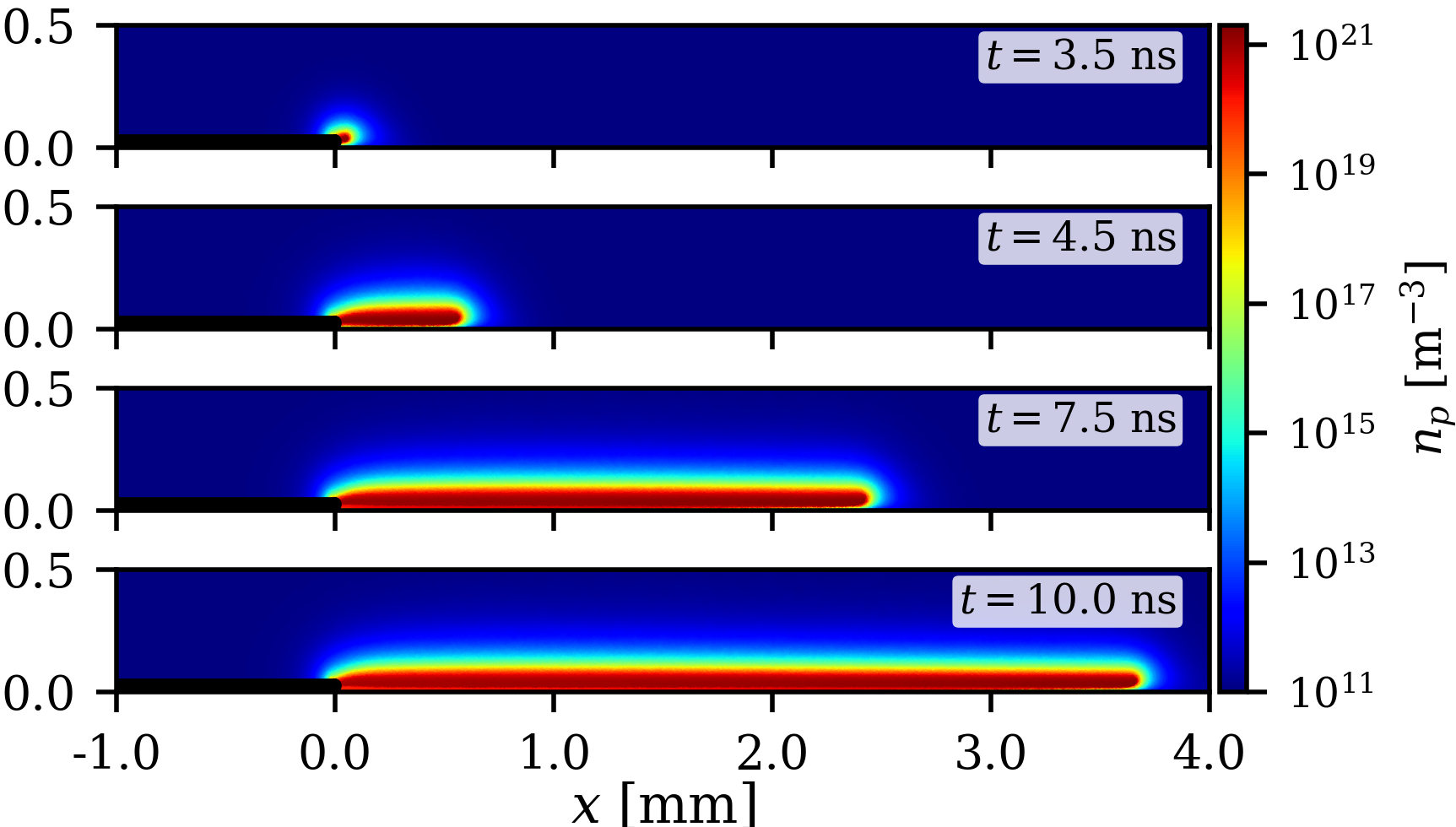}
    \caption{Positive ion density contours at $t = 3.5$, $4.5$, $7.5$, and $10\,$ns for the nanosecond SDBD actuator simulation using the Scharfetter-Gummel (SG) scheme. The domain shown spans $x \in [-1, 4]\,$mm and $y \in [0, 0.5]\,$mm, restricted to the gas region.}
    \label{fig:n_pIon_SDBD_contour}
\end{figure}

\begin{figure}[!htbp]
    \centering
    \includegraphics[scale=1]{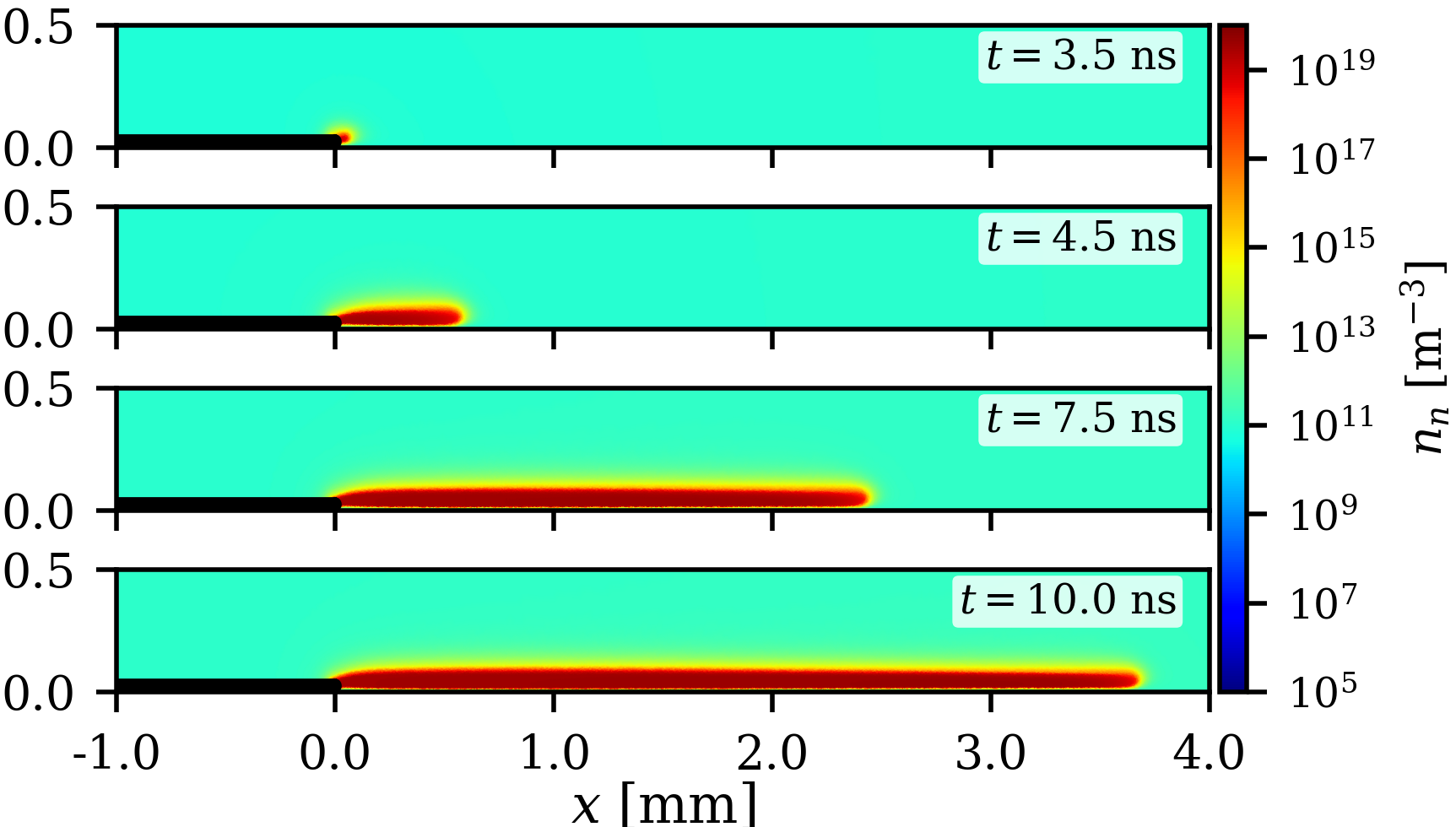}
    \caption{Negative ion density contours at $t = 3.5$, $4.5$, $7.5$, and $10\,$ns for the nanosecond SDBD actuator simulation using the Scharfetter-Gummel (SG) scheme. The domain shown spans $x \in [-1, 4]\,$mm and $y \in [0, 0.5]\,$mm, restricted to the gas region.}
    \label{fig:n_nIon_SDBD_contour}
\end{figure}

\begin{figure}[!htbp]
    \centering
    \includegraphics[scale=1]{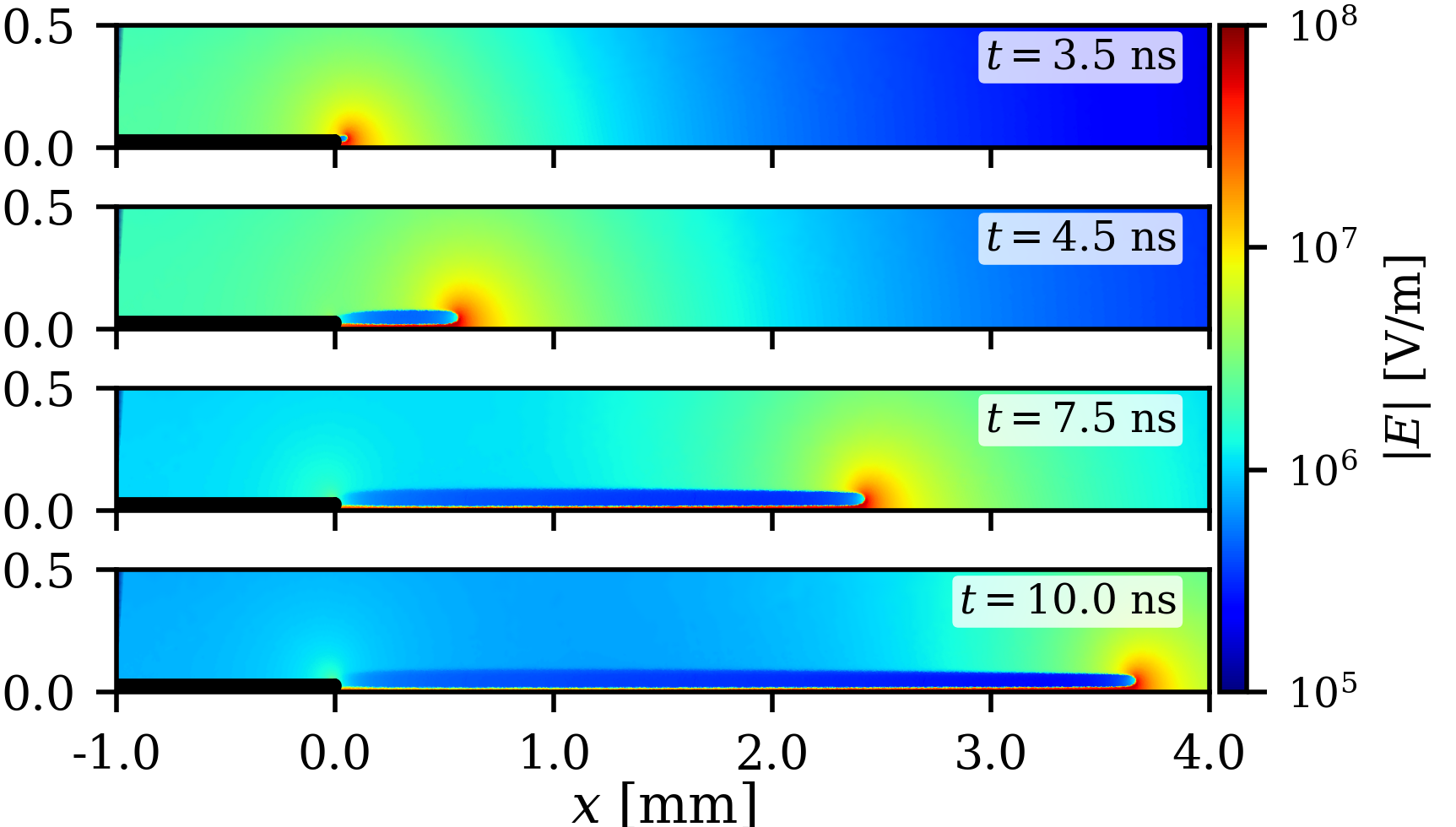}
    \caption{Electric field magnitude contours at $t = 3.5$, $4.5$, $7.5$, and $10\,$ns for the nanosecond SDBD actuator simulation using the Scharfetter-Gummel (SG) scheme. The domain shown spans $x \in [-1, 4]\,$mm and $y \in [0, 0.5]\,$mm, restricted to the gas region.}
    \label{fig:E_SDBD_contour}
\end{figure}

\Cref{fig:n_e_SDBD_contour,fig:n_pIon_SDBD_contour,fig:n_nIon_SDBD_contour,fig:E_SDBD_contour} show the electron density, positive ion density, negative ion density, and electric field magnitude at four time instances, from the streamer inception until it has propagated approximately $4\,$mm, using the SG scheme. As expected, the streamer initiates near the tip of the powered electrode, where the electric field is highest. Concentrations of negative and positive charge accumulate in this region, initiating the shielding process. Due to a small offset of the positive charge at the streamer tip, a positive streamer then propagates parallel to the dielectric surface. The maximum electron and positive ion densities reach approximately $2 \times 10^{21}\,$m$^{-3}$, while negative ions peak around $2 \times 10^{19}\,$m$^{-3}$. A sheath of positive ions forms close to the dielectric surface, where their density significantly exceeds that of electrons. In this region, between the streamer body and the dielectric, the electric field drives positive ions toward the surface while electrons are repelled in the opposite direction. Calculated values of electron density and electric field magnitude (in the order of $10^7 - 10^8\,$V/m at the streamer head) are consistent with those reported in the literature for similar cases of surface streamer propagation \cite{zhu2017nanosecond, kourtzanidis2017three, kourtzanidis2021self}.

\begin{figure}[!htbp]
    \centering
    \includegraphics[scale=1]{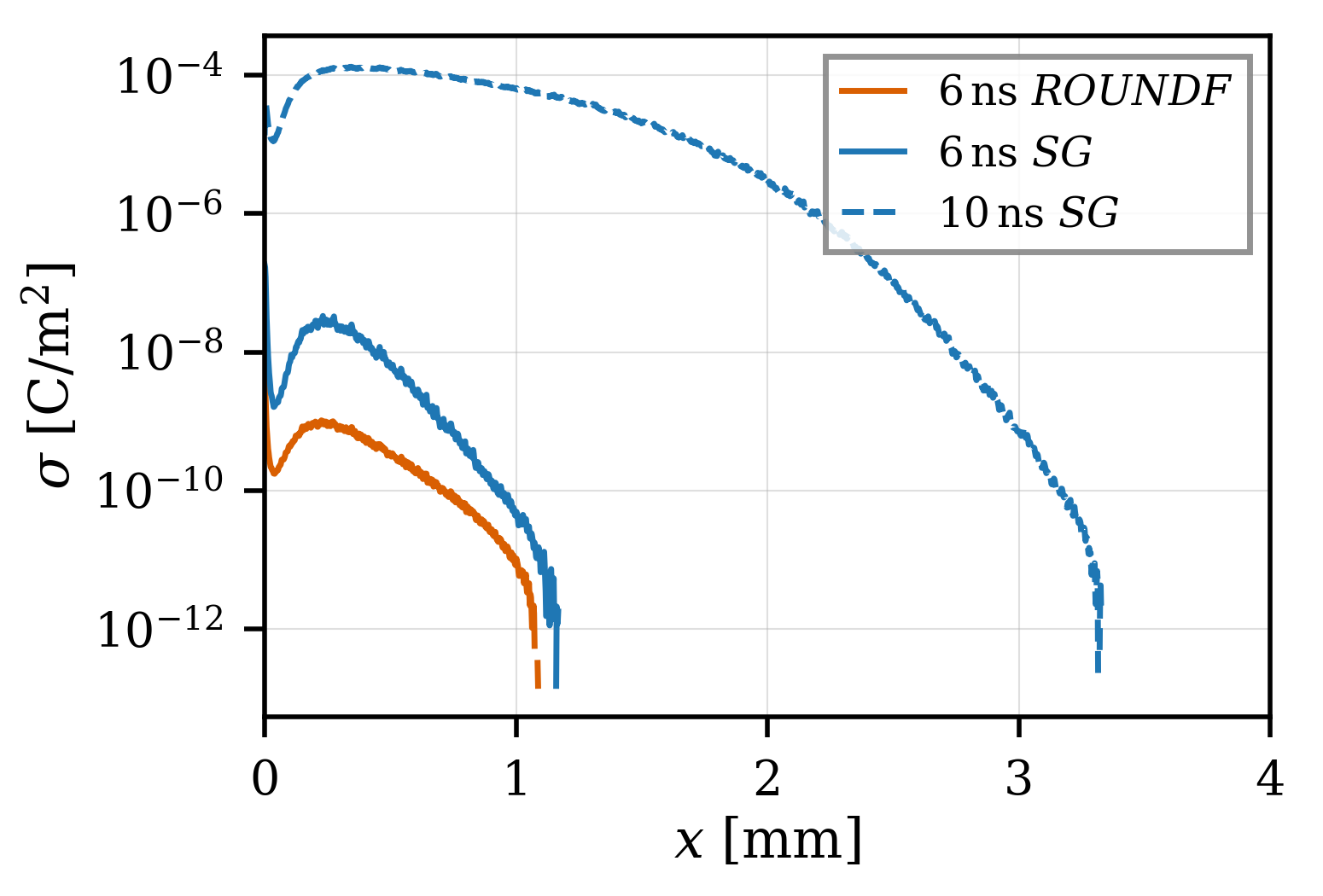}
    \caption{Surface charge accumulation on the dielectric surface for the nanosecond SDBD actuator simulation at $t = 6\,$ns and $t = 10\,$ns, shown in logarithmic scale. The blue line corresponds to the SG scheme and the orange line to the Gauss ROUNDF scheme, the latter shown only at $t = 6\,$ns. Negative values, being small in magnitude, are set to zero in the logarithmic representation.}
    \label{fig:surfCharge_comparison_SDBD}
\end{figure}

This process leads to progressive positive charging of the dielectric as the streamer propagates. This surface charge, together with the ion sheath, generates an electric field component opposing the streamer's normal field component, reducing the total electric field normal to the dielectric surface. The charge accumulated on the dielectric is shown in \Cref{fig:surfCharge_comparison_SDBD}, in logarithmic scale (C/m$^2$), along the dielectric surface for two time instances of the SG scheme. As time progresses, the surface charge increases due to the growing positive ion density near the dielectric and the relaxation phase of the streamer, and will eventually reach a point where the normal field is fully shielded, leading to polarity inversion between the electrode voltage and the charged dielectric surface. The importance of surface charging in NTP dynamics is well-known and has been demonstrated numerically in nanosecond pulsed and AC-SDBDs\cite{kourtzanidis2021self, kourtzanidis2017three} as well as volume DBDs\cite{kourtzanidis2023full}. The same figure includes the result of the ROUNDF scheme at $t = 6\,$ns, which shows surface charge values,approximately one order of magnitude lower than that of the SG scheme at the same instant. This is consistent with the more diffusive nature of the SG scheme, which enhances the transport of positive ions toward the sheath and the dielectric surface.

\section{Parallel scalability analysis} \label{scalability}

Before concluding the paper, it is important to analyze the scalability of the developed solver. Plasma simulations are computationally demanding problems, given the wide range of time and spatial scales involved, as well as the number of equations solved per time step in a multi-species problem. Good parallel scaling is therefore essential for solvers of this kind.

Here we focus on single-node performance. Tests were carried out on two different systems, whose specifications are summarized in \Cref{tab:hardware_specs}: a workstation (referred to as WS) and a high-performance computing node (referred to as HPC).

\begin{table*}[h!]
\centering
\caption{Hardware specifications of the two systems used for the 
scalability analysis.}
\label{tab:hardware_specs}
\begin{tabular}{lll}
\hline
\textbf{Specification}   & \textbf{WS}                        & \textbf{HPC} \\
\hline
CPU                      & AMD Ryzen 9 5900X                  & AMD EPYC 7543P \\
Cores / Threads          & 12 cores                           & 32 cores \\
Max CPU frequency        & 4.95\,GHz                          & 2.79\,GHz \\
L3 cache                 & 64\,MiB                            & 256\,MiB \\
RAM                      & 32\,GB DDR4 @ 2133\,MT/s (2-channel)  & 54\,GB DDR4 (8-channel) \\
OS                       & Ubuntu 22.04.5 LTS                 & Ubuntu 22.04.5 LTS \\
MPI                      & OpenMPI 4.1.2                      & OpenMPI 4.1.2 \\
Compiler                 & GCC 11.4.0                         & GCC 11.4.0 \\
\hline
\end{tabular}
\end{table*}

\subsection{Strong scaling analysis}

\begin{figure*}[h!]
    \centering
    \includegraphics[scale = 1]{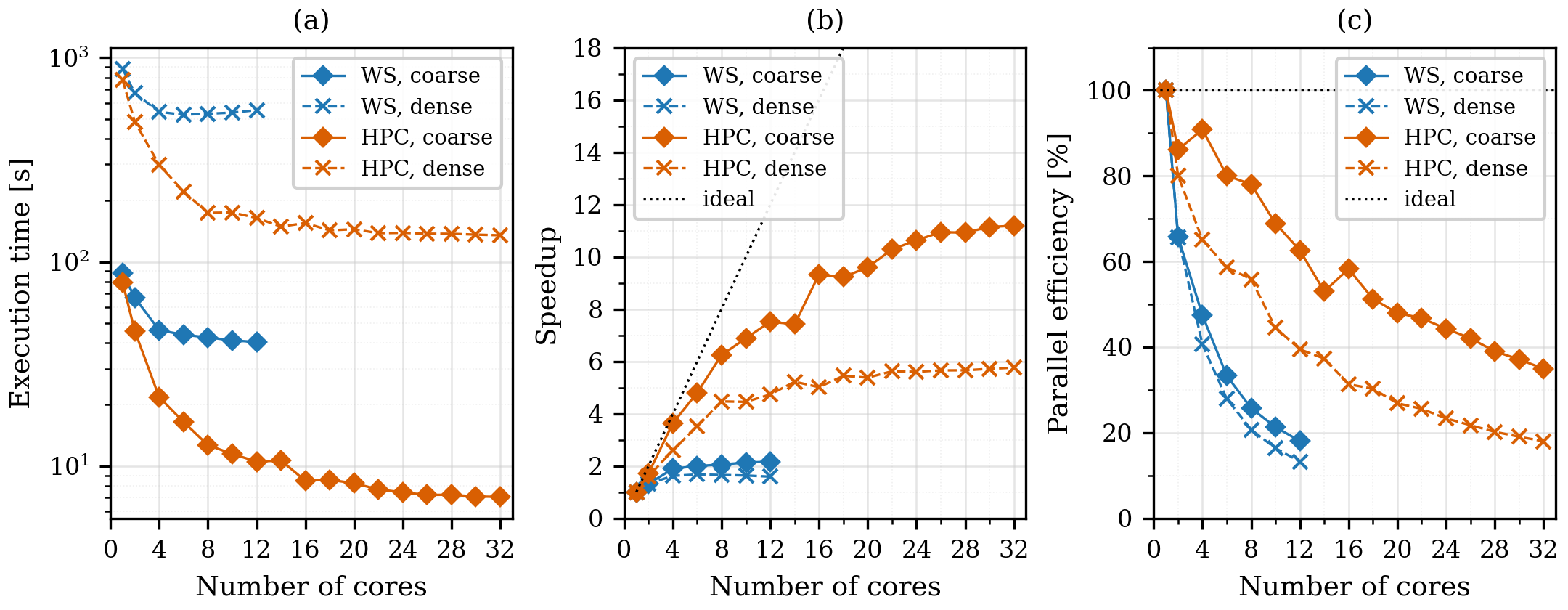}
    \caption{Strong-scaling analysis of the positive streamer Case~A for 
    the coarse ($449{,}413$ cells) and dense ($3{,}472{,}738$ cells) meshes 
    on the WS (Ryzen 9 5900X) and HPC (EPYC 7543P) systems: 
    (a) wall-clock execution time over 100 time steps, 
    (b) speedup $S = T_1/T_n$ with the ideal linear scaling for reference, 
    and (c) parallel efficiency $E = S/n$ with the ideal $100\%$ line.}
    \label{fig:scaling_analysis}
\end{figure*}

The first benchmark is a strong-scaling test of the positive streamer Case~A, 
performed on two meshes: a coarse mesh of $449{,}413$ cells and a dense mesh 
of $3{,}472{,}738$ cells. For each configuration, the solver was run for 
$100$ time steps and the wall-clock execution time was recorded as the number 
of MPI processes was varied. The results, summarized in 
\Cref{fig:scaling_analysis}, are reported as execution time, speedup 
$S = T_1 / T_n$, and parallel efficiency $E = S/n$.

For the WS, the scaling saturates very early on both meshes. The coarse mesh 
reaches a maximum speedup of about $2.2\times$ at 12 cores, with efficiency 
dropping below $50\%$ already at 4 cores and to $18\%$ at 12 cores. The dense 
mesh is even worse: the execution time stops decreasing past 6 cores and 
actually \emph{increases} from 8 cores onwards, with the speedup peaking at 
roughly $1.7\times$. This behaviour is the signature of memory-bandwidth 
saturation. OpenFOAM, like virtually every cell-centred FVM solver, is a 
memory-bound code: the dominant kernels (sparse matrix-vector products, 
gradient and Laplacian assembly, field copies) are limited by the rate at 
which data can be streamed from main memory rather than by floating-point 
throughput. The WS has only two DDR4 channels at 2133\,MT/s and a relatively 
small 64\,MiB L3 cache, so once more than a couple of cores actively pull 
data, the memory controllers become the bottleneck and adding cores yields 
diminishing or even negative returns.

The HPC node tells a different story. The coarse mesh achieves a speedup 
of $11.2\times$ at 32 cores, with parallel efficiency staying above $75\%$ up to 8 cores and around $35\%$ at full node usage. The improvement over the WS is twofold: first, the EPYC 7543P has eight DDR4 memory channels, which 
relieves the bandwidth bottleneck that crippled the WS; second, its much 
larger aggregate L3 cache means that, as the per-core working set shrinks 
with increasing core count, larger portions of the mesh and its associated 
fields effectively fit in cache, reducing main-memory traffic. The dense 
mesh, on the other hand, plateaus around $5.8\times$ speedup beyond $\sim$16 
cores, since its per-process working set never fully fits in cache, and the 
solver is bound by the bandwidth of the (still 8-channel) memory subsystem.

Overall, these results confirm that the solver scales as well as expected for 
a memory-bound FVM code: scaling is governed almost entirely by the memory 
hierarchy of the host system, with the number of memory channels and the 
relative size of the L3 cache compared to the per-process working set being 
the decisive factors.

\subsection{MPI profiling analysis}

\begin{figure*}[h!]
    \centering
    \includegraphics[width=\textwidth]{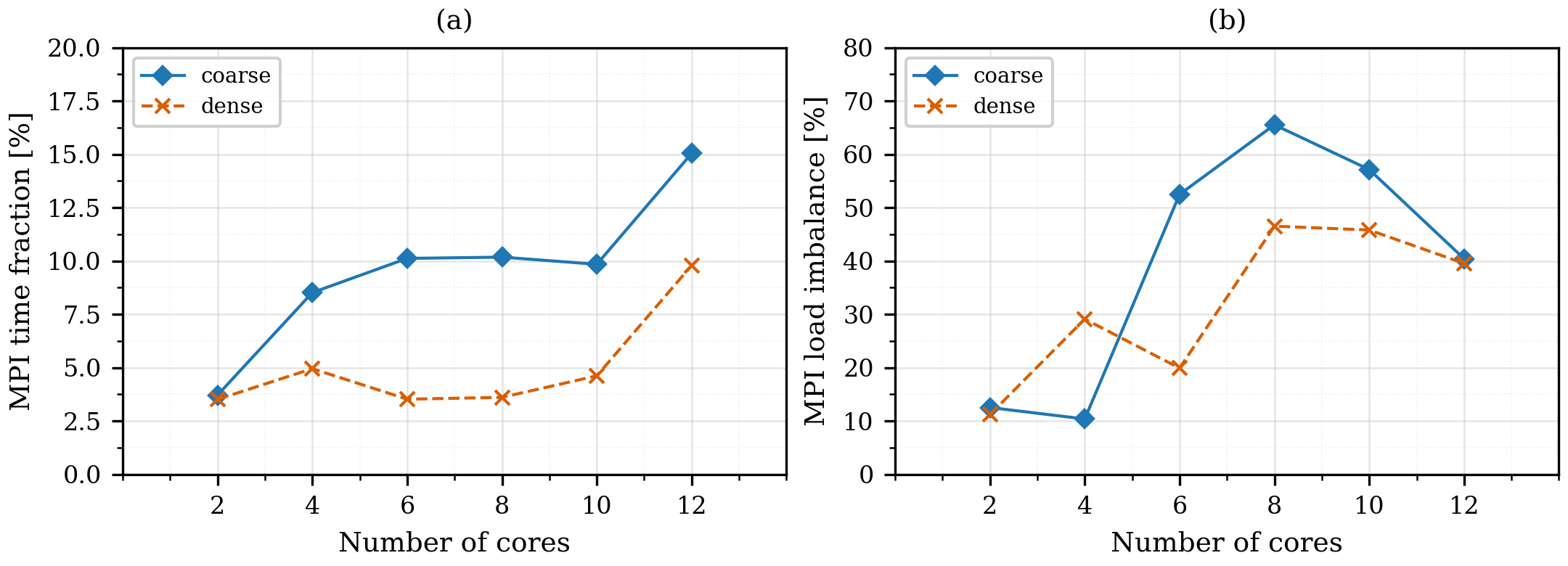}
    \caption{MPI profile statistics on the WS (Ryzen 9 5900X) for the 
    positive streamer Case~A: (a) MPI time as a fraction of the average 
    per-rank elapsed time, and (b) MPI load imbalance, defined as 
    $(t_{\max} - t_{\min})/t_{\max}$ over the cumulative MPI time of each 
    rank.}
    \label{fig:ws_mpi_fraction_imbalance}
\end{figure*}

To verify that the scaling stall observed on the WS originates from the 
memory subsystem rather than from MPI overhead, MPI traces were collected 
with AMD uProf for the coarse and dense meshes at $\{2, 4, 6, 8, 10, 12\}$ 
ranks. The two most informative aggregate metrics, average per-rank MPI 
time fraction and load imbalance, defined as $(t_{\max} - t_{\min})/t_{\max}$ where $t_i$ is the cumulative MPI time of rank $i$, are shown in \Cref{fig:ws_mpi_fraction_imbalance}.

The first observation is that MPI is \emph{not} the bottleneck. The MPI 
fraction of per-rank elapsed time stays below $15\%$ on the coarse mesh and 
below $10\%$ on the dense one across the entire range of core counts. The 
dense mesh, with its much larger per-process working set, shows an even 
smaller MPI percentage because the compute cost dominates. In absolute 
terms, the total MPI time per rank stays nearly flat with the number of 
cores, while the compute time barely decreases past 4-6 cores, which a direct consequence of the memory-bandwidth saturation discussed previously. The jump at 12 ranks is attributable to OS and background processes competing for resources, as 12 represents the maximum number of physical cores available on the Ryzen 9 5900X.

The second observation is the rapid growth of load imbalance. The 
difference between the slowest and fastest rank in cumulative MPI time 
rises from approximately 10\% at 4 ranks to 50-65\% at 6-10 ranks on the 
coarse mesh, with similar behavior on the dense one. The positive streamer 
is a strongly localized phenomenon, and the partitions containing the 
streamer head perform substantially more work than those in the quiescent 
regions. Static domain decomposition through scotch cannot 
anticipate this localization, and the resulting work imbalance accumulates 
as idle time on under-loaded ranks. Inspection of the per-routine 
breakdown confirms this. MPI\_Waitall alone accounts for 
85-94\% of the total MPI time across all configurations, meaning that 
most of the reported MPI cost is synchronization with the slowest peer 
rather than genuine communication overhead. The remaining 
MPI\_Allreduce contribution (5-10\%) originates from the global 
reductions inside the PBiCGStab/CG iterations of the Poisson solver, 
while MPI\_Bcast and the point-to-point exchanges together 
account for less than 10\%.

\subsection{Solver component breakdown}

Beyond the overall scaling and MPI behavior, it is informative to break 
down the per-time-step cost of the solver into its main physical 
components in order to identify which parts of the algorithm dominate the 
runtime. The wall-clock time of each major block of the time loop is 
measured separately and reported as a percentage of the total solver time 
in Table~\ref{tab:cost_breakdown}. Five categories are distinguished:
\begin{enumerate}
    \item \textbf{Poisson solve and electric field computation.} Assembly 
    and solution of the Poisson equation for the 
    electric potential, followed by the electric field evaluation, 
    boundary correction, and computation of the reduced field $E/N$.
    \item \textbf{Drift-diffusion transport.} Assembly and solution of the 
    species continuity equations.
    \item \textbf{Transport coefficient update.} Evaluation of the 
    mobilities and diffusivities through lookup tables as functions of $|\mathbf{E}|$.
    \item \textbf{Chemistry.} Evaluation of the source terms as functions of $|\mathbf{E}|$.
    \item \textbf{Charge updates and other.} Volume charge 
    density updates, time-step adjustment, reporting and remaining overhead.
\end{enumerate}
 
\begin{table}[!ht]
\centering
\caption{Wall-clock time breakdown of the main solver components for the 
positive streamer case. Times correspond to the maximum across all MPI 
ranks.}
\label{tab:cost_breakdown}
\small
\begin{tabular}{l r}
\hline
\textbf{Component} & \textbf{\%} \\
\hline
\textit{Poisson solve and electric field} &  \textit{66.9} \\
\quad Build potential equation       &   9.0 \\
\quad Solve potential equation       &  18.7 \\
\quad Compute $E_{flux}$              &    7.0 \\
\quad Compute $\mathbf{E}$          &  28.4 \\
\quad Correct $\mathbf{E}$ BCs      &    2.1 \\
\quad Compute $|\mathbf{E}|$        &     1.2 \\
\quad Compute $E/N$                 &     0.5 \\
\hline
\textit{Drift-diffusion transport}   & \textit{28.5} \\
\quad Build transport equations      &  23.8 \\
\quad Solve transport equations      &    4.7 \\
\hline
\textit{Transport coefficient update} &  \textit{2.6} \\
\hline
\textit{Chemistry}                   &    \textit{0.4} \\
\hline
\textit{Charge updates and other}    &  \textit{1.6} \\
\hline
\end{tabular}
\end{table}

The dominant cost is the electromagnetic block, which accounts for 
approximately two-thirds of the total runtime. Within this block, the 
reconstruction of the face-normal electric flux field $\phi_E$ from the 
surface-normal gradient of the potential is the single most expensive 
operation at 28.4\%. This cost could potentially be reduced by computing 
the electric field from the cell-centered gradient of the potential 
instead, which would bring the electromagnetic share down to approximately 
4.0\%. The Poisson linear solve itself accounts for 18.7\%, while the 
transport equation assembly dominates the drift-diffusion block at 23.8\%, 
with the actual linear solve contributing only 4.7\%. Chemistry appears 
negligible at 0.4\% because in the present implementation the source terms 
are evaluated explicitly rather than through an implicit ODE integration, 
which avoids the per-cell stiff solver cost that would otherwise arise in 
cases with more complex reaction mechanisms. It should be noted that both 
the drift-diffusion transport and the chemistry costs will grow with the 
complexity of the problem, since the transport block scales with the number 
of mobile charged species, each requiring its own drift-diffusion equation, 
while the chemistry cost scales with the number of reactive species and 
reactions in the mechanism.

\section{Discussion and future directions}
\label{Discussion}

The general framework, validation cases and scaling performance of our new solver, as presented in the previous sections, show great promise and lay the path for multidimensional, multiphysics plasma simulations. Nevertheless, we'd like to provide here, a discussion on various optimization pathways and additional functionalities we aim to integrate in the near future as well as current limitations. First, focusing on necessary plasma-focused extensions, we will focus on the implementation of the Local mean Energy Approximation (LMEA/LEA) through the solution of the electron energy equation, photoionization, field emission boundary conditions, and circuit models for predicting discharge voltage and current (with the implementation of Sato's equation\cite{morrow1999discharge} for the latter).  Each of these features will be thoroughly validated through dedicated benchmark cases, accompanied by convergence and stability studies.

Another major development chapter for this solver is the incorporation of 
chemistry. Our main aim is to couple OpenFOAM with Cantera \cite{cantera}, 
an open-source suite of tools for problems involving chemical kinetics, 
thermodynamics, and transport processes. Since Cantera is written in C++, 
the integration with OpenFOAM does not appear to be particularly challenging, and works following this path have already been reported in the literature \cite{OpenFOAM-Cantera}. This stage of development will present clear challenges, as the resulting systems of equations will be inherently stiff. In OpenFOAM, such problems are typically addressed using operator splitting, where the transport problem is solved separately from the chemistry ODEs, an approach already followed in the native chemistry solvers of OpenFOAM. At this step, a thorough analysis will be needed to determine which approach best fits our requirements, exploring different operator splitting schemes, or even a fully coupled solution strategy using the block matrix capabilities of OpenFOAM. With operator splitting, all OpenFOAM ODE solvers become available, and additionally, since PETSc integration has already been established in this work, the PETSc ODE solvers can also be leveraged.

Another important aspect that was not addressed in this paper is the tuning of PETSc solvers and preconditioners to achieve optimal computational performance and scalability on large HPC systems. As discussed in the body of the paper, the developers of petsc4foam \cite{Petsc4Foam} have reported excellent HPC scalability for OpenFOAM using their library, and therefore a dedicated study focused on preconditioner and solver tuning represents an important step toward making the solver computationally efficient. Furthermore, petsc4foam supports parallel execution on GPUs, which was not explored here, and could provide significant additional acceleration.

OpenFOAM in general offers substantial benefits: it is a finite volume 
framework well established in the CFD community, open-source, thoroughly 
tested, and supported by a rich ecosystem of integrations and tools that 
can directly benefit our solver. For example, a natural future direction 
is the development of a coupled plasma-fluid solver, in order to study the 
mutual influence between the flow field and the plasma. OpenFOAM already 
provides well-established incompressible and compressible flow solvers, 
making such an integration relatively seamless. The main challenge lies 
in the coupling strategy rather than the integration itself. The ultimate 
goal is a unified plasma-fluid solver capable of studying flow patterns in 
plasma, electrohydrodynamic (EHD) forces acting on the fluid, and energy 
exchange processes between the plasma and the bulk fluid.

Nevertheless, alternatives should be considered where OpenFOAM may not be 
optimally suited for specific tasks. For instance, faster dedicated Poisson solvers exist, and whether their use is necessary will need to be evaluated. For such integrations, preCICE \cite{OpenFOAMpreCICE} would be a valuable tool, as it enables the coupling of different solvers across different numerical frameworks (e.g., finite volume with finite elements). preCICE also represents an alternative approach to the multi-region problem. For example, if the solver is to focus exclusively on the gas-phase region, different domains such as the dielectric regions can be coupled externally through preCICE. This would further extend the multiphysics capabilities of the framework, enabling, in the longer term, the simulation of plasma interactions with both solids and fluids.

A further promising direction is the integration of machine learning, which 
is entering the scientific computing community at a rapid pace. The stiff 
chemistry that will be introduced is a prime target, where trained neural 
networks can serve as fast surrogates for the chemical source terms and 
replace expensive ODE integration, an approach already demonstrated within 
the OpenFOAM and Cantera environment \cite{DeepFlame}. The Poisson solve 
offers a second opportunity, since machine-learned preconditioners and 
network-predicted initial guesses have been shown to cut the iteration 
counts of the resulting linear systems \cite{PoissonCNN}. Both of these can 
be embedded through the existing C++ interfaces of libraries such as 
libtorch, keeping the entire workflow inside the OpenFOAM framework.

\section{Conclusions} \label{Conclusions}
\label{sec:conclusions}
This paper introduced \texttt{SoPlasmaFoam}, an open-source multi-region 
plasma-dielectric solver built on OpenFOAM and integrated with PETSc, 
blastAMR, and the ROUND scheme library. Beyond presenting the solver, this 
work delivered three methodological contributions that are transferable to
other plasma-fluid codes: a systematic benchmark of convective flux schemes
for streamer transport, a quantitative study of the Poisson-transport
coupling requirements, and a drift-robust wall boundary condition
formulation. These are complemented by a thorough validation across
distinct plasma regimes and a detailed analysis of computational
performance and parallel scaling.

The solver was validated against canonical benchmark cases, including a
low-pressure DC glow discharge and positive streamers in dry air at
different pre-ionization background levels, showing good agreement with
established fluid-model plasma solvers from the literature. The
multi-region capability was demonstrated through a surface dielectric
barrier discharge simulation.

A detailed assessment of flux discretization schemes was carried out on a
stiff scalar transport problem and on the positive streamer case. Among
all schemes tested, ROUNDF emerged as the clear best choice, outperforming
all standard TVD limiters in both accuracy and robustness. The
Scharfetter-Gummel scheme proved highly stable but introduced excess
numerical diffusion on coarser meshes, with deviations diminishing upon
mesh refinement and confirming convergence to the correct solution. The
superiority of ROUNDF over both standard limiters and Scharfetter-Gummel
makes it the recommended scheme for streamer transport simulations.

The Poisson-transport coupling was investigated through a corrector loop
sensitivity study, revealing that even a semi-implicit Poisson treatment
does not eliminate the need for outer correction loops. For Courant numbers
below 1.0, one to two correction loops per time step were found sufficient,
though this remains application-dependent and should be verified on a
case-by-case basis. A drift-robust wall boundary condition was also
developed and validated, overcoming the P\'{e}clet-dependent failure of the
conventional mixed-boundary formulation that is commonly employed in
plasma-fluid codes.

The solver exhibited strong parallel scaling up to 6-12 cores on a single
node, beyond which memory bandwidth saturation reduced further speedup,
confirming the memory-bound nature of the finite volume algorithm. On the
static positive streamer benchmark, \texttt{SoPlasmaFoam} completed the
dense mesh case in a few hours and the coarse mesh in as little as 1.2
hours. With AMR enabled, the same case completed in 8 to 40 minutes
depending on core count, competitive with the fastest reported plasma codes
while keeping the cell count low. These results show that
\texttt{SoPlasmaFoam} is competitive with the fastest reported plasma codes on the streamer
benchmark and the methodological findings reported here provide
practical guidance for the broader plasma simulation community, paving the
way towards multidimensional, strongly coupled, multiphysical simulations
of emerging and contemporary plasma applications.

\section*{Author Contributions}

\textbf{Rention Pasolari:} Data curation, Formal analysis, Investigation, Methodology,
Software, Validation, Visualization, Writing - original draft, Writing - review \& editing.

\textbf{Konstantinos Kourtzanidis:} Conceptualization, Data curation, 
Formal analysis, Funding acquisition, Investigation, Methodology, Software, Supervision, Validation, Writing - review \& editing.

\section*{Acknowledgements}
This work was supported by the European Union Project CAIPIRINH3A, under the GA Number 101191768. Views and opinions expressed are however those of the author(s) only and do not necessarily reflect those of the European Union or CINEA. Neither the European Union nor the granting authority can be held responsible for them. The work of K. Kourtzanidis is also supported by the H.F.R.I. project CO2-SPLIT. The research project is implemented in the framework of H.F.R.I call “4th Call for H.F.R.I.’s Research Projects to Support Postdoctoral Researchers” (H.F.R.I. Project Number: 28223). Open access publishing was supported by HEAL-Link (Hellenic Academic Libraries Link).

\bibliographystyle{elsarticle-num}
\bibliography{bibliography}

\end{document}